\documentclass[final,times,11pt]{elsarticle} 
\usepackage{amssymb}
\journal{The Foundational Questions Institute}

\newcommand{\be}{\begin{equation}}
\newcommand{\ee}{\end{equation}}
\newcommand{\simless}{\lower.5ex\hbox{$\; \buildrel < \over \sim\;$}}
\newcommand{\simgreat}{\lower.5ex\hbox{$\; \buildrel > \over \sim\;$}} 
\newcommand{\starmass}{ M_\star } 
\newcommand{\mpro}{m_{\rm p}}

\newcommand{\rhov}{\rho_{\scriptscriptstyle\Lambda} }

\newcommand{\omegab}{\Omega_{\rm b}}
 
\newcommand{\omegad}{\Omega_{\rm dm}} 
\newcommand{\mplanck}{M_{\rm pl}}

\newcommand{\umass}{ m_{\rm u}}
\newcommand{\dmass}{ m_{\rm d}}

\newcommand{\emass}{ m_{\rm e}}

\newcommand{\rhoeq}{\rho_{\rm eq}}

\newcommand{\conlum}{{\mathcal{C}_{\scriptscriptstyle{\kern-0.1em\star}}}} 
\newcommand{\metal}{\raisebox{0.2ex}{--}\kern-0.55em Z}
\newcommand{\diproton}{{^2{p}}} 
\newcommand{\quni}{{q_{\infty}}} 

\begin{document}

\begin{frontmatter}

\title{{\bf The Degree of Fine-Tuning Needed 
for a Viable Universe: Not Fragile?}} 

\author{Fred C. Adams$^{1,2,3}$}

\address{$^1$Physics Department, University of Michigan, 
Ann Arbor, MI 48109, USA} 

\address{$^2$Leinweber Institute  for Theoretical Physics,
University of Michigan, Ann Arbor, MI 48109, USA} 

\address{$^3$Astronomy Department, University of Michigan, 
Ann Arbor, MI 48109, USA}

\address{\rm email: fca@umich.edu $\qquad$ ORCID: 0000-0002-8167-1767}

\begin{abstract}
In order for the universe to develop astrophysical structures and
ultimately support life, the fundamental constants that determine the
laws of physics and the cosmological parameters that specify cosmic
properties must fall within a range of values. The main goal of this
review is to delineate these ranges and highlight the physical
processes that provide the strongest constraints.  We start with the
premise that multiple universes can exist and that they can sample
different realizations of the laws of physics. Although the total
number of parameters is large, it appears that a much smaller subset
has important consequences for habitability. This treatment focuses on
the coupling constants that determine the strength of the fundamental
forces $(\alpha,\alpha_G,\alpha_{\rm s},\alpha_{\rm w})$ and the
masses of the particles $(\umass,\dmass,\emass)$ that make up atomic
matter.  In addition, we consider astrophysical/cosmological
parameters, including the energy density parameter $\Omega$, the dark
energy density $\rhov$, the baryon-to-photon ratio $\eta$, the dark
matter contribution $\delta$, the amplitude $Q$ of primordial density
fluctuations, and the number ${\cal D}$ of large spatial dimensions.
These quantities are constrained by the need for the universe to
emerge from its epoch of nucleosynthesis with an acceptable chemical
composition, live for a long time, and ultimately produce galaxies,
stars, and planets. Stellar lifetimes must be long enough and surface
temperatures must be high enough to support life.  The planets must be
massive enough to support atmospheres as well as complex biospheres.
These requirements place constraints on the fundamental constants and
cosmological parameters.  This overview discusses several classic
instances of possible fine-tuning in stars, including the triple-alpha
reaction, stable diprotons, and unstable deuterium. For each
fine-tuning issue, we estimate the range of parameter space for which
viable universes exist. Finally, we note that for universes with
significantly different parameters, a variety of astrophysical
processes can generate energy, drive nucleosynthesis, and potentially
support habitability.

\end{abstract}

\begin{keyword}
Fine-tuning \sep Multiverse \sep Fundamental Constants \sep Cosmology 
\sep Stellar Evolution \sep Nucleosynthesis \sep Habitability 
\end{keyword}

\end{frontmatter}

\vskip1.0truein

\vskip1.0truecm

{\bf Table of Contents}

\medskip 

1. Introduction \dotfill 3

\medskip 

2. Fundamental and Cosmological Parameters \dotfill 7

\medskip 

3. Omega and Cosmic Geometry \dotfill 12

\medskip 

4. Big Bang Nucleosynthesis \dotfill 15

\medskip 

5. Dark Energy and the Cosmological Constant \dotfill 19

\medskip 

6. Variations in the Fluctuation Amplitude $Q$ \dotfill 23

\medskip 

7. Galaxies and their Propertiess \dotfill 27

\medskip

8. The Existence of Stars \dotfill 29

\medskip

9. The Triple-Alpha Process \dotfill 32

\medskip

10. Stable Diprotons \dotfill 36

\medskip

11. Unstable Deuterium \dotfill 39

\medskip

12. Weak Force \dotfill 43

\medskip

13. Planets \dotfill 46

\medskip

14. Stable Atoms and Nuclei \dotfill 49

\medskip

15. Quarks \dotfill 53

\medskip

16. Number of Space-time Dimensions \dotfill 56

\medskip

17. Anthropic Arguments \dotfill 57

\medskip

18. Summary \dotfill 59

\medskip

Appendix A. Probability Distributions \dotfill 61 

\medskip

References \dotfill 65

\newpage

\bigskip 
\section{Introduction} 
\label{sec:intro}

During the past century, our understanding of astrophysics, cosmology,
and fundamental physics has developed to the point where we can
describe the evolution of our universe and its contents with accuracy
and precision. Starting from a high-energy birth-like moment, the
universe expands, cools, and eventually produces all of the structures
that are necessary to support biological operations. We are here, and
thus our universe contains observers.

During its first few minutes of existence, the cosmos forges light
nuclei, so that the first nontrival structures come into existence.
Some time later, on the largest scales, galaxies and clusters condense
out of the expanding background space-time of the universe. Their deep
gravitational potential wells gather, organize, and store the raw
materials necessary for the later development of life.  Stars form
within the galaxies, with a wide range of properties, and synthesize
almost the entire periodic table of elements. Planets form alongside
their stellar hosts and provide a wide variety of potentially
habitable environments. Through the production of these building
blocks, our universe acquires the right properties to allow for the
genesis of life. And in at least one instance, on one planet, life has
successfully emerged.

Significantly, the laws of physics have the right properties to form
and maintain all of the complex ingredients necessary for the
developement of life, including a long-lived universe, galaxies,
stars, planets, and complex nuclei. This convergence of fortune leads
us to ask: How much can the laws of physics be changed and still allow
for the universe to support life? Many authors have argued that small
changes would render the cosmos devoid of life so that the universe
must be fine-tuned (compare \cite{barnes2012,bartip,carr,davies2006,
donoghuethree,hogan,lewbarn,liviorees2018,reessix,schellekens}).
{\sl One goal of this review is to make a quantitative assessment of
this issue by finding the ranges of physical parameters that allow for
a working universe.}

Ultimately, we would like to find the constraints on the laws of
physics necessary for a given universe --- and hence a given version
of physics --- to support life. Unfortunately, at the present time, we
do not have a standard list of ingredients for life, nor a
well-defined theory that describes how life arises. In the absence of
such development, this review considers universes to be successful if
they can produce complex structures, including nuclei, planets, stars,
and galaxies. In practice, these structures, along with the universe
itself, must also be sufficiently long-lived. For purposes of this
review, these successful universes will also be described as
`habitable' or `viable'.

In order to delineate the properties required for a universe to be
viable in the sense outlined above, first step is to specify the
parameters that are allowed to vary (see Section
\ref{sec:parameters}). The next step is to determine the ranges of
those parameters that allow for a universe to successfully develop
complex structures (where this task is the focus of subsequent
sections).

Once the conditions required for a successful universe are understood,
and quantified, we would like to assess how much fine-tuning is
required for a universe to develop complexity. Regarding this issue,
previous definitions of fine-tuning come in different varieties.
First note that the colloquial meaning of `tuning' is the adjustment
of some quantity, like the frequency of a radio receiver. One type of
fine-tuning thus occurs when a parameter must be adjusted to within a
small tolerance in order for the universe to be viable.  In some
cases, relatively small changes to the parameter can lead to vastly
different outcomes. For example, small changes to the strength of the
strong force can render the deuterium nucleus unstable, which in turn
necessitates alternate paths of nucleosynthesis.\footnote{Note that
thee smallness of required changes can be measured with respect to the
observed parameter values in our universe and/or relative to a larger
range of possible values.}  Another type of fine-tuning arises when a
parameter has a value that is vastly different from its expected
value. The cosmological constant represents an example of this type of
hierarchy, as the observed value of energy scale is smaller than the
Planck scale by $\sim30$ orders of magnitude.\footnote{Here the
expected scale is assumed to be the Planck scale. Often this
discrepancy is expressed in terms of energy density, given by the
fourth power of the energy scale, so that the hierarchy becomes 120
orders of magnitude.}  Hierarchy problems are often considered as
instances of fine-tuning, even though the value of the parameter in
question need not fall within a narrow range. Tuning arises in this
situation when small values of a parameter can be attained (even when
larger values are expected) if a collection of large terms in the
equations nearly --- but not quite --- cancel. The cancellation
process can be rather sensitive to the sizes of the large terms and
thus requires tuning (for further discussion see
\cite{dine2015,hossenfelder,thooft,wellstune}). Both types of
fine-tuning -- small tolerances and large hierarchies -- arise
in this review. 

\begin{figure}[tbp]
\centering 
\includegraphics[width=0.95\textwidth,trim=0 150 0 100]{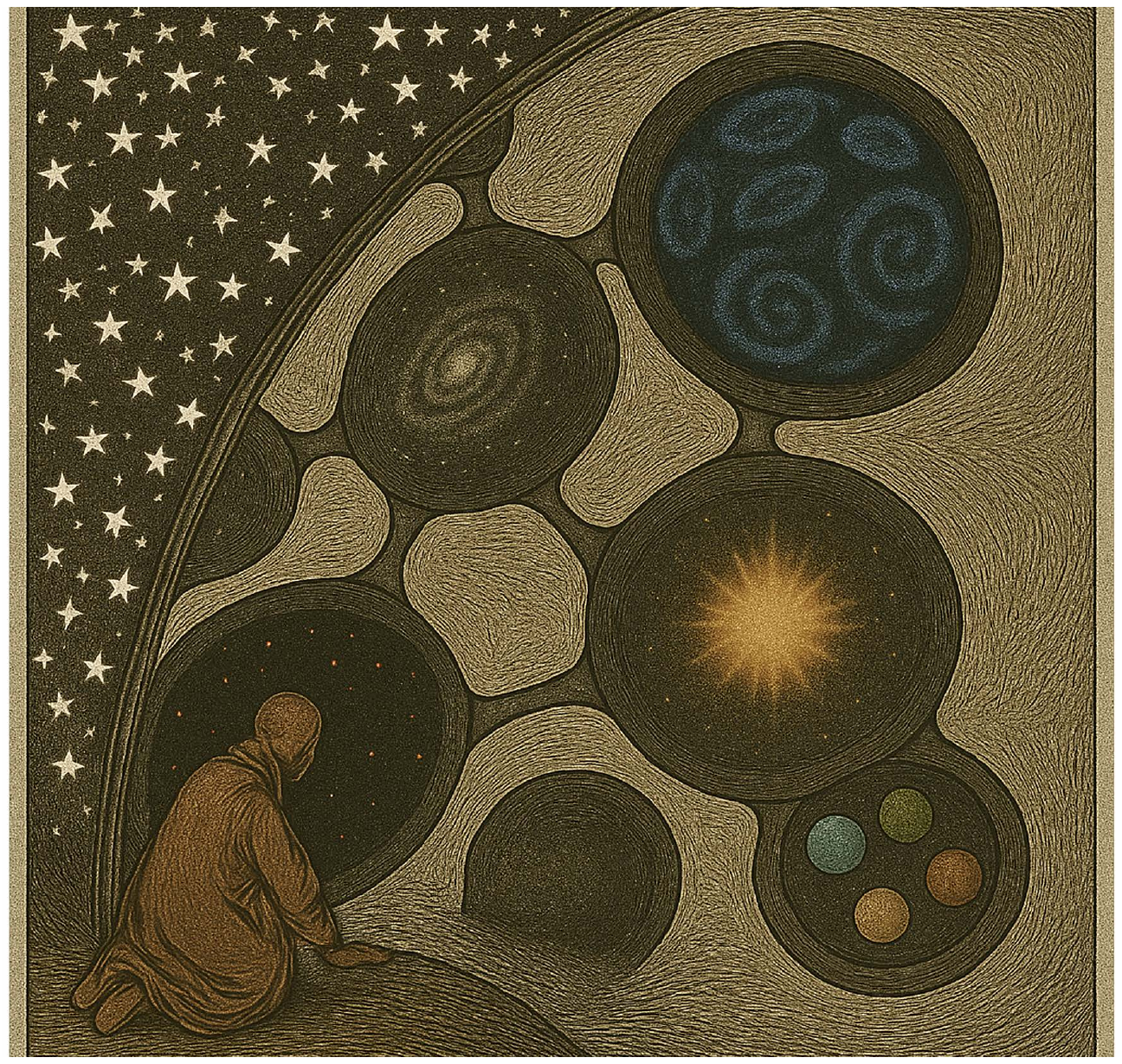} 
\vskip85pt
\caption{Diagramatic representation of a small portion of the
multiverse.  Within the larger ensemble, each individual universe
(represented as a separate bubble) could have a different realization
of the laws of physics, different values for the cosmological
parameters, and different cosmic inventories.  The number of
theoretically expected universes is vastly larger than the number
shown here. The different regions of space-time are, in general,
causally disconnected so that the observer views them only in a
conceptual sense. [This AI-generated figure will be replaced by
an artist-generated picture in the published version of the article.] }  
\label{fig:multiverse} 
\end{figure} 

\begin{figure}[tbp]
\centering 
\includegraphics[width=0.9\textwidth,trim=0 150 0 150]{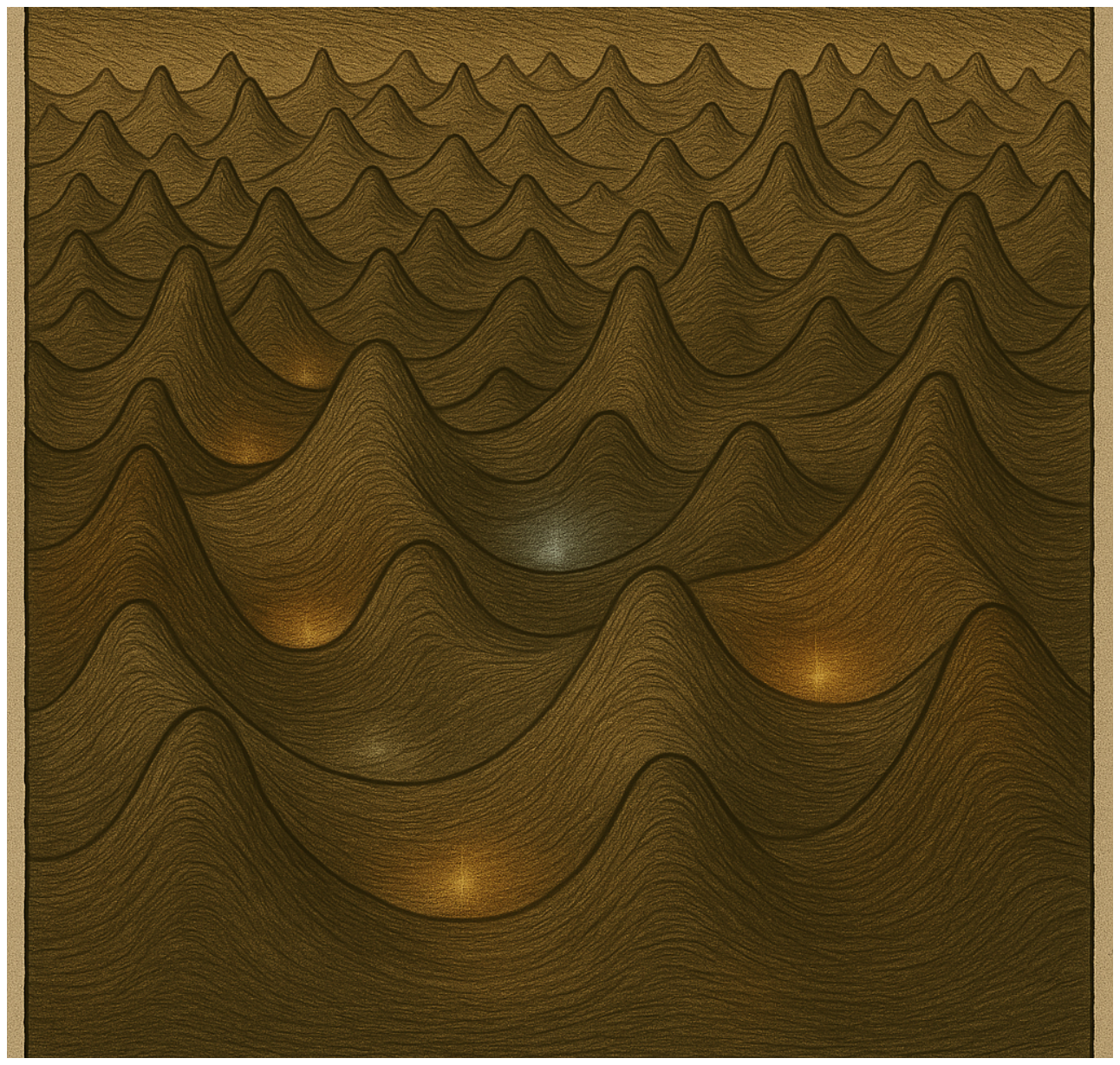} 
\vskip100pt
\caption{Diagramatic representation of the landscape of possible
vacuum states of the universe. As the universe evolves from its
high energy beginnings to lower temperatures, the vacuum state
settles into a local minimum of its potential. In principle, each
local minimum could result in a particular realization of the
fundamental constants of the corresponding region of space-time,
so that each universe could have its own version of physical law.
[This AI-generated figure will be replaced by an artist-generated
  picture in the published version of the article.] } 
\label{fig:landscape} 
\end{figure}

With the allowed ranges of the fundamental constants specified, a
logical next step would be to estimate the fraction of all possible
universes that could develop structure and contain observers, i.e.,
estimate the probability that a universe could be habitable. In order
to make such assessments, however, one would need to know not only the
values of the constants that allow for working universes, but also the
underlying probability distributions from which the parameters are
selected. Unfortunately, at the present time, these underlying
probability distributions remain unknown 
\cite{donoghuethree,hartlesrednicki,hossenfelder}. This issue is
complicated by several factors. First, the form of the probability
distributions for sampling the the parmeters makes an enormous
difference in the estimates for the probability of a working
universe. Second, the task of determining the underlying probability
distributions rests on a different footing than that of finding the
allowed ranges of parameters. To determine the allowed range of the
gravitational constant $G$ that allows for working stars, for example,
one constructs stellar models by solving the equations of stellar
structure with different values of $G$. The relevant equations are
known and vetted.  Constrast this situation with the question of
finding the underlying probabiltiy distribution from which $G$ is
chosen. Not only is the distribution unknown, we have no theoretical
framework from which to attempt to estimate the distribution.  Despite
these difficulties, the issue of probability distributions is
important, and is discussed further in the Appendix.

The region of space-time that makes up our universe need not be a
unique occurrence.  Within the context of current cosmological
theories, our universe could represent one member of a vast assemblage
of universes that make up a much larger entity known as the
``multiverse'' \cite{carrellis,davies2004,deutsch,donoghuethree,
ellis2004,garriga2008,hallnomura,lindemultiverse,reesbefore}.
Figure \ref{fig:multiverse} shows a schematic representation of this
possibility: We can contemplate a vast collection of different
universes, each with a different realization of the laws of physics,
where each region develops its own inventory of cosmic
contents.

One way for different universes to attain different laws of physics is
motivated by recent developments in string theory and its
generalizations. The vacuum structure of the underlying theory could
contain an an enormous number of minima. As the universe evolves from
its energetic beginning to lower temperatures, it settles into one of
the many vacuum states, which defines its version of physical law
\cite{boussopolchinski,halverson,hogan2006,kachru,lindevanchurin,
schellekens,susskind}. Figure \ref{fig:landscape} presents a schematic
representation of the potential energy function for this type of
configuration space, where each minimum could lead to different
features for the low-energy state of the universe. The following
picture emerges: Each distinct region of space-time (one universe
within the multiverse) samples the distribution of possible vacuum
states by settling into a particular minimum of the potential. The
vacuum state determines, in part, the laws of physics, which would
vary from region to region within the entire ensemble of regions that
make up the multiverse. Note that this scenario remains controversial
and that the landscape of configuration space described here is still
under study \cite{banks2012,banksdinegorb,bena,gleiser,schellekens2008}.

In the formulation considered here, our particular universe is only
one portion of a much larger space-time manifold, and our region is
governed by one particular realization of many possible versions of
physical law.  Other regions could have different vacuum states,
resulting in elementary particles with different masses, forces with
different coupling strengths, as well as different cosmological
properties. These possible variations are reviewed next. The remainder
of this manuscript discusses and constrains the range of possible
values that these parameters can have and allow for a given universe
to develop complex structures, including nuclei, planets, stars, and
galaxies (see \cite{adamsreview} for a more in-depth review of these
issues). Finally, the question of probability distributions is taken
up in the Appendix.

Note that it makes sense to contemplate variations in the fundamental
constants of nature --- even without invoking the landscape as a
mechanism to specify their values and whether or not the other
universes actually exist. If the different regions of space-time do in
fact sample a distribution of fundamental parameters, then our
considerations would apply to the multiverse. But even if other
regions of space-times are purely hypothetical, or if they all must
have the same constants, the thought experiment that considers
parameter variations still provides us with important information
about how our universe operates.

\bigskip 
\section{Fundamental and Cosmological Parameters} 
\label{sec:parameters}

One premise of fine-tuning arguments is that the laws of physics could
have different forms, with different realizations in distinct regions
of space-time that we consider to be separate universes. In this
context we need to specify {\it how} the laws of physics could vary
from universe to universe. To fully address this issue, one would have
to recalculate the formation and evolution of the universe, stars,
planets, atoms, and nuclei for all possible versions of physical law.
Given the difficulty of this undertaking, the usual approach is to
consider the general form of the laws of physics to remain the same
(e.g., all universes support some form of Maxwell's equations) but
allow the constants or parameters to assume different values (e.g.,
the fine-structure constant $\alpha$ could have different values).
Adopting this less ambitious approach, we still need to specify what
parameters are allowed to vary across the multiverse.

The parameters that determine the laws of physics can be conceptually
divided into two categories. From the point of view of fundamental
physics, the starting point is the Standard Model (SM) of Particle
Physics \cite{gaillard,kanebook}.  This highly successful paradigm is
described by a Lagrangian with 19 parameters in the limit where
neutrinos are massless \cite{hogan}, and the number of parameters
increases to 26 with the inclusion of neutrino physics \cite{tegmark}
(see also \cite{cahn}).  Cosmology has its own standard model, one
that is becoming more solidified with increasingly precise data, and
is described by $\sim10$ additional parameters
(e.g., \cite{huterer,liddle2004, lahavliddle}). Here we review the
parameters from both particle physics and cosmology, and identify the
subset that play the most important role in the possible fine-tuning
of our universe -- and others.

The list of input SM parameters can be summarized as follows. The
first nine parameters are provided by the masses of the six quark and
three leptons, or equivalently the Yukawa coupling constants that
specify their masses.  The Higgs mechanism accounts for the masses of
particles and requires (at least) two parameters, a mass and a vacuum
expectation value. The quark mixing matrix, which determines the
strength of flavor-changing interations involving the quarks, is
specified by three mixing angles and a phase, and thereby contributes
another four parameters. The neutrino sector requires three additional
Yukawa coupling constants to specify the masses, three mixing angles
for neutrino interactions, and an additional phase (seven more
parameters). Finally, the model includes a phase angle for the QCD
vacuum and three coupling constants for the gauge group $U(1)\times
SU(2) \times SU(3)$. These numbers thus sum to 26. Even with this
abundance of parameters, the Standard Model remains incomplete, as it
does not include dark matter or gravity.

The parameters that appear in the Standard Model Lagrangian are not
necessarily in one-to-one correspondence with their counterparts that
appear in the equations of structure and equations of motion for
astrophysical objects. The three gauge group constants can be written
in terms of the strong and weak coupling constants $\{g_{\rm s},g_{\rm
w},\theta_W\}$, where $\theta_W$ is the Weinberg angle. These
quantities must be evaluated at a particular energy scale (as they are
energy dependent). Including the Higgs vacuum expectation value
${\cal V}$, the masses of the intermediate vector bosons can be
written in the form $m_W = {\cal V} g_{\rm w}/2$ and $m_Z = {\cal V}
g_{\rm w}/(2 \cos\theta_W)$.  The electromagnetic coupling constant is
then given by $e=g_{\rm w}\sin\theta_W$ (which is evaluated at the
scale $m_Z$).  The strength of the electromagnetic interaction is thus
controlled by the parameter $\alpha = e^2/4\pi \approx1/128$, where
the value is that appropriate for the weak energy scale. In the zero
energy limit, the fine-structure constant approaches the familiar
value $\alpha\approx1/137$. The strength of the weak interaction and
the strong interaction are then given by $\alpha_{\rm w} = g_{\rm
w}^2/4\pi$ and $\alpha_{\rm s} = g_{\rm s}^2/4\pi$. Finally, the Fermi
constant takes the form $G_F = 1/(\sqrt{2}{\cal V}^2)$ =
(293GeV)$^{-2}$.

Although all of the parameters could be important in principle, some
play larger roles in practice. For the sake of definiteness, this
present discussion focuses on only seven of the fundamental parameters
outlined above. The strengths of the four forces are clearly
important. Although it does not appear in the SM, gravity must be
included and its strength is specified by the structure constant 
\be
\alpha_G \equiv G m_{\rm p}^2 = {m_{\rm p}^2 \over \mplanck^2} 
\approx 6 \times 10^{-39} \,,
\ee
in units where $\hbar=1=c$ and $\mplanck$ is the Planck mass. 
Since the heavier particles tend to decay into the lighter ones, the
first generation of particles becomes more important.\footnote{For
completeness, note that changes in the properties of the first
generation particles are likely to be accompanied by changes to the
other generations, which in turn could have an impact.}  We thus consider
the masses of the lightest quarks $\{\umass,\dmass\}$ and the mass of
the electron $m_e$. Following tradition, the electron mass is 
specified here by its ratio relative to the proton, 
\be
\beta \equiv {m_{\rm e} \over m_{\rm p}} \approx {1 \over 1836} \,, 
\ee
where the numerical value corresponds to our universe.\footnote{Note
that the parameter $\beta$ depends on the mass of the proton, which
in turn depends on the masses of the up and down quarks, as well as
the strong and electromagnetic coupling constants. As a result,
when $\beta$ changes, it could encompass changes in these other
parameters. }
With these choices, the relevant parameters from fundamental physics
that vary from universe to universe are reduced to the set 
\be
\Big\{ \alpha_G, \alpha_{\rm w}, \alpha, \alpha_{\rm s}, \beta,
\umass, \dmass \Big\} \,.
\label{smparameter} 
\ee

\begin{table} 
\centerline{\bf Fundamental and Cosmological Parameters} 
\medskip
\centering
\def\arraystretch{1.15} 
\begin{tabular}{lcc}
\\
\hline 
\hline 
Quantity & Symbol & Observed value \\
\hline 
\hline 
Up quark mass & $\umass$ & 2.3 MeV \\ 
Down quark mass & $\dmass$ & 4.8 MeV \\ 
Electron-to-proton mass ratio & $\beta$ & 1/1836\\  
\hline 
Gravitational constant & $\alpha_G$ & $6\times10^{-39}$ \\ 
Weak coupling constant & $\alpha_{\rm w}$ & $10^{-5}$ \\ 
Fine-structure constant & $\alpha$ & 1/137 \\
Strong coupling constant & $\alpha_{\rm s}$ & 15 \\ 
\hline 
Fluctuation amplitude & $Q$ & $10^{-5}$ \\ 
Baryon-to-photon ratio & $\eta$ & $6\times10^{-10}$\\
Dark matter abundance & $\delta$ & $3\times10^{-9}$\\
Vacuum energy scale & $\lambda$ & 0.003 eV \\
Cosmic density parameter$^\dagger$ & $\Omega$ & 1 \\
\hline
Number of large spatial dimenions & ${\cal D}$ & 3 \\ 
\hline 
\hline 
\end{tabular}
\caption{Working subset of the parameters that are allowed to vary
from universe to universe. The list includes only the parameters that
are considered in this review. The third column lists the values that
are observed in our universe. \hskip4.0truecm 
$^\dagger$Note that although the value of the density parameter
can vary, most of this review only considers universes where
$\Omega\rightsquigarrow1$ (perhaps due to inflation or its counterpart). 
Notice also that habitability requires a long-lived universe,
which is facilitated by the required value of $\Omega$.}
\label{table:parameters} 
\end{table}

In the standard model of cosmology, the properties of our universe can
be described by a collection of quantities \cite{huterer,kolbturner},
summarized as follows: One important characteristic is the spatial
curvative, which is specified by the constant $k$ (see
Section \ref{sec:omega}). In our universe the geometry is observed to
be (nearly) spatially flat \cite{planck2018}, so that $k\approx0$ and
the total energy density of the universe corresponds to $\Omega=1$
(cf. \cite{divalentino}). As argued below, we expect most successful
universes to have similar properties, and thus we consider universes
with $\Omega=1$ for most of this review. One way to realize this value
is for a universe to be driven to this state by some dynamical
mechanism (see Section \ref{sec:omega} for further discussion).

With the total energy density determined, the contributions of the
various components must be specified. These components include the
energy density in the vacuum, the density of dark matter, the density
of baryonic matter, and the density in radiation. The different types
of energy redshift differently as the universe expands. To leading
order, however, the ratio $\eta$ of the number of baryons to photons
remains constant. As a result, the baryonic component is specified by
$\eta$, which has the value $\eta\approx6\times10^{-10}$ in our
universe. One can define an analogous parameter $\delta$ that
specifies the contribution of dark matter.  In our universe, the ratio
of dark matter to baryonic matter is about 6:1 by mass, although the
nature of the dark matter remains unknown
\cite{baer,feng,jungman,steffen}.  As another cosmic component, the
radiation energy density is given by the current temperature of the
cosmic microwave background radiation, with $T_{\rm cmb} \approx 2.7$
K. This value corresponds to an energy contribution of only
$\Omega_R \sim 10^{-4}$ at the present epoch. To complete the cosmic
inventory, we must include the contribution from the vacuum or
equivalently the dark energy component. Although ths vacuum energy
need not be constant in time, the simplest approach is to characterize
it by a single energy density or energy scale, which has the value
$\lambda\approx0.003$ eV in our universe at the present
epoch \cite{frieman,riess}.

\begin{figure}[tbp]
\centering 
\includegraphics[width=0.80\textwidth,trim=0 150 0 150]{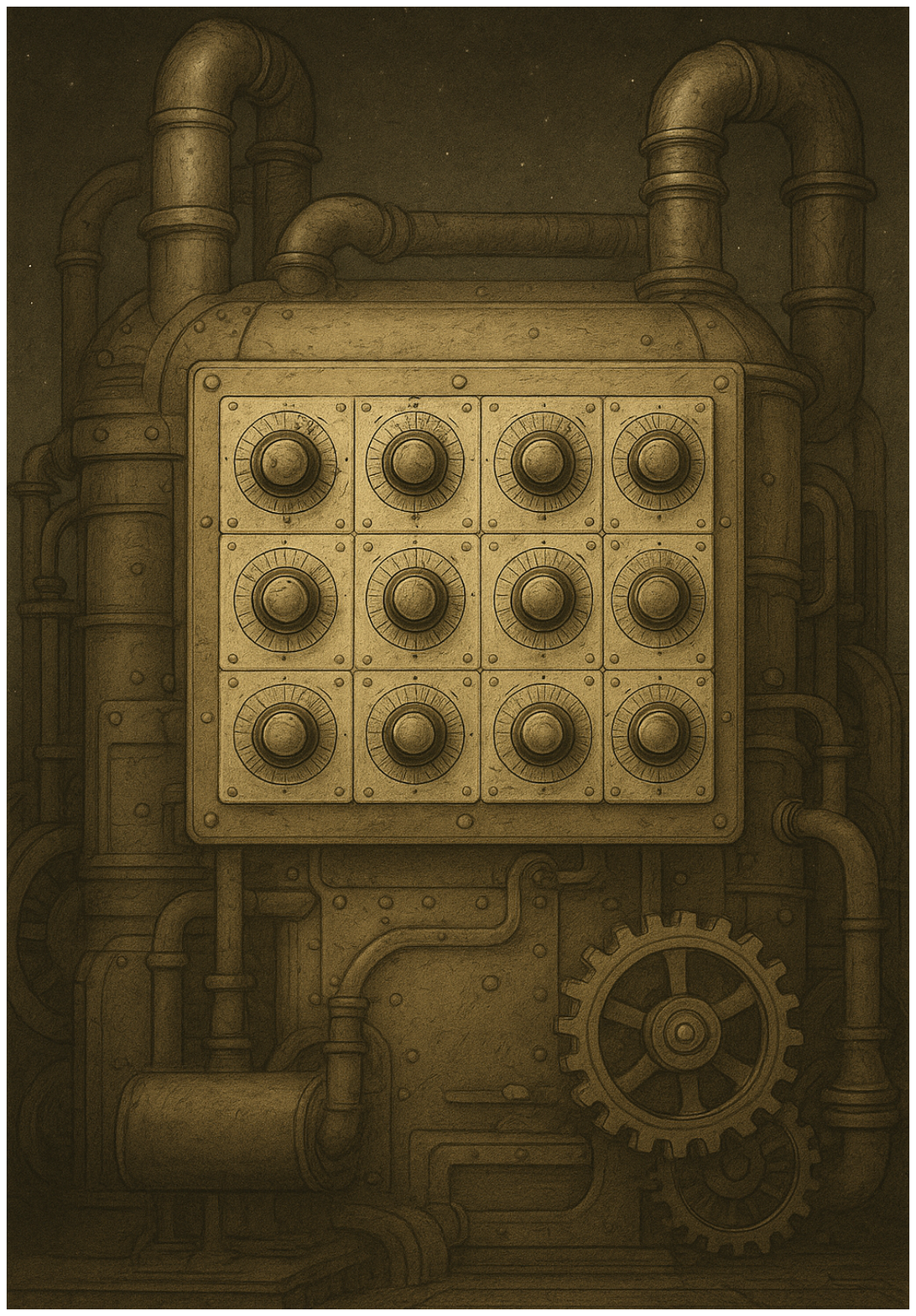} 
\vskip80pt
\caption{The universe has at least a dozen parameters that determine
its ability to live for a long time, and form complex structures such
as galaxies, stars, planets, and heavy nuclei. One can imagine that
each universe inherits values for each of the relevant constants
during its epoch of formation, as depicted here by the settings of the
dials. A given universe can become habitable --- in that it can produce 
galaxies, stars, planets, and nuclei --- when the dials are set within
the proper ranges. Keep in mind that the parameter values could be
correlated so that the dial settings are not necessarily independent.
[This AI-generated figure will be replaced by an artist-generated
  picture in the published version of the article.] } 
\label{fig:machine} 
\end{figure}  

In standard cosmology, the current value of the Hubble parameter plays
an important role as it sets the current cosmic age. For
considerations of other universes, however, we are not concerned with
the formation of structure at any particular epoch, only whether or
not such structures can form at any epoch. For completeness, however,
we define the Hubble parameter $H={\dot a}/a$, where $a(t)$ is the
scale factor of the universe. We denote its present-day value as
$H_0$, where $H_0 \approx 70$ km s$^{-1}$ Mpc$^{-1}$.

Although the universe is mostly homogeneous and isotropic, some
departures from this ideal state are present. Moreover, some
departures are necessary in order for the universe to generate large
scale structures, including galaxies, which in turn allow for the
formation of stars and planets. The seeds of structure formation are
described by a spectrum of density fluctuations, which are observed to
have a nearly scale-invariant spectrum, i.e., their amplitude does not
vary appreciably with the length scale of the perturbation. In
addition to `ordinary' density fluctuations, the universe can also
support fluctuations in energy density due to gravitational waves,
which are often called tensor modes (with the density fluctuations
described earlier being scalar modes). These tensor modes in our
universe are expected to also have a nearly scale-invariant spectrum,
but their amplitude is subdominant. As a result, a full specification
of the fluctuations requires a spectral shape and an amplitude for
both the density and the tensor modes. In practice, however, we can
characterize the fluctuations with a single amplitude corresponding to
that of the dominant contribution.\footnote{Note that structure
formation can occur with a range of different distributions for the
density fluctuations \cite{bardeenbig,blumenthal}.} In our universe,
this amplitude has the value $Q\approx10^{-5}$. As a result of these
density fluctuations, the temperature of the cosmic microcwave
background has corresponding fluctuations
$(\Delta T)/T={\cal O}(Q)\sim10^{-5}$, which can be observed
\cite{planck2014,planck2016,copi,muir,cobe,wmap}. With this
simplification, the minimal collection of cosmological parameters
that must be considered is given by the set  
\be
\Bigl\{ \Omega, \eta, \delta, \lambda, Q \Bigr\} \,.
\label{cosparameter} 
\ee
Although it rests on a somewhat different footing, the number
$\cal D$ of large spatial dimensions is sometimes considered as
another parameter that can vary from universe to universe 
\cite{barrow1983,ehrenfest1,reessix,tegmarkdimension}. In our
universe, $\cal D$ = 3, but theories of quantum gravity often require
more spatial dimensions, some of which can be macroscopic (thereby
allowing ${\cal D}\ne3$). 

At the present time, we cannot calculate these cosmological quantities
from {\it a priori} considerations. The hope is that further
theoretical developments will provide more clarity. For example, the
theory of the inflationary universe generically results in
$\Omega\rightsquigarrow1$ and some value of $Q$
\cite{albrechtstein,guthpi,guth2000,guth2007,linde1983,lindeinflate}. 
Our goal is to understand the physics at the relevant high energy
scale (at about the GUT scale) well enough to make inflation (or its
counterpart) fully predictive. Processes that violate conservation of
baryon number could determine the baryon-to-photon ratio $\eta$, and
extensions of the Standard Model to include a dark matter sector could
specify the dark matter contribution $\delta$. Determining the dark
energy contribution (and hence specifying $\lambda$) requires a
solution to the cosmological constant problem (see Section
\ref{sec:lambda}).  Although these cosmological quantities might
someday be derived from more fundamental considerations, their
specification requires physics beyond the Standard Model (e.g., baryon
number violation \cite{langacker81, nath2007}, the dark matter
sector \cite{feng,jungman}, and the cosmological constant
\cite{weinberg89}). This additional physics will be described
by additional parameters appearing in the SM. Until such extensions
are developed, we can use the parameter set given by equation
(\ref{cosparameter}).

The subsequent sections of this review discuss how variations in the
parameters outlined above lead to changes in the universe, galaxies,
stars, planets, and nuclei. A full accounting of the available
parameter space represents an enormous task. As a result, most of the
following discussion considers the variation of only one or two
parameters at a time and estimates the ranges needed for a universe to
be viable. One should keep in mind that all of the parameters can
vary. On the other hand, the variations in multiple parameters could
be correlated, as suggested by many scenarios for physics beyond the
Standard Model.  Finally, we note that in the considerations of
fine-tuning, it is the dimensionless constants that matter, rather
than their numerical values in any particular system of units
(although we discuss quark masses in MeV for convenience, instead of
the corrsponding Yukawa coupling constants).

\medskip 
{\bf Bottom line:} {\sl Although we don't know all the ways in which 
the laws of physics can vary from universe to universe, approximately
one dozen parameters control the formation and evolution of structure
(as summarized in Table \ref{table:parameters} and depicted in
Figure \ref{fig:machine}).}

\bigskip 
\section{Omega and Cosmic Geometry} 
\label{sec:omega} 

Our universe is remarkably simple in some respects.  In particular,
the universe is nearly homogenous and isotropic on large scales.
Given these advantageous features, the expansion of the universe can
be described by the Friedmann equation, 
\be
\left({ {\dot a} \over a} \right)^2 = {8\pi G \over 3}
\rho - {k \over a^2} \qquad {\rm where} \qquad
\rho = \rho_{\rm R0} a^{-4} + \rho_{\rm M0} a^{-3} +
\rho_{\Lambda0} a^{-3(1+w)} \,,
\ee
where $a(t)$ is the scale factor, $\rho$ is the total energy density,
and $k$ is a constant that specifies the curvature (where by
convention $k=0,\pm1$ for universes that are flat, positively curved,
and negatively curved; note that we have taken $c=1$ ). The second
equation specifies the dependence of the constituent energy densities
on the scale factor, where $w=p_\Lambda/\rho_\Lambda$ is an equation
of state parameter for dark energy.  It is highly covenient that the
expansion history of the universe can be described by such a simple
differential equation. However, this equation poses one of the classic
problems of fine-tuning in cosmology -- the issue generally known as
the flatness problem.\footnote{For completness, we note that other
tensions exist in the standard cosmological model, such as
discrepancies in the measured value of the Hubble constant
\cite{kamionkowski}.}

To illustrate this issue, we start with the standard definitions,
\be
H \equiv { {\dot a}\over a} \,, \qquad
\rho_{\rm crit} \equiv {3H^2 \over 8\pi G} \,, \qquad
{\rm and} \qquad
\Omega_j \equiv {\rho_j \over \rho_{\rm crit}} \,,
\ee
where the index $j$ labels the contribution of the various types
of energy to the universe including matter, radiation, and
vacuum energy. The total density parameter is then given by
\be
\Omega = \Omega_R + \Omega_{\rm b} + \Omega_{\rm dm} +
\Omega_\Lambda \,,
\ee
where we have divided the matter contribution into baryons and dark
matter. The Friedmann equation for the evolution of the scale factor
can be rewritten in the form 
\be
\Omega (t) = {1 \over 1 - \chi(t)} \qquad {\rm where} \qquad
\chi(t) \equiv {3k\over8\pi G\rho a^2} \,. 
\ee
Note that in general the density parameter $\Omega$ is a function of
cosmic time, or equivalently a function of the scale factor. For a
radiation dominated universe, the energy density $\rho\sim a^{-4}$ so
that $\chi\sim a^2$; for a matter dominated universe, $\rho \sim
a^{-3}$ so that $\chi \sim a$. In either case, the parameter $\chi$
grows as the universe expands. For $k>0$ ($k<0$) this behavior results
in the density parameter $\Omega$ growing (decreasing) rapidly. For
the intermediate case with $k=0$, corresponding to a spatial flat
geometry, the density parameter does not evolve and takes the value
$\Omega=1$. Our own particular universe has a spatial geoemtry that
is close to flat, with $\Omega\approx1$ \cite{planck2018}. 

The time dependent behavior of the parameter $\chi$ poses a
problem. For the case of postive curvature, with $k>0$, the density
parameter $\Omega$ of the universe increases with time and eventually
instigates a reversal so that the universe recollapses. For the case
of negative curvature, with $k<0$, the density parameter decreases and
leads to an empty universe. Both scenarios are problematic for
habitability. The closed and recollapsing universe does not
necessarily allow for sufficient time for life to develop. The open
and empty universe suppresses the formation of large scale cosmic
structure. Only the intemediate case, a spatially flat universe with
$k=0$, allows for the density parameter $\Omega$ to remain nearly
constant.

The above discussion argues that universes favorable for the
development of life have $\Omega \sim 1 \sim constant$, which in turn
requires that the parameter $\chi$ remain small. But if $\chi$ is
small at later epochs in cosmic history when structure formation
occurs, then $\chi$ must be even smaller at earlier times. To leading
order, for the limit where $\chi\ll1$, the departure of the density
parameter from unity is given by 
\be
\big|\Omega - 1\big| = {\cal O}(\chi) \,.
\ee
In order for our universe to have $\Omega\sim1$ at the present epoch,
as observed, then the value of $\chi$ at the Planck epoch would have
to be $\chi(t_{\rm pl})\sim10^{-60}$. In other words, the value of the
density parameter at the beginning of our cosmic timeline must be
unity to a precision of about 1 part in $10^{60}$. This level of
tuning represents the flatness problem. In the absence of a dynamical
mechanism to drive $\Omega$ toward unity, equivalently to drive $\chi$
to small values, the universe would need rather special initial
conditions to evolve into its present-day form.

The inflationary universe paradigm provides a dynamical mechanism that
can alleviate the flatness problem. As outlined above, the problem is
that we need the parmaeter $\chi$ to have a small value, say
$\chi\sim10^{-60}$, at the early cosmic times.  Since
$\chi \propto \rho^{-1} a^{-2}$, the energy density of the universe
must decrease sufficiently slowly with increasing scale factor $a$ in
order for $\chi$ to decrease. Although this condition can be achieved
within a large number of scenarios, the simplest version allows the
energy density of the universe to be dominated by vacuum energy so that
$\rho$ = $\rho_\Lambda \sim contant$. In this case, the parameter
$\chi\propto{a}^{-2}$ so that it {\it decreases} rapidly with
increasing scale factor. In order to reduce the value of $\chi$ by
60 orders of magnitude, the inflationary epoch must allow for the
universe to be dominated by vacuum energy while it expands by about
30 orders of magnitude. This condition is usually written in terms of
the required number $N$ of e-foldings of the inflationary epoch, so
that the constraint takes the form 
\be
{\rm e}^{2N} \simgreat 10^{60} \qquad \Rightarrow \qquad
N \simgreat 30 \ln(10) \approx 70\,.  
\ee
This estimate is approximate. The number of e-foldings depends on
the energy scale at which inflation takes place and other properties,
although most theories require $N \sim 60 - 70$ \cite{kolbturner}. 

Although an early epoch of inflationary expansion effectively drives
the universe toward flatness, with $\Omega\approx1$, the paradigm
poses new issues. The universe must enter into the inflationary epoch
when the energy is dominated by the vacuum and then it must
transition into a radiation dominated state. Although many scenarios
have been put forth to show how our universe enters inflation, exits
inflation, and then reheats the universe by filling it with radiation,
no single model is universally agreed upon. A vast literature explores
the issues that any universe must face to successfully execute an
inflationary epoch \cite{brandenberger,carroll2014,carroll2010,corichi,
gibbons,hawkingpage,hollandswald,kaloper,schiffrin,steinhardt2011,turok2002}.
The estimated prospects for success vary widely, from expected
\cite{guth2013} to exponentially unlikely \cite{ijjas}. Finally, 
some authors argue that flatness is not a problem \cite{helbig2012}.

For completeness, we note that the inflationary epoch affects other
cosmic properties. Our own universe is observed to be isotropic.
Specifically, the temperature of the cosmic microwave background has
the same temperature to a precision of $\sim1$ part in $10^5$, as
obsereved from all directions. In the absence of inflation, portions
of the universe that are in opposite sides of the sky were not
previously in causal contact --- the horizon of the universe is just
now, at the present epoch, large enough to encompass regions that are
so widely separated. As a result, the observed isotropy represents a
potential probem --- generally known as the horizon problem. It turns
out that the early epoch of inflationary expansion alleviates this
problem as well: The same rapid expansion allows for regions that were
in causal contact before inflation to be blown up to enormous sizes
so that they are just now re-entering the horizon. The amount of
inflation required to account for the flatness problem is comparable
to the amount of inflation that leads to an isotropic universe. As a
result, universes that live for a long time tend to be spatially flat
and be close to isotropic \cite{baumann,kolbturner}. Note that the
requirement of a long cosmic lifetime (flatness) is more necessary
than isotropy for habitability, but the two features tend to go
together (at least in the context of inflationary solutions).
Inflation also dilutes the number density of unwanted relics such a
magnetic monopoles \cite{thooft1974,preskill}, where this issue was
one of the original motivations for the model \cite{guth1981} (see
\cite{kolbturner} for a textbook discussion). Yet another consequence
of an inflationary epoch is that quantum fluctuations can be imprinted
on the otherwise smooth background \cite{bardeenbig,bardeen,guthpi}.
These fluctuations have important ramifications for structure
formation and are considered in Section \ref{sec:qvary}.

\medskip 
{\bf Bottom line:} {\sl In order for universes to live long enough to
become habitable, some process --- either inflation or its replacement
--- must drive the universe to become sufficiently flat and isotropic.}

\bigskip 
\section{Big Bang Nucleosynthesis} 
\label{sec:bbn}

The epoch of Big Bang Nucleosynthesis (BBN) processes about one fourth
of the baryonic mass in our universe into light nuclei. In addition to
its transformative effect on the cosmic inventory, this epoch provides
a crucial test --- and an important point of confirmation --- for
theories of cosmic evolution \cite{coctwo,kawanocode,wagoner,walker}.
In order for a universe to become habitable, it must successfully
produce complex nuclei through some mechanism. Since stars provide an
efficient pathway for the production of heavy elements, nuclear
processing through BBN is not necessary for a universe to become
habitable. On the other hand, BBN can render a universe lifeless, in
principle, by processing all of the protons into heavier nuclei,
thereby leaving no hydrogen for water or stellar fuel at later
times.\footnote{For completeness, note that nuclei could be broken
apart (after BBN) by a number of different astrophysical processes,
although this issue has not received much attention.}  Unfortunately,
the minimum abundance of hydrogen necessary for a universe to remain
viable is not known. Fortunately, however, BBN tends to be
inefficient, so that unprocessed protons are left behind except under
the most extreme circumstances. This section briefly reviews different
BBN scenarios and shows that a broad range of parameter space allows
the universe to leave behind an ample supply of protons.

Before considering alternate scenarios for BBN, it is useful to
review some of the basic features that operate in our universe
\cite{huterer,kolbturner}. The epoch of BBN starts when the cosmic
age $t\sim1$ sec and the background temperature $T\sim1$ MeV. This
crossover point corresponds to two important milestones: The weak
interactions freeze out, so that the conversion rate $\Gamma_w$
between protons and neutrons becomes slower than the cosmic expansion
rate $H$.  In our universe, the temperature at this point of equality,
$\Gamma_w=H$, takes the approximate form 
\be
T_f \sim G_F^{-2/3} \mplanck^{-1/3} \,, 
\ee
where $G_F$ is the Fermi constant and $\mplanck$ is the Planck mass. 
In addition, the energy at this epoch becomes low enough that nuclei
can survive if they are able to form. Note that the freezing of weak
interactions results from the competition between the weak force and
gravity, whereas the binding energy of nuclei results from competition
between the strong force and electromagnetism. As a result, all four
forces play a role in defining the start of the BBN epoch and its
subsequent evolution.

Before the weak interactions freeze out, the universe is able
to maintain Nuclear Statistical Equilbrium (NSE), which implies that
the ratio of protons to neutrons tracks its equilibrium value, 
\be
{n \over p} = \exp\left[ - { {\Delta m} \over T} \right] \,, 
\ee
where $\Delta{m}=m_n-m_p\approx1.29$ MeV is the mass difference
between neutrons and protons. At the time of freeze out, the ratio
$n/p\sim1/5$, equivalently $n/(n+p)\sim1/6$, so that protons
significantly outnumber the neutrons.

With the above preliminaries in place, the basic outcome of BBN can be
understood: The $^4$He nucleus is tightly bound, and has a low
atomic number, so that the natural outcome is to process as much
material into helium as possible. The neutrons readily interact with
protons, but the protons repel each other, so the net result is to
lock up essentially all of the neutrons into $^4$He. The resulting 
estimate for the mass fraction of helium is then given by 
\be
Y_4 = {2 (n/p)_{\rm bbn} \over 1 + (n/p)_{\rm bbn}} \,. 
\ee
The first guess would be to use $n/p$ = 1/5, the value at freeze-out,
so that $Y_4\approx1/3$. In practice, however, the production of
helium takes some time, namely a few minutes,\footnote{This time
delay, sometimes called the deuterium bottleneck \cite{kolbturner},
arises because of the small baryon-to-photon ratio. Notice also that
$n/p$ is a continuous function of time, so that using particular
values is approximate. }  so that additional neutrons decay into
protons before the above formula becomes applicable. At this somewhat
later time, $n/p\sim1/7$, so that the estimate becomes
$Y_4\approx1/4$, which is close the values calculated from detailed
numerical treatments. In contrast to this substantial amount of
$^4$He, only trace amounts of deuterium, lithium, and other elements
are produced. These additional nuclei are important for the study of
the early universe, but have relatively little bearing on the
potential habitability of other universes.

Note that the mass difference $\Delta{m}$, the freeze-out temperature
$T_f$, and the binding energy of light nuclei $E_b$ are all of order 1
MeV, a scale comparable (in order of magnitude) to the mass $m_e$ of
the electron. This coincidence allows for interesting BBN to take
place, i.e., for the early universe to produce light elements, but
leave behind unprocessed protons. If the binding energies were much
lower, for example, then all of the neutrons would decay before the
universe is cool enough to allow for long-lived nuclei. If the binding
energy were much higher, then it would be energetically favorable for
the universe to fuse all of its protons into larger nuclei.
Additional variations are discussed below.

The abundance of baryons is determined by the baryon-to-photon ratio,
which has a value $\eta\approx6\times10^{-10}$ in our universe.  The
standard cosmological practice is to vary the value of $\eta$, which
produces different yields for the abundances of the light elements. By
comparing the theoretical yields with observational estimates, one can
determine the value of $\eta$ as quoted above \cite{burst,burst2,
kawanocode,kolbturner}. The yield of $^4$He increases with increasing
$\eta$, whereas the abundance of deuterium decreases. The larger
values of $\eta$ allow for the reaction rate to increase, so that the
universe efficiently locks up all of the neutrons into helium, with
$Y_4\to1/3$. The deuterium abundance decreases with increasing $\eta$
as it becomes converted into helium.  Even with the increased
interaction rate, however, the helium nuclei cannot combine to make
larger elements, which only have trace abundances.  Going the other
way, if one considers smaller values of $\eta$, the reaction rate
decreases, and less helium is produced. The abundance of deuterium
increases, but only because the $^2$H nuclei are not processed into
helium.

Next one can consider the effects of increasing the gravitational
constant $G$ \cite{kawanocode}, which determines (in part) the
expansion rate of the background universe ($H\propto G^{1/2}$). As the
value of $G$ increases, the expansion rate increases, and the freezing
out of the weak interactions occurs earlier. As a result, more
neutrons remain in play, and the yield of helium increases. An
increased expansion rate also causes the nuclear reactions themselves
to freeze out earlier, thereby shutting down helium production.  With
standard values for the other cosmological parameters, helium yields
increase until $G$ is about 100 times larger than the measured value
and then decrease with even larger values \cite{adamsreview}. 

\begin{figure}[tbp]
\centering 
\includegraphics[width=0.9\textwidth,trim=0 250 0 250]{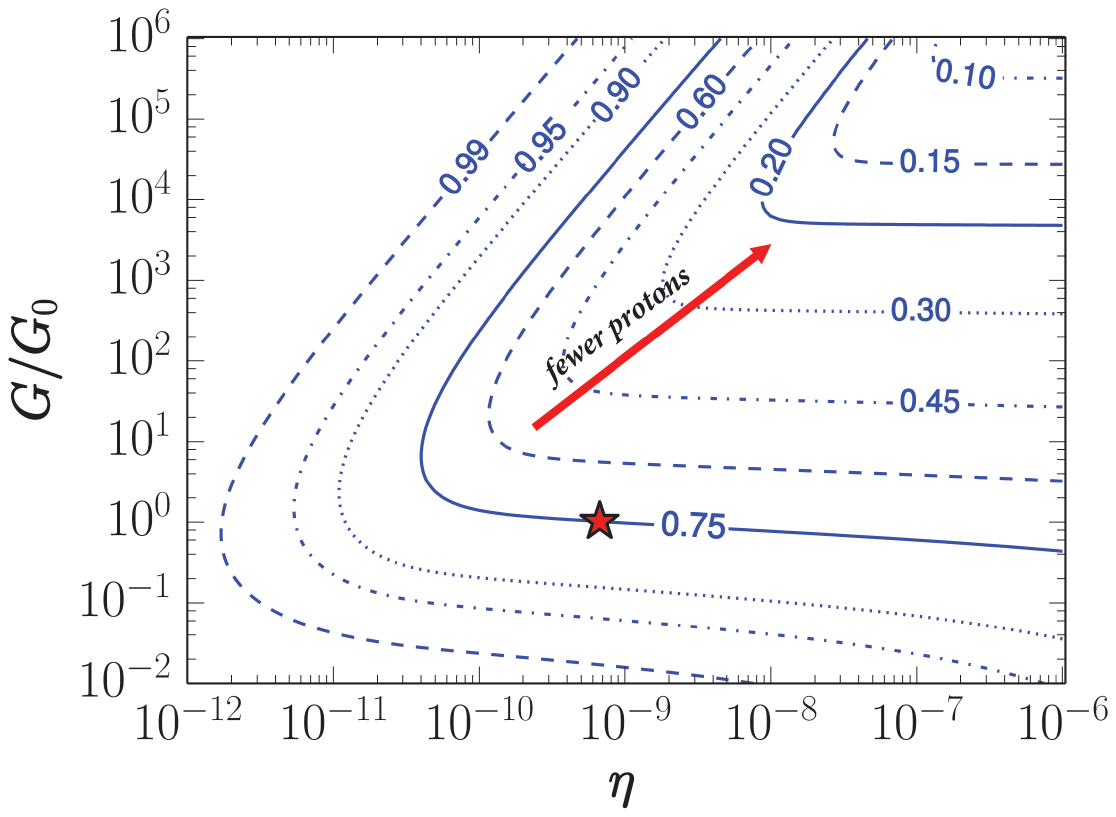} 
\vskip20pt
\caption{Remaining hydrogen (proton) mass fractions after Big Bang
Nucleosynthesis for varying gravitational constant $G$ and baryon to
photon ratio $\eta$. The red star marks the location of our universe
in the diagram. Note that even in the upper right corner of the 
$\eta$--$G$ plane, the universe retains $\sim$10\% of it mass in protons 
(adapted from \cite{adamsrhovac}, using the code from
\cite{burst,burst2}). }
\label{fig:bbnplane} 
\end{figure}  

Figure \ref{fig:bbnplane} shows the fraction of the universe remaining
in protons as both the gravitational constant and the baryon-to-photon
ratio are varied by 8 and 6 orders of magnitude respectively
\cite{burst,burst2,adamsrhovac}. Smaller values of both $G$ and
$\eta$ leave even more of the baryons in the form of protons, so that
such universes remain habitable. As both $G$ and $\eta$ increase, the
mass fraction in protons decreases. But even with $\eta$ larger by a
factor of $10^4$ and $G$ larger by a factor of $10^6$, about 10
percent of the mass in baryons remains in the form of protons (upper
right corner of the figure). Extreme conditions are thus required to
burn up all of the protons and leave behind a sterile universe. 

Variations in the fine-structure constant $\alpha$ work in the
opposite direction. If the value of $\alpha$ is increased, the
increased efficacy of Coulomb repulsion inhibits nucleosynthesis.
Most of the neutrons thus end up in deuterium, with trace amounts of
$^4$He and essentially no other elements. In the limit of small
$\alpha$, all of the neutrons are efficiently processed into helium,
which approaches the benchmark abundance $Y_4 \to 1/3$.

One can also consider variations in the neutron lifetime, which
arise from the corresponding variations in the strength of the weak
force \cite{grohsweakless,weakless,howeweakful}, and which can affect
BBN yields. The neutron lifetime can be written in the form 
\be
\tau_n^{-1} = c_n G_F^2 m_e^5 \,,
\label{neutronlife} 
\ee
where the dimensionless constant $c_n\approx0.07$ \cite{kolbturner}. 
Note that increasing the neutron lifetime corresponds to decreasing
strength of the weak force. As the neutron lifetime decreases, the
(stronger) weak force is able to compete with the expansion of the
universe down to lower temperatures so that the freeze-out occurs
later and the NSE abundance of neutrons is small. A stronger weak
force thus shuts down BBN. In the opposite case, with the weak force
even weaker and a longer neutron lifetime, the temperature of
freeze-out is high and the neutron to proton ratio aproaches unity. In
this limit, with nearly equal numbers of protons and neutrons, most of
the universe ends up in the form of helium. If the neutron lifetime is
increased to $\tau_n=10^6$ sec, about 1000 times larger than in our
universe, the helium abundance increases to $Y_4\approx0.90$, so that
10 percent of the mass is left in protons. For even longer neutron
lifetimes, however, the helium abundanace approaches unity, and
such universes would not be habitable (in that no hydrogen would
be available for stellar fuel or for water).  

Even in the limit of no weak force, where $\tau_n\to\infty$, some
universes could thrive (see \cite{clavelli,grohsweakless,weakless}).
If the abundance of baryons, specified by the baryon-to-photon ratio
$\eta$ has a low enough value, then even the neutrons cannot interact.
In our universe, the nuclear reaction rate for neutrons is much faster
than the expansion rate at the BBN epoch, so that almost all of the
neutrons interact before they decay. If the neutron interaction rate
is small enough, so that 
\be
\eta n_\gamma \langle\sigma{v}\rangle < H \,, 
\ee
then the neutrons will not be able to interact fast enough to become
fully depleted (where $n_\gamma$ is the number density of photons and
$\langle\sigma{v}\rangle$ specifies the cross section for neutron
interactions).  In the weakless limit ($\tau_n\to\infty)$, one needs
$\eta\sim4\times10^{-12}$ (about 100 times smaller than that of
standard cosmology) to maintain the same proton abundance as our
universe \cite{grohsweakless,weakless}.

\medskip 
{\bf Bottom line:} {\sl The early universe leaves behind an ample
supply of protons over a wide range of the available cosmological
parameter space. As a result, it is difficult for BBN to render
the universe lifeless.}

\bigskip 
\section{Dark Energy and the Cosmological Constant} 
\label{sec:lambda}

Dark energy currently provides the dominant component of our cosmic
energy inventory. This counterintuitive quantity represents the energy
level of the vacuum or a geometric property of space-time (see
\cite{boussovac,martin,weinberg89} and references therein). Although
the vacuum is often equated with empty space, it can have a nonzero
energy density. The existence and properties of this cosmic component
lead to the classic issue known as the cosmological constant problem.


A nonzero energy density for empty space poses two coupled problems:
First, in order for dark energy to allow the universe to form
structure and also have an influence at the present epoch, its energy
density must be roughly comparable to the current critical density.
The resulting value is much smaller than theoretically expected
values, so the ordering of energy scales represents a hierarchy
problem (see the discussion below). Second, if the dark energy can
affect the cosmic expansion, then it tends to suppress the formation
of large scale structure. If the universe contains too much dark
energy, then the resulting acceleration suppresses the gravitational
collapse of galactic halos and other structures.

Although the presence of dark energy is often called the cosmological
constant problem, the vacuum energy density of the universe is not
necessarily constant in time. Our universe displays hints of time
dependence \cite{desiresult}, but more data are required. Nonetheless,
time dependence of the dark energy can arise through a number of
mechanisms, and these could be realized in other universes. For
example, the dark energy could be the vacuum energy of background
scalar fields, which evolve smoothly and generally decrease with time
\cite{sahni,quintessence}. In contrast, the vacuum energy density
could change through cosmological phase transitions
\cite{coleman,sakharov1984,voloshin,zeldovich2}, and thereby
evolve rapidly with time. Although other universes could support
various forms of time dependent behavior, the present discussion
considers only the simplest case of constant dark energy.

{\sl The Problem:} The cosmological constant problem arises from the
General Theory of Relativity (GR), where the field equations can be
written 
\be
R_{\mu\nu} - {1\over2} R g_{\mu\nu} + \Lambda g_{\mu\nu} = 
8 \pi G T_{\mu\nu} \,, 
\label{grfield} 
\ee
where the symbols have their usual meanings (with $c=1$).\footnote{In
approximate terms: The left side of the equation, which includes the
Einstein tensor and the cosmological constant, determines the
structure of the gravitational field. The right side includes the
stress energy tensor, which acts as the source term for the matter fields.}
The cosmological constant is given by the parameter $\Lambda$, which
can be nonzero in the most general version of the theory. Without
further input, the value of $\Lambda$ is unspecified. One can make an
argument based on dimensional analysis. Noting that the gravitational
constant $G=\mplanck^{-2}$ sets the scale in GR, the expected value of
the cosmological constant should be given by
\be
\Lambda/G = \rhov = {\cal O} \left( \mplanck^4 \right)
\sim 10^{112} \, {\rm eV}^4 \,. 
\ee
In our universe, however, the value of $\rhov$ determined from
cosmological observations is much smaller, only $\rhov\sim10^{-10}$
eV$^4$. Although the argument based on dimensional grounds is clearly
incorrect, the physics that determines the value of $\rhov$ remains
unknown. This issue constitutes the cosmological constant problem
\cite{boussovac,martin,weinberg89}. 

One part of the cosmological constant problem is that the expected
value of $\rhov$ is many orders of magnitude larger than the observed
value. The number of orders of mangtiude depends on whether one
considers the energy density $\rhov$ or the energy scale $\lambda$.
Since the two quantities are related by $\rhov\sim\lambda^4$, the
observed value of $\rhov$ is too small by 120 orders of magnitude,
whereas the observed value of $\lambda$ is too small by `only' 30
orders of magnitude. 

Aside from the dimensional considerations given above, classical GR
does not provide a preferred value of the cosmological constant
$\Lambda$. In quantum field theory, however, the vacuum is
predicted to support an energy density $\rho_V$ of the form 
\be
\langle T_{\mu\nu} \rangle_V = - \rho_V g_{\mu\nu}\,.
\label{qftrhov} 
\ee
The left side of the equation represents the vacuum contribution to
the energy-momentum tensor, which is proprotional to the metric due to
Lorentz invariance (given by the right side). The energy density of
the vacuum $\rho_V$ is then the ratio of the two terms. By convention,
the sign is chosen so that $\rho_V\sim\rhov$ (where
$\Lambda\equiv8\pi{G}\rhov$). 

Quantum field theory predicts that $\langle T_{\mu\nu}\rangle\ne0$.
In a manner roughly analogous to the zero-point energy of a quantum
harmonic oscillator, the modes for the free fields produce a
zero-point energy that contributes to the energy density of the
vacuum. In standard treatments \cite{carrollpress,padmanabhan}, the
expected contribution has the form 
\be
\big|\langle T_{\mu\nu} \rangle\big| \propto \big| \rho_V \big| 
\sim M_{\rm max}^4 \,, 
\label{zeropoint} 
\ee
where $M_{\rm max}$ is the maximum energy scale or cutoff scale.
This treatment is approximate. In order for the theory to obey
Lorentz invariance and produce the well-known equation of state
$\langle{p}\rangle$ = $-\langle\rho\rangle$ for the vacuum, a
more nuanced approach is necessary. 

The cutoff scale $M_{\rm max}$ sets the size of the expected vacuum
contribution, but its value is not known. In the context of quantum
gravity, the Planck mass $\mplanck$ is expected to be the relevant
scale. Using $M_{\rm max} = \mplanck$ leads to the $\sim120$ order of
magnitude discrepancy seen before. Since quantum gravity is not fully
established, field theory considerations might not apply at such high
energies. However, we know that quantum field theory works well up to
the energy explored in particle accelerators, so the that cutoff
energy scale should be larger than the Higgs mass. If we then use
$M_{\rm max}\sim100$ GeV, the vacuum energy density would be
$\rho_V \sim 10^8$ GeV$^4$, which is still too large by about 54
orders of magnitude.

Note that both sides of the field equation (\ref{grfield}) could
contribute to the vacuum energy density. For the case where
$\Lambda\ne0$, which appears on the left side of the equation, the
contribution is usually called the cosmological constant. For the case
where $\langle{T_{\mu\nu}}\rangle\ne0$ due to contributions from
quantum fluctuations, which appears on the right side, the
contribution is usually called a vacuum energy density. For a
homogeneous and isotropic universe, both cases contribute the same way
to the cosmic energy inventory, and hence the evolution of the
universe. Current experiments indicate that $\rhov\ne0$, but do not
distinguish between energy density from a cosmological constant
$\Lambda$ and the expectation value $\langle T_{\mu\nu}\rangle$. 

Although many ideas have been put forth, the cosmological constant
problem does not have a definitive solution at the present time
\cite{picon,garriga2001,martin,peacock2020,rubakov,sahni,sola,
tyewassarman,weinberg89}.  Given the large number of mechanisms by
which our universe could attain a nonzero vacuum energy density, we
expect other universes within the multiverse to sample different
values for the energy density $\rhov$ of the dark energy.

{\sl Bounds from Structure Formation:} Although the energy density
$\rhov$ of the vacuum cannot be derived from first principles at the
present time, its value can be constrained by astrophysical
considerations. For example, in order for the universe to produce
structure, it must enter into a matter dominated era, and this
requirement places an upper bound on the energy scale of the vacuum, 
\be
\lambda \simless \eta m_{\rm p} {\Omega_M \over \Omega_{\rm b}}
\sim 4\, {\rm eV}\,, 
\ee
where the numerical value uses the parameters
$(\eta,\Omega_M,\Omega_{\rm b}$) of our universe (where $\Omega_{\rm b}$
is the density parameter for baryons and $\Omega_M$ is that for all
matter).  Given its observed value $\lambda \sim 0.003$ eV, our
universe satisfies this constraint by a factor of $\sim1000$ in energy
scale and a factor of $\sim10^{12}$ in energy density.

Stronger constraints on the vacuum energy result from the need for
galaxy formation to take place before the accelerating expansion
shuts down structure formation. Pioneering constraints of this type
\cite{banks,daviesunwin,sakharov,weinberg87} required that nonlinear
strucuture (namely quasars) form by redshift $z=4.5$ and found the 
upper limt  
\be
\rhov < {500 \over 729} \rhoeq Q^3 \,,
\label{weinbound} 
\ee
where $\rhoeq$ is the density at the epoch of matter/radiation
equality and where the numerical coefficient is derived using a
particular model for galaxy formation \cite{peebles67}. Using the
amplitude of density fluctuations $Q=10^{-5}$ observed in our
universe, one obtains the upper limit $\rhov\simless200\rhov(obs)$.

This bound on $\rhov$ assumes that the other cosmic properties are the
same as in our universe. The constraint becomes much weaker if the
amplitude $Q$ of primordial density fluctuations is allowed to
vary \cite{adamsrhovac,aguirretegmark,garriga2000,liviorees,martel,mersini}.
As shown by equation (\ref{weinbound}), the limit is proportional to
$Q^3$. For a universe with $Q=10^{-3}$, which is 100 times larger than
in our universe but still smaller compared to unity, the bound on
$\rhov$ becomes a million times weaker. Notice also that larger values
of $Q$ are readily realized \cite{garriga2006}.  Recent studies
\cite{oh2022,sorini2024} have explored structure formation in universes
with different values for the cosmological constant and the
fluctuation amplitude using detailed numberical simulations; they find
that the halo mass function and star formation history are influenced
by the value $\Lambda$, with results broadly consistent with previous
work.

The value of $Q$ cannot become arbitrarily large, as structure
formation must produce galaxies with habitable properites.  As the
amplitude $Q$ increases, structure formation occurs earlier in cosmic
history when the universe is denser, and the background density
informs the density of the structures that form. Larger values of $Q$
lead to denser galaxies, which can compromise habitability by
scattering planets \cite{tegmarkrees,tegmark} and by producing strong
background radiation fields \cite{coppess}. In spite of these
complications, however, the fluctuation amplitude can be as large as
$Q\sim10^{-2}$. This increase in $Q$ allows the bound on $\rhov$ to
become a billion times less constraining. One thus obtains a limit of
the form $\rhov\simless10^{11}\rhov(obs)$.

\begin{figure}[tbp]
\centering
\vskip-1.0truein
\includegraphics[width=0.9\textwidth,trim=0 150 0 150,clip]{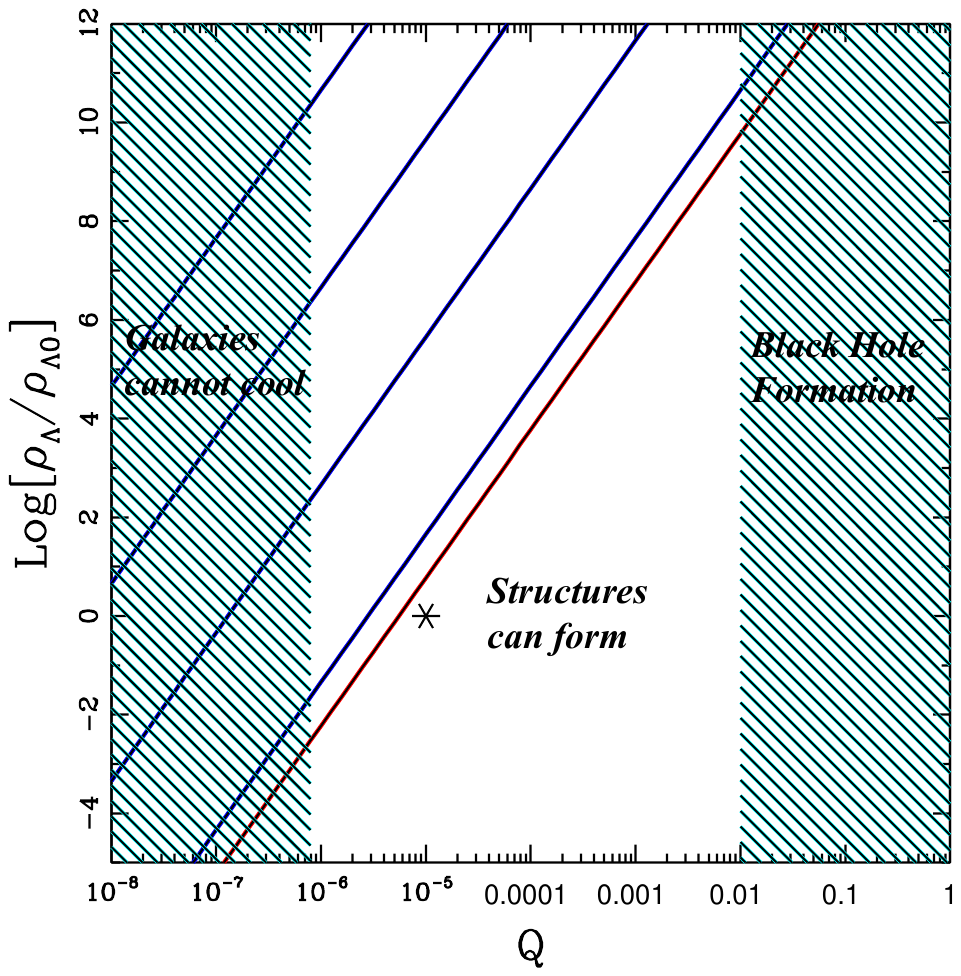}
\vskip-24pt
\caption{Constraints on the vacuum energy density $\rhov$ as a function
of amplitude $Q$ of the primordial fluctuations.  The allowed region
lies between the shaded regions and below the lines. Constraints are
shown for varying values of $\eta$, from that in our universe
$\eta=6\times10^{-10}$ (red curve), to larger values $\eta$ =
$10^{-9}$, $10^{-8}$, $10^{-7}$, and $10^{-6}$ (blue, from bottom to
top). Our universe is marked by the star symbol. Large values of $Q$
are ruled out because black holes are overproduced \cite{bhrees},
whereas small values of $Q$ are ruled out because gas in galaxies
would be too rarefied to cool \cite{tegmarkrees,tegmark}.  Large
values of $\rhov$ are ruled out because structure formation is
compromised by the accelerating expansion.}  
\label{fig:qvlambda} 
\end{figure}

The bounds presented above assume that the universes transition from
radiation dominated to matter dominated at the same epoch as in our
universe. Although the baryon-to-photon ratio $\eta$ is constrained,
it can be significantly larger and still allow the universe to emerge
from its BBN epoch with an ample supply of protons (Section
\ref{sec:bbn}). In universes with larger $\eta$, matter domination
occurs earlier, and density fluctuations have more time to grow.
As a result, larger values of $\eta$ allow a universe to have larger
vacuum energy and the generalized bound on $\rhov$ takes the form
\be
\rhov < (2 A^3 a_R) Q^3 \eta^4 \mpro^4 
\left({\Omega_M \over \Omega_b}\right)^4 \,,
\label{genweinbound} 
\ee
where the dimensionless constant $A$ and the radiation constant
$a_R$ are of order unity. The baryon-to-photon ratio $\eta$ could
be larger by a factor of $\sim1000$ (Figure \ref{fig:bbnplane}),
so that the bound on $\rhov$ becomes about a trillion times
less constraining \cite{adamsrhovac}.

Figure \ref{fig:qvlambda} depicts the bounds discussed here. The
maximum denergy density $\rhov$ is plotted as a function of the
amplitude $Q$ of the density fluctuations. The allowed range of $Q$ is
constrained: For $Q\simgreat10^{-2}$, catastrophic black hole
formation can occur \cite{bhrees}. For $Q\simless10^{-6}$, gas cannot
cool fast enough to condense and collapse into galactic
structures \cite{reesost,tegmarkrees}. Within the allowed range for
$Q$, vacuum energy density $\rhov$ is bounded from above.  The red
curve shows the constraint for universes with the same matter content
$(\eta,\omegad/\omegab)$ as our universe. With all of the other
properties of our universe held fixed, $\rhov$ can only increase by a
modest factor.  Much larger values are possible for larger fluctuation
amplitudes $Q$, where $\rhov$ can be larger than the observed value by
a factor of $\sim10^{10}$ and still allow for structure formation. If
the baryon-to-photon ratio $\eta$ is increased, the bound on $\rhov$
becomes substantially weaker, as shown by the blue curves in
Figure \ref{fig:qvlambda}.

For universes with larger $\rhov$ and larger $Q$, galaxy formation
occurs earlier and results in denser structures. As a result, some
fraction of the galactic real estate would cease to be habitable, as
discussed in Section \ref{sec:galaxies}. In addition, as $\rhov$
increases, the total mass contained within the late-time cosmological
horizon decreases, so that the number of stars within the horizon
becomes smaller. For the largest allowed values of $\rhov$, the number
of stars in the late-time horizon is estimated to be as small as
$\sim10^{12}$, roughly the number within a large galaxy
\cite{adamsreview}.  

The above constraints apply to positive values $\rhov>0$. Since
contributions to the vacuum energy density can arise form both
geometric and matter sources, the net result could be a cosmological
constant term with $\rhov<0$. In this case, the vacuum energy slows
down cosmic expansion and does not suppress structure formation.
Instead, the entire universe could collapse to a singularity in a
finite time. The total lifetime $t_{\rm s}$ for a flat universe
containing matter and a negative cosmological constant is given by 
\be
t_{\rm s} = {2\pi \sqrt{3}\over3} 
\left( 8\pi G |\rhov|\right)^{-1/2} 
\sim \left( G |\rhov| \right)^{-1/2}
= {\mplanck\over\lambda^2} \, .
\ee
For a universe to be habitable, it must live for a sufficiently
long time \cite{bartip,weinberg87}. A minimum of time scale of
$\sim1$ Gyr is often invoked as a requirement for life to arise
\cite{chyba,mckay,orgel}, although this quantity is not well
determined. A lower limit on the cosmic lifetime implies an upper
limit on the magnitude of the vacuum energy density.  The choice
$t_{\rm s}=1$ Gyr leads to a bound on negative vacuum energy of the
form $|\rhov| \simless 100\,\rho_{\rm crit} \approx
100\,\rho_{\scriptscriptstyle{\Lambda}0}$. Negative values of the
cosmological constant are thus more constrained than positive values.

\medskip 
{\bf Bottom line:} {\sl The cosmological constant (or current dark
energy density) can vary by many orders of magnitude and still allow
for structure formation and habitability. On the other hand, the
allowed values are much smaller than that given by the Planck scale.}

\bigskip 
\section{Variations in the Fluctuation Amplitude $Q$} 
\label{sec:qvary}

One of the striking features of our universe is the presence of
galaxies -- vast collections of stars, gas, and dark matter that serve
as the primary sites of cosmic structure. The possibility of a working
universe with complex chemistry, planetary systems, and ultimately
life depends on the existence of galaxies. However, galaxies with the
right properties to support habitability are not guaranteed
outcomes. Their formation \cite{benson2010} relies on the growth of
tiny fluctuations in the matter density in the early universe,
irregularities that act as seeds for later cosmic structures.

These primordial fluctuations, perhaps imprinted during an early
inflationary epoch, are characterized by an amplitude $Q$.  The value
of $Q\sim10^{-5}$ in our universe, but could have other values in
other regions of space-time. This parameter has profound consequences
for the habitability of the universe.  If $Q$ is too small, galactic
structures form late with low densities, and cannot cool enough to
make large numbers of stars.  Conversely, if $Q$ is too large, then
galactic structures form early with high densities, resulting in 
environments too compact and violent to support habitable planetary
systems.

As a result, viable galaxies arise only in universes where the
fluctuation amplitude $Q$ falls within the proper range.  Galaxies
must not only form, but they must also exhibit the right properties:
dense enough to cool and condense into stars, yet rarefied enough to
allow solar systems to have long-term stability.  The amplitude $Q$ of
primordial density fluctuations must allow for this balance to allow
for the possibility of complexity and life.

If $Q$ is smaller than the observed value, galactic structures still
form, but they are less dense. Galaxies in such a universe would not
be able to radiate heat efficiently. Gas with galaxies needs to cool
in order to collapse into stars, but the cooling rate depends
sensitively on density: Only when gas clouds reach sufficient
densities can atomic collisions and radiative processes carry away
thermal energy quickly enough. In a low-$Q$ universe, proto-galactic
gas clouds remain too diffuse to cool efficiently and star formation
is suppressed. The minimum fluctuation amplitude is estimated to be
$Q_{\rm min} \approx 6 \times10^{-7}$
\cite{tegmark,tegmarkrees,adamsreview}. 

If $Q$ is lowered even further, galaxies become not only inefficient
as stellar factories but also have shallow gravitational potential
wells. Elements heavier than helium are produced by stellar
nucleosynthesis, and these manufactured nuclei must be retained within
the galaxy. If the potential wells of galaxies are too shallow, these
heavy elements escape into intergalactic space and cannot enrich
subsequent generations of stars and planets. Without such enrichment,
planets -- especially rocky planets like Earth -- cannot form. The
minimum fluctuation amplitude that allows galaxies to have sufficient
gravity is about $Q_{\rm grav} \sim 10^{-7}$. The bound is not sharp,
as galaxies with shallow potential wells can retain some fraction of
their enriched gas, but the fraction decreases with lower $Q$.

As the amplitude of the density fluctuations grows larger, the density of
galaxies increases. To leading order, the density scale of galaxies is
determined by the density of the background universe at the epoch when
galactic collapse occurs. This consideration
\cite{kolbturner,martel,peacockbook,peacock,peebles67,schecter,
tegmarkrees,tegmark} leads to the characteristic density scale 
\be
\rho_c = \left({12\pi^2\over5f_{\rm vir}^2}\right) (\mpro\eta)^4
\left({\Omega_M\over\Omega_b}\right)^4 Q^3\,,
\ee
where the virial parameter $f_{\rm vir}\sim1$. 

Overly dense galaxies are problematic for habitability in two ways.
First, their high stellar densities make interactions between
planetary systems and passing stars more frequent, with the result
that planetary orbits become too readily disrupted. With a dramatic
decrease in the lifetimes of planetary orbits, the prospects for
habitability decrease accordingly \cite{coppess,tegmarkrees}. Even
though planet formation could proceed as usual, the resulting bodies
might not survive in orbit long enough for life to develop. In
approximate terms, planet formation takes place over millions of
years \cite{armitage}, whereas habitability requires billions of
years \cite{lunine,scharf}. On the other hand, life could be more
resilient than is sometimes imagined \cite{sloan}. 

Second, dense galaxies produce environments with strong background
radiation fields \cite{coppess}, analogous to the destructive
environments of dense star clusters in our universe \cite{thompson}.
Close-packed stars enforce a bright background sky of electromagnetic
radiation and frequent supernovae flood the interstellar medium with
energetic cosmic rays. For any planets that manage to persist, such
intense radiation fields would erode atmospheres and pose severe
challenges for the stability of their surface environments (e.g.,
see \cite{sandora2022b,sandora2025b}).  The cosmic rays (particles)
could be attenuated, and magnetic fields can slow down their diffusive
transport. As a result, the photons are likely to provide the largest
destrutive effect. 

\begin{figure}[tbp]
\centering 
\includegraphics[width=0.9\textwidth,trim=0 150 0 150]{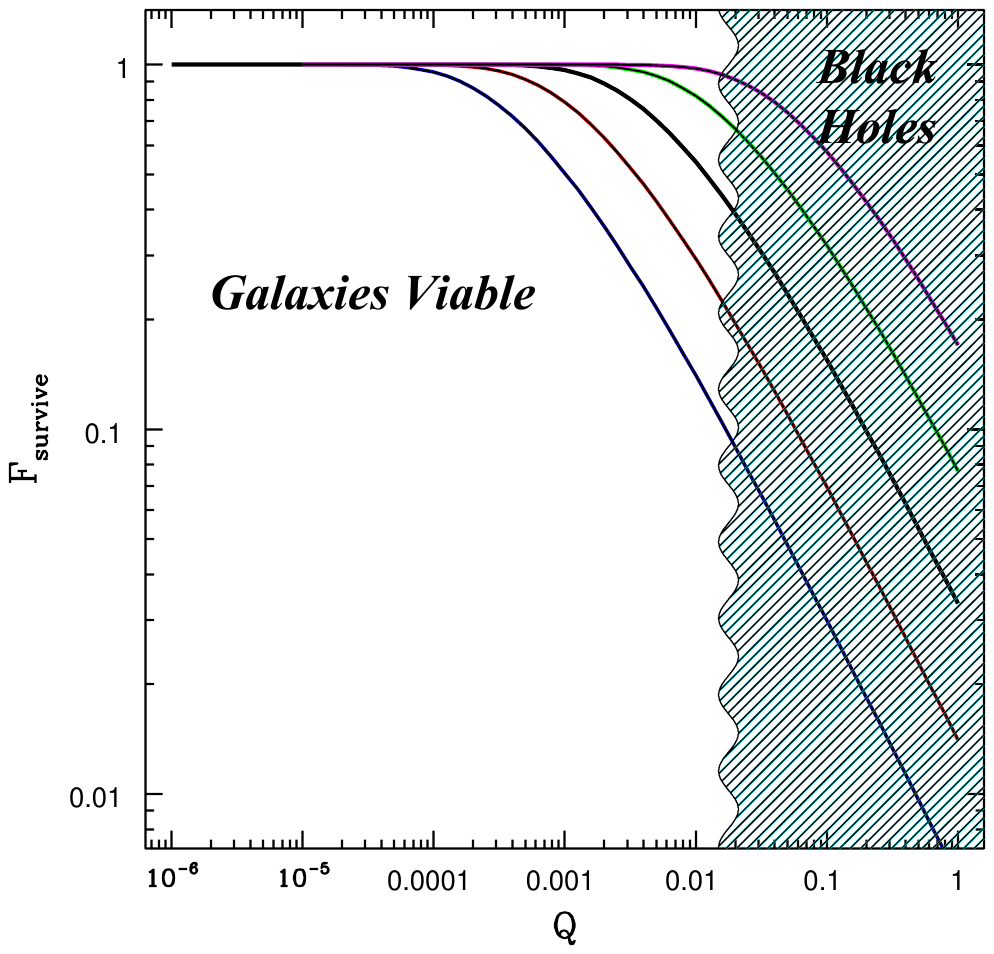} 
\vskip-12pt
\caption{Fraction of galactic real estate that remains habitable as a
function of the amplitude $Q$ of primordial density fluctuations
(adapted from \cite{adamsreview,coppess}). The curves show the results
for a collection of different galaxy masses, specifically for
$\log_{10}[M/M_\odot]$ = 10 (blue), 11 (red), 12 (black), 13 (green),
and 14 (magenta). The shaded region on the right side of the diagram
is disallowed because large $Q$ values can result in catastrophic
black hole formation. }
\label{fig:galsurvive} 
\end{figure}

Galaxies are not uniform in density. They exhibit a wide range of
densities, with inner regions far more crowded than their outskirts.
As the mean galactic density increases, the fraction of the galaxy
that remains at sufficiently low density to support life decreases
accordingly. Nonetheless, some fraction of stars reside in the outer
regions of galaxies where the density is low enough for habitability.
Figure \ref{fig:galsurvive} shows the fraction of the stellar
population that resides in galactic regions with low enough density to
survive, without being disrupted by planetary scattering; the fraction
is plotted as a function of the fluctuation amplitude $Q$ (adapted
from \cite{coppess}).  The curves are shown for a collection of
different galactic masses, as the mean density varies inversely with
galaxy mass. As shown in the figure, sizable fractions of the stellar
population remain potentially habitable even for fluctuation
amplitudes $Q$ that are 100 or even 1000 times larger than in our
universe. The shaded region at the right hand side of the figure
depicts the expected result that sufficiently large amplitudes $Q$
lead to catastrophic black hole formation \cite{bhrees} and -- most
likely -- the loss of viability. Although the maximum allowed value
for the fluctuation amplitude depends on what fraction of the galactic
real estate remain viable, a good working estimate is $Q_{\rm
max}\approx$ $10^{-2}$. This finding, combined with the minimum values
outlined above, suggest the allowed range 
\be
Q_{\rm min} \approx 6 \times 10^{-7} \simless Q \simless 10^{-2}
\approx Q_{\rm max}\,.
\ee
The fluctuation amplitude can thus vary over more than four
orders of magnitude. 
 
\begin{figure}[tbp]
\centering 
\includegraphics[width=0.8\textwidth,trim=0 150 0 150]{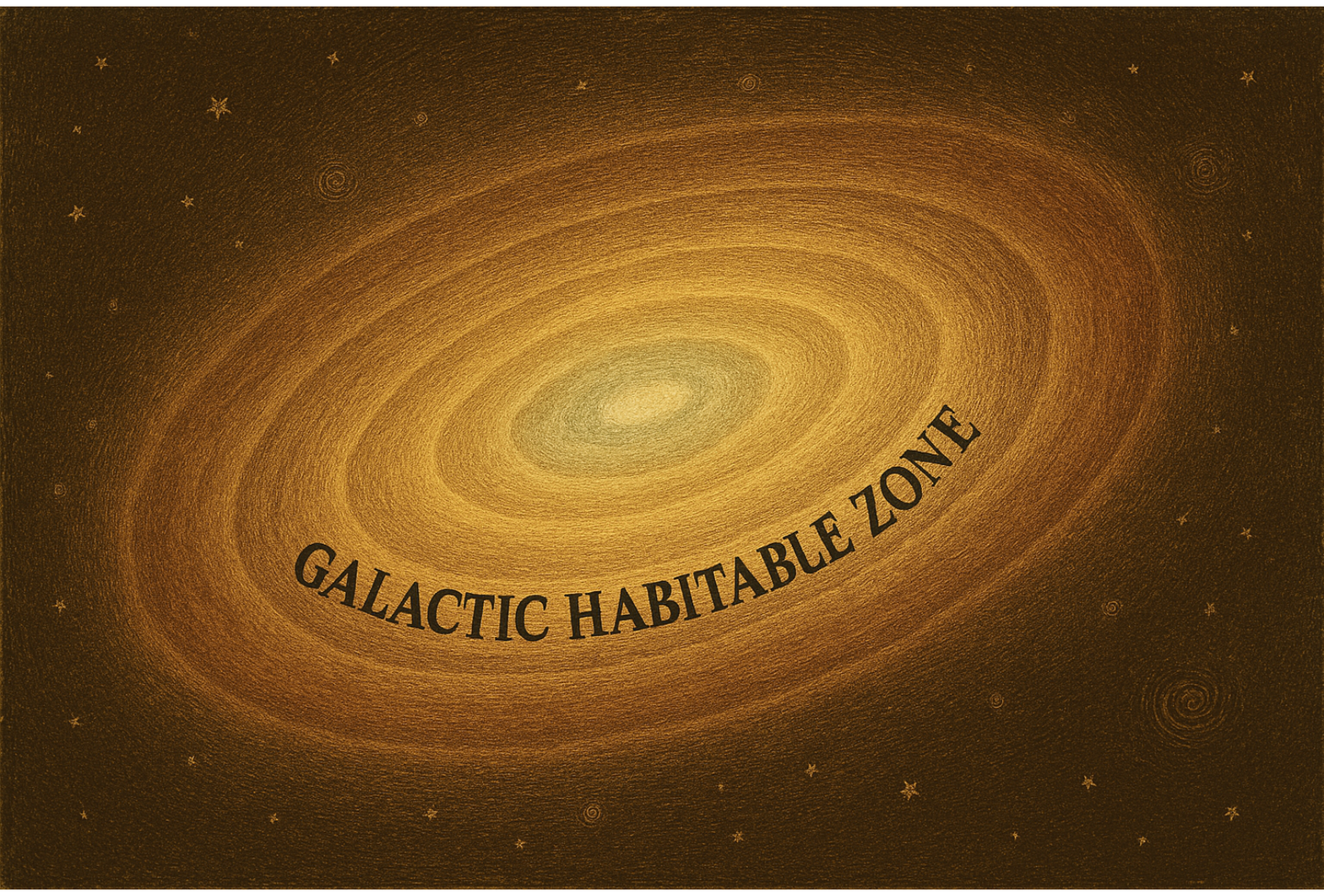} 
\vskip60pt
\caption{The Galactic Habitable Zone. In universes with moderately
large $Q\sim10^{-3}$, galaxies are denser and support zones where the
night sky is comparable in brightness to the day sky on Earth. In
these zones, most planets can have the right surface temperature to
support liquid water environments. [This AI-generated figure will be
replaced by an artist-generated picture in the published version of
the article.] }
\label{fig:ghz} 
\end{figure}

Figure \ref{fig:galsurvive} shows the survival fractions for stars and
their planetary systems in the face of increased scattering
disruption.  The corresponding curves for survivability in the face of
strong radiation fields have a similar
shape \citep{adamsreview,coppess}.  In such galaxies, the radiation
fields in the inner regions are so bright that the surface
temperatures of planets are expected to be too high to support liquid
water \cite{kaib,thompson}.  The outer regions are cold, so that planets
must be heated by their host stars. In the intermediate regions,
however, the night sky of the galaxy could be comparable in brightness
to the day sky on Earth. In this region, which can be called the
Galactic Habitable Zone \cite{coppess}, planets in almost any orbit
can be habitable.  This possibility is depicted in Figure
\ref{fig:ghz}. Under the most favorable conditions, corresponding to
fluctuation amplitude $Q\sim10^{-3}$, the number of planets in the
Galactic Habitable Zone can be large enough that the galaxies would
have more potentially habitable planets than those in our universe.

\medskip 
{\bf Bottom line:} {\sl The fluctutation amplitude $Q$ can vary by
several orders of magnitude. Values larger than that of our universe,
where $Q\sim10^{-5}$, are easier to realize in inflationary theories
and can potentially allow for more habitable planets. }

\bigskip 
\section{Galaxies and their Properties} 
\label{sec:galaxies}

The mass scale of galaxies can be derived from the requirement that
the gas cooling time is comparable to the collapse time for
galaxy-like structures \cite{reesost}. This condition defines a
mass scale of the form 
\be
M_{gal} = \alpha_G^{-2} \alpha^5 \left( {\mpro\over m_e} \right)^{1/2}
\mpro = \alpha_G^{-1/2} \alpha^5 \left( {\mpro\over m_e} \right)^{1/2}
\starmass \,, 
\ee
where the final equality is written in terms of the characteristic 
mass scale for stars \cite{phillips} (see also Section \ref{sec:stars}): 
\be
\starmass = \alpha_G^{-3/2} \mpro\,.
\label{starmass} 
\ee
This mass scale is essentially the Chandrasekhar mass \cite{chandra}
without the numerical constants. The number of stellar mass scales
contained within a galactic mass scale is thus given by
$N\sim\alpha^5/\alpha_G^{1/2}$.  As a result, if the fine-structure
constant becomes too small, then galaxies would contain few stars.
In practice, this constraint sets an approximate lower limit
$\alpha \simgreat \alpha_0/100$, where $\alpha_0$ is the value in our
universe.

Galactic properties depend on the relative abundances of baryonic
matter and dark matter. The baryonic component is specified by
the baryon-to-photon ratio $\eta$. We can define an analogous
quantity $\delta$ to specify the dark matter component, i.e.,
\be
\delta \equiv \eta {\Omega_M \over \Omega_b}\,.
\ee 
A number of considerations constrain the abundances $\eta$ and
$\delta$ (see \cite{adamsreview} and referfences therein). If the
matter content of the universe is too large, then the epoch of matter
domination occurs before Big Bang Nucleosynthesis.  While this
inversion of events would not necessarily be catastrophic, as BBN
generally does not prevent the universe from becoming habitable
(Section \ref{sec:bbn}), it would result in a rather different type of
universe. Since the temperature of BBN is given by nuclear energy
scales, which in turn are comparable to the electon mass, a rough
estimate for this requirement takes the form 
\be
(\eta + \delta) \simless {m_e \over \mpro} \sim 10^{-3}\,.
\ee
In our universe, the sum $\eta+\delta\sim4\times10^{-9}$, so
that the matter content would have to increase by a factor of
$\sim2.5\times10^5$ for such early matter domination. 

Another constraint arises from the requirement that structure
formation occurs after the epoch of decoupling. If collapsing galaxies
and other large scale structures become nonlinear while the baryons
remain coupled to the background radiation, then structure formation
would be altered. In this context, the first structures with galactic
masses become nonlinear at a temperature scale $T_{nonlin} \approx 9 Q
(\eta+\delta) \mpro$ \cite{tegmark} and the temperature of
recombination has the approximate form $T_{\rm rec} \approx \alpha^2
m_e/100$. This latter result follows from evaluation of the Saha
equation and the fact that the baryon-to-photon ratio $\eta\ll1$. With
these scales defined, the constraint that recombination occurs before
structure formation takes the form 
\be
(\eta+\delta) \simless \alpha^2 Q^{-1} {m_e \over 900 \mpro}
\sim 3 \times10^{-6}\,. 
\ee
This value is larger than $\eta+\delta$ in our universe by
a factor of $\sim750$.  

In our universe, matter domination takes place before recombination
and decoupling. If the matter content is small enough, however, these
transitions occur in the opposite order. Although gravitational
collapse of structures would still take place if the order is
reversed, the growth would proceed differently because baryons and
dark matter would collapse together. The boundary between these two
possibilities occurs when matter domination and recombination occur at
the same time, which requires
\be
(\eta+\delta) = {\alpha^2 \over 100} {m_e\over\mpro} \sim
3 \times 10^{-10}\,. 
\ee
Our universe is thus relatively close to this transition point. 

Whereas the above considerations show how the total matter content
(as specified by the sum $\eta+\delta$) can change the formation of
galactic structures, the relative abundance also has potentially
interesting consequences. The dark matter component of a galaxy is
collisionless and does not take part in small scale collapse. In
contrast, the baryonic component in galactic disks fragments into
molecular clouds and eventually into stars. The stability of galactic
disks and the efficacy of star formation thus depend on the relative
amount of dark matter. If the dark matter component is large, so that
$\delta\gg\eta$, then disks are stable, and star formation is
suppressed. If the dark matter component is small, so that
$\delta/\eta\ll1$, then galactic disks can become violently unstable
and star formation can occur rapidly. In addition, Silk damping will
be enhanced in the pre-collapse phases and can alter galactic
structure \cite{silkdamp}. These possible changes to galaxy properties
do not necessarily prevent habitable environments, so that precise
limits on the ratio $\delta/\eta$ are not well determined. Nonetheless,
previous work \cite{tegmark} has suggested that the range 
\be
1 \simless {\delta\over\eta} \simless 300 
\ee
would allow galactic disk evolution and star formation to proceed as
it does in our universe. Smaller (larger) values of the ratio
$\delta/\eta$ would lead to more rapid (slower) star formation rates,
but would not necessarily prevent habitability. 

\medskip 
{\bf Bottom line:} {\sl Whereas galaxy formation requires the proper
values of the cosmological constant $\Lambda$ and the fluctuation
amplitude $Q$, galactic properties depend on the relative abundances
of baryons $\eta$ and dark matter $\delta$. A wide range of values
result in viable galaxies: The sum $\eta+\delta$ can be larger by a
factor of $\sim10^5$ and the range of the ratio $\delta/\eta$ is at
least 300.}

\bigskip 
\section{The Existence of Stars} 
\label{sec:stars}

The operations of our universe, including the development of life,
require the existence of stars. These engines of the cosmos play two
vital roles, in providing energy to the universe \cite{fukugita} and
in synthesizing essentially all nuclei beyond helium \cite{timmes,trimble}.
As a result, stars contribute both the power supply and the raw
materials necessary for the universe to be interesting.

In order for a star to operate, all four fundamental forces of nature
must work in concert \cite{chandra,clayton,hansen,kippenhahn,phillips}.
Gravity holds the star together. Electromagnetic radiation allows the
star to emit energy and power the universe, and electromagnetic
interactions control the transport of radiation through the stellar
interior. The star is held up against collapse by ordinary thermal
pressure, described by the ideal gas law, but the requisite high
temperatures are maintained by thermonuclear fusion. The fusion
process itself is driven by the strong force, but the weak force is
required to turn protons into neutrons, which is necessary to produce
helium. In order for stars to form and function, the strengths of the
forces must fall within ranges that are considered here.

We first emphasize that stars can be produced over a range of masses,
from roughly $\sim0.1M_\odot$ to 100 $M_\odot$, spanning a total
factor of $\sim1000$. Moreover, the mass of a star has an enormous
impact on stellar evolution. Nonetheless, it useful to define a
characteristic stellar mass scale (see \cite{phillips} and equation
[\ref{starmass}]), given by  
\be
\starmass = \alpha_G^{-3/2} \mpro = \mplanck^3 
\mpro^{-2} \approx 1.85 M_\odot \,. 
\ee
This mass scale is essentially the Chandrasekhar mass \cite{chandra},
up to the dimensionless coefficient. Stars in our universe thus form,
live, and die with masses that are within a factor of $\sim30$ of the
mass scale required for self-gravity to overcome quantum mechanical
degeneracy pressure.

The lower limit on stellar masses is determined by the neccessity for
nuclear fusion to occur, which in turn requires that the stellar core
becomes sufficiently hot. For stars with too little mass, quantum
mechanical degeneracy pressure prevents the self-gravity of the star
from crushing the stellar core to achieve the necessary temperature
(see \cite{adams,bartip,phillips} and many others). This
consideration leads to a lower limit of the form
\be
M_{\ast{\rm min}} = 6(3\pi)^{1/2}(4/5)^{3/4} \left(
{T_{\rm nuc} \over m_e} \right)^{3/4} \alpha_G^{-3/2} \mpro \,,
\label{minmass} 
\ee
where $T_{\rm nuc}\sim10^7$ K is the temperature required for nuclear
fusion. One can make an estimate for the nuclear burning temperature
by taking into account the quantum tunneling necessary for protons
to overcome their Coulomb repulsion and undergo fusion (e.g., see
\cite{adamsreview,phillips}), so that the above expression simplifies
to the form  
\be
M_{\ast{\rm min}} \approx {1\over4}
\left({\alpha\over\sqrt{\beta}}\right)^{3/2}
\starmass \approx 0.04 \starmass \,.
\label{minmasstwo} 
\ee

At the other end of the mass range, stars become unstable if the
pressure contribution from their internal radiation greatly exceeds
that provided by the ideal gas law \cite{clayton,kippenhahn,phillips}.
Stars of ever larger mass need correspondingly larger luminosities to
provide the temperatures needed for pressure support. But radiation
pressure $P_R\sim T^4$ and thus increases faster than gas pressure
$P=nkT$, so that a crossoover occurs. The resulting maximum mass
scale can be written in the form
\be
M_{\ast{\rm max}} \approx 50 \alpha_G^{-3/2} \mpro \approx
50 \starmass \,.
\label{maxmass} 
\ee

One can immediately see that the possible range for stellar masses
places an interesting limit on the fine-structure constant.  The
minimum mass scales as $M_{\ast{\rm min}}\sim\alpha^{3/2}$, whereas
the maximum mass is independent of $\alpha$. In our universe, the
maximum mass is about 1000 times larger than the minimum mass, so that
increasing $\alpha$ by a factor of 100 would close the gap. More
rigorous constraints can be derived by solving the equations of
stellar structure and finding the values of the fundamental constants
that allow for solutions \cite{adams,adamsnew,adamsreview} and hence
working stars. The results show that the fine-structure constant does
indeed have an upper bound $\alpha \simless 100 \alpha_0$. Protons in
stellar cores have to tunnel quantum mechanically in order to fuse,
and tunneling is exponentially suppressed. As $\alpha$ becomes too
large, the nuclear reaction rate becomes too small to maintain the
high temperatures necessary for pressure to hold up the star. Going in
the other direction, stars also impose a lower bound on $\alpha$. If
the fine-structure constant is too small, then protons can experience
nuclear fusion without tunneling, and the star would essentially become
a nuclear bomb.\footnote{Note that the formation of stellar objects
would be quite different and perhaps more difficult in this sceanrio.}
As a result, the value of $\alpha$ cannot be more than $\sim100$ times
smaller than that in our universe. For the case with all other
constants fixed, the fine-structure constant can vary by a factor of
$\sim100$ upward or downward and still allow for working stars.

\begin{figure}[tbp]
\centering 
\includegraphics[width=1.0\textwidth,trim=0 150 0 150]{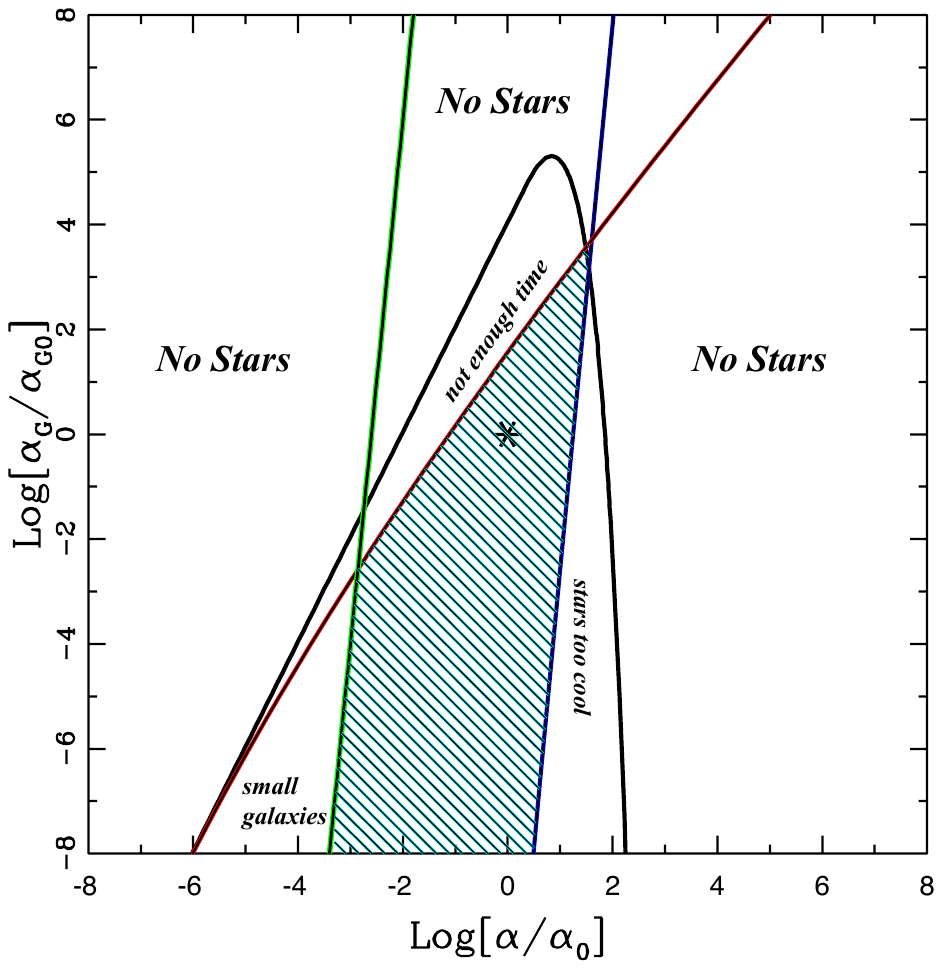} 
\caption{Allowed region in the $\alpha$-$\alpha_G$ plane for the existence
of stars (adapated from \cite{adamsnew}). Solutions to the stellar
structure equations exist only for the region of parameter space below
the black curve. The red curve delineates the regime where the main
sequence lifetimes are long enough for life to evolve. The blue curve
delineates the regime where the photospheric temperature is high
enough to drive chemical reactions. The green curve marks the regime
where galaxies become too small in mass to support stars and star
formation. }
\label{fig:starplane} 
\end{figure} 

If both the gravitational constant $G$ and the fine-structure constant
$\alpha$ can vary, the corresponding allowed region in the paremeter
space is shown in Figure \ref{fig:starplane}. In order for long-lived,
stable, nuclear burning stars to exist, the points in parameter space
must fall below the black curve. On the right side of the diagram,
where $\alpha$ is large, Coulomb repulsion is so strong that nuclear
reactions are highly suppressed. On the left side of the digram, for
small $\alpha$, solutions to the stellar structure equations do not
exist because all of the proton fuel reacts at once. As the
gravitational constant increases, so that gravity is stronger, nuclear
reactions must be more robust to provide the requisite pressure, and
the range of allowed parameter space shrinks. For sufficiently large
$G$, more than $\sim10^5$ times the value in our universe, gravity
becomes so strong that stars can no longer operate.

The discussion thus far only concerns the existence of stars as
long-lived nuclear furnances. Additional constraints reduce the
allowed region of parameter space (as also shown in Figure
\ref{fig:starplane}). Although we do not know exactly what
stellar properties are required for a universe to be habitable, some
basic requirements have emerged \cite{kasting1993,kasting2003}.

First, the stars must live long enough for life to evolve. Since
biology involves chemical reactions, which are determined by atomic
properties, the basic time scales must depend on the fine-structure
constant. For example, the atomic oscillation time scale varies
according to the relation $t_A \approx 1/(\alpha^2m_e)$.  We
also do not know how much total time is required. Current
considerations of astrobiology suggest that a time scale of 1 Gyr is a
good benchmark \cite{lunine,scharf}. If we use the requirement that
the longest-lived star burns its nuclear fuel over a time scale of at
least 1 Gyr, scaled as $\alpha^{-2}$, then stellar parameters must
fall below the red curve in Figure \ref{fig:starplane} (see 
\cite{adamsnew,adamsreview}). 

Next we require that stellar surfaces are hot enough to drive chemical
reactions. For large values of $\alpha$, nuclear reactions are highly
suppressed by enhanced Coulomb repulsion, so that stellar luminosities
and surface temperatures are lower. The typical photon energy
$E_\gamma=\nu\sim T_\ast$ must be comparable to the energy of
chemical reactions, where this energy is a fraction $\epsilon_c$ of
the atomic energy scale $\alpha^2m_e$. These considerations imply
the constraint 
\be
T_\ast > \epsilon_c \alpha^2 m_e \,,
\ee
which is shown as the blue curve in Figure \ref{fig:starplane}. Since
stellar temperatures decrease rapidly with increasing $\alpha$, this
boundary in parameter space is sharp \cite{adamsnew,adamsreview}. 

\medskip 
{\bf Bottom line:} {\sl Stars exist over a wide range of parameter
space, with the fine-structure constant, the gravitational constant,
and the nuclear cross sections varying by several orders of magnitude.
Moreover, such stars can be sufficiently hot and long-lived. }

\bigskip 
\section{The Triple-Alpha Process}  
\label{sec:triplealpha}

One of the classic instances of possible fine-tuning of the universe
is the triple-alpha process, which is the mechanism that produces
carbon in steller interiors \cite{clayton,kippenhahn}. After stars
have burned all of the hydrogen in their cores into helium, the next
stage of stellar nucleosynthesis is to use the helium nuclei as fuel.
In our universe, however, the seemingly natural reaction of fusing two
helium nuclei (alpha particles) into beryillium cannot operate because
the required product nucleus $^8$Be is unstable. The production of
carbon is thus more complicated.

Despite the somewhat unfavorable energetics and the short half-life of
only $\sim10^{-16}$ sec, some $^8$Be nuclei are synthesized. As a
result, evolved stellar cores contain a standing population of these
transient particles. During its brief existence, a $^8$Be nucleus can
interact with an additional alpha particle \cite{salpeter} and produce
carbon (specifically $^{12}$C). Under ordinary circumstances, with the
density and temperature conditions of stellar interiors, the carbon
production rate is too slow to account for the carbon abundance in our
universe. As posited by Hoyle \cite{hoyle} and subsequently measured
in the laboratory \cite{dunbar,fowler}, however, the carbon $^{12}$C
nucleus has an energy level at the proper energy to allow the
reactions to proceed under resonant conditions. The presence of this
resonance allows carbon production to take place more rapidly and 
accounts for the observed carbon production in stars.

The nuclear reaction rate for carbon production is exponentially
sensitive to the energy of the resonance. More specifically, the
reaction rate depends on the energy difference
\be
E_R \equiv E_{12}^\star - 3 E_4\,,
\ee
where $E_{12}^\star$ is the excited energy level of the carbon-12
nucleus and $E_4$ is the ground state energy of helium.  The 0$^+$
resonance provides the dominant contribution and has an energy level
of 7.6444 MeV, which results in a resonance energy of $E_R \approx
379.5$ keV.  The net reaction rate for fusing three alpha particles
into carbon then takes the form 
\be
R_{3\alpha} = C n_\alpha^3 \left({1\over|E_4|T}\right)^3
\Gamma_\gamma \exp\left[-{E_R\over T}\right] \,,
\ee
where $\Gamma_\gamma$ is the radiative width of the Hoyle excited
state, $n_\alpha$ is the number density of $\alpha$ particles, and
$C=3^{3/2}8\pi^3$ is a dimensionless constant.  The key feature is
that the reaction rate is exponentially sensitive to the resonance
energy $E_R$. As a result, small changes in the resonance energy
$\Delta E_R = E_R - (E_R)_0$ could lead to large changes in the
reaction rates that produce carbon. Moreover, relatively small changes
to the fundamental parameters ($\alpha,\alpha_s,\dots$) can readily
change the resonance energy \cite{epelbaum2003,epelbaum2009,
epelbaum2011,epelbaum2012,epelbaum2013,lahde,meissner,meissner2} (see
also \cite{deboer} for a review of C$+\alpha\to$O reaction). This
sensitivity has led many previous authors to assert that the triple
alpha process must be fine-tuned in order for the universe to produce
carbon (e.g., \cite{barnes2012,bartip,carr,hogan,aguirre}).

The first stellar evolution calculations to address this issue
\cite{livio} considered stars with masses $M_\ast=20M_\odot$ and
found that an increase in the $0^+$ energy level by $\Delta{E}_R$ =
$+60$ keV did not substantially alter carbon production, but an
increment of $\Delta{E}_R$ = $+277$ keV substantially reduced carbon
yields.  Subsequent studies found qualitatively similar results, but
showed that the carbon yields depend on stellar mass for a given change
in the resonance level \cite{csoto,oberhummer,schlattl}.

\begin{figure}[tbp]
\centering 
\includegraphics[width=1.0\textwidth,trim=0 150 0 150]{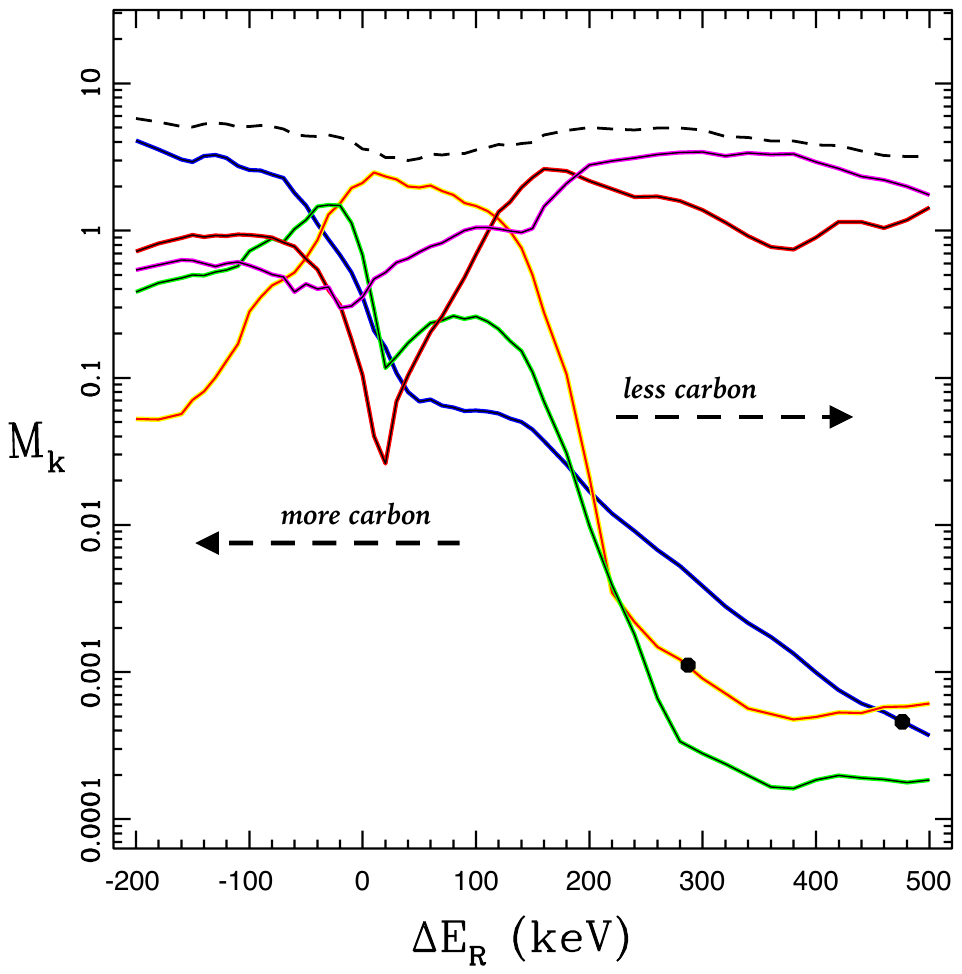} 
\vskip-12pt
\caption{Yields of alpha elements (composed of multiple helium
nuclei) from nucleosynthesis in massive stars as a function of the
change in the energy level of the $0^+$ resonance in the carbon
nucleus (adapated from \cite{huang}).  The yields are given for a
collection of stars in the mass range $M_\ast=15-40M_\odot$, weighted
by the initial mass function. The curves correspond to different alpha
elements, including carbon (blue), oxygen (orange), neon (green),
magnesium (red), and silicon (magenta). The upper dashed curve
(black), which is roughly constant, shows the total yield from all of
the alpha elements. The stellar population produces a net positive
carbon output for changes in the resonance energy up to approximately
$\Delta E_R \approx$ 480 keV. }
\label{fig:metals} 
\end{figure}

Since stars of different masses alter their carbon production in
different ways as the resonance level changes, one must consider a
range of stellar masses. An example is shown in Figure
\ref{fig:metals}, which plots the carbon yields \cite{huang} (see
also \cite{sandora2022a}) for a collection of massive stars with a
standard initial mass function \cite{salpeterimf}, where the stellar
evolution simulations use the {\sl MESA} computational package
\cite{mesaone,mesatwo}. The blue curve shows the net carbon yield
as a function of the change in resonance energy $\Delta E_R$. Note
that if the resonance energy decreases, then the carbon production
increases. The carbon yields decrease with increasing $\Delta E_R$,
but continue to produce a net positive carbon contribution until the
level reaches nearly $\Delta E_R=+500$ keV. More quantitatively,
changing the resonance level by $\sim300$ keV results in an order of
magnitude increase (decrease) when the level is moved downward
(upward).

Since allowed range extends down to $\Delta E_R = -300$ keV,
Figure \ref{fig:metals} shows that stars can produce carbon over the
range of resonance energies from $-300$ to $+500$ keV. This range can
be compared to two benchmarks. First note that although the $^8$Be
nucleus is unstable, it only fails to be bound by an energy increment
of $\sim92$ keV. As a result, the changes to nuclear physics required
so that stars can no longer make carbon through the triple-alpha
process are much larger (by almost an order of magnitude) than the
changes required to make $^8$Be stable. This turn of events would
obivate the need for the triple-alpha process altogether. In a
universe with stable $^8$Be, stars readily fuse $^4$He into $^8$Be in
their first post-main-sequence phase. The resulting nuclei can then
produce carbon, either as a later stage of nuclear burning in the same
star, or in subsequent stellar generations \cite{agalpha}.  

As another benchmark, the resonance energies for nuclear reactions are
determined by the energy levels of the excited states of the nuclei,
and these levels have typical spacings of $\sim1$ MeV. For carbon
nuclei, the excited states have energies at $E\approx4.4$, 7.7, 9.6,
13, and 15 MeV, so that the spacing is $\sim2.7$ MeV.  Significantly,
the allowed range of $\Delta E_R$ of 800 keV = 0.8 MeV is a nontrivial
fraction of the spacing interval. This spacing implies a one in four
chance of being sufficiently near a resonance to produce carbon.

With the ranges of allowed energy levels for the triple-alpha process
estimated, one would like to place constraints on the more fundamental
constants (coupling parameters and quark masses) that determine those
energies. The dependence is complicated, as the energy levels depend
on multiple parameters (quark masses and the fine-structure constant),
and because the calculations still have some unknowns (e.g., see 
\cite{epelbaum2013,epelbaumtwo}). For example, for some coupled
variations in the fundamental parameters, the leading order variation
in the resonance energy can be made to vanish.\footnote{More
specifically, Ref. \cite{epelbaum2013} states ``Given the current
theoretical uncertainty in $A_s$ and $A_t$, our results remain
compatible with a vanishing $\partial E_R/\partial m_\pi$, in other
words with a complete lack of fine-tuning.'' Here the $A_k$ are
scheme-dependent quantities in the lattice calculation and $m_\pi$ is
the pion mass.}  In the generic case, however, the simulations show
that both the light quark masses and the fine-structure constant must
change by $\sim2-3\%$ to alter the carbon resonance energy by 100
keV \cite{epelbaum2013}. These estimates, combined with the results
from stellar structure (see Figure \ref{fig:metals}), imply allowed
variations of order 10\% in the quark masses.  The same calculations
provide corresponding estimates for the changes in the binding energy
of $^8$Be. The derivatives that specify how much the binding energy
changes with input parameters are comparable --- but somewhat smaller
than --- those for $E_R$. As a result, smaller changes to the quark
masses (and/or $\alpha$) are needed to enforce a 92 keV change in
binding energy $E_8$ (making $^8$Be stable) than are needed to change
the carbon resonance enough to compromise the triple-alpha process.

Figure \ref{fig:metals} shows an additional trend of interest. As the
resonance level increases and the carbon yields decrease, the
abundance of oxygen increases at first. For higher resonance levels,
the triple-alpha processes continues to make carbon, but it operates
at a higher temperature. As a result, the carbon nuclei are
subsequently fused into oxygen. For still higher values of the
resonance energy, the operating temperature continues to increase, and
the oxygen is burned into neon. As a result, we see that stars tend to
make alpha elements of increasing atomic number. The relative
abundances of the alpha elements depend on the energy level of the
triple-alpha resonance (which sets the operating temperature) but the
total yield in alpha elements is nearly constant, as shown by the
dashed black curve in Figure \ref{fig:metals}. This finding opens
up yet another channel for carbon production. If massive stars make
oxygen, for example, instead of carbon, then spallation reactions
can break apart the oxygen nuclei to produce carbon, either in the
interstellar medium, including supernova remnants \cite{spallsnr}, 
or in the disks surrounding forming stars \cite{adamsalpha}.  

\medskip 
{\bf Bottom line:} {\sl The production of carbon is not as sensitive
to the energy level of the carbon nucleus as is commonly thought. The
allowed range of resonance energy (about 800 keV) is an appreciable
fraction of the spacing between nuclear energy levels (about 3
MeV). In additon, the changes to nuclear physics required to render
stars incapable of producing carbon (about 500 keV) are much larger
than the changes required to make beryllium-8 stable (92 keV), which
would obviate the need for the triple-alpha process. }

\bigskip 
\section{Stable Diprotons} 
\label{sec:diproton} 

In our universe, deuterium is the only stable nucleus with atomic mass
number $A=2$. With modest changes to the strong interaction, diprotons
would have a stable bound state. This change can be brought about by
increasing the strong coupling constant by $\sim10\%$
\cite{davies1972,pochet,reessix,rees2003} and/or by changes to quark
masses \cite{barrkhan,beane}. In this scenario, protons could undergo
fusion in stellar cores by means of the strong interaction.  In
contrast, in our universe, protons must first fuse into deuterium,
which requires a proton to transform into a neutron, and this process
requires the weak interaction.  Moreover, the nuclear reaction cross
section using strong interactions is estimated to be larger than the
usual reaction by a factor of $\sim10^{18}$ \cite{dyson1971}. Because
of this enormous enhancement, many authors have claimed that the
existence of stable diprotons would render the universe lifeless
\cite{bartip,bradford2009,davies1972,dyson1971,hogan,reessix}.
Specifically one might worry that the enormous cross section would
lead to short stellar lifetimes. As it turns out, however, stars are
remarkably self-regulating and can last for trillions of years, even
with these greatly enhanced cross sections.

The expected exhancement in the cross section can be understood as
follows. Although the existence of stable diprotons allows for a large
number of nuclear pathways, one of the fundamental reactions that
produces diprotons has the form 
\be
p + p \to \diproton + \gamma \,. 
\ee
This reaction is the analog of the fusion reaction for deuterium
in our universe, 
\be
p + d \to {^3{\rm He}} + \gamma\,.
\label{dburning} 
\ee
The standard argument is that the two reactions are the same except
for the additional neutron in the second case, but the neutron plays
only a spectator role, so that the cross sections should be similar
(see the discussion of \cite{adamsdiproton} and references therein). 
The deuterium burning reaction has a larger cross section than its
counterpart involving the weak interaction ($p+p\to d+e^{+}+\nu_e$)
by the factor of $\sim10^{18}$. 

Given that deuterium burning occurs in our universe, one can
immediately see that the large enhancement in the cross section need
not be catastrophic. When stars form, they are too large in radius and
too cool in their cores to sustain nuclear reactions \cite{sal1987}.
Instead, during the earliest phases of stellar evolution, the
luminosity is governed by gravitational contraction (as well as
accretion luminosity if the star is still gaining mass). As
contraction proceeds, the stellar radius decreases and the core
temperature increases.  When the temperature reaches about one million
kelvin, deuterium burning takes place through the reaction of equation
(\ref{dburning}). This temperature ($T_c \sim10^6$ K) is well below
the typical hydrogen burning temperature in stellar cores
($T_c \sim1.5\times10^7$ K). This order of magnitude lower temperature
largely compensates for the 18 orders of magnitude increase in the
cross section -- see the discussion below -- so that the stellar
luminosity is roughly comparable during deuterium burning and hydrogen
burning. The deuterium burning phase only lasts about $10^5$
years \cite{stahlerbirth}, instead of $\sim10^{10}$ years for hydrogen
burning, because the deuterium abundance is only $\sim10^{-5}$. After
the star runs out of deuterium fuel, it contracts further so that the
core reaches the temperature of $T_c\sim15$ million kelvin required
for hydrogen burning.

The extreme temperature dependence of the nuclear reaction rates arises
because of quantum mechanical tunneling: For the conditions present in 
stellar cores, with temperature of millions of kelvin, the classical
turning point for protons is orders of magnitude larger than the range
of the strong force. As result, the protons not only have to tunnel in
order to fuse, but they have to tunnel a long distance. This ordering
of scales makes sense, as the typical proton lives for billions of
years in the core of a sun-like star, so that the fusion probability
must be low. The velocity-averaged cross section for nuclear reactions 
in a star can be written in the form 
\be
\langle\sigma{v}\rangle = {8\over9} \left({2\over3E_Gm_R}\right)^{1/2}
{\cal S}_{\rm eff} \Phi^2 \exp[-\Phi] \,,
\label{avesigma} 
\ee
where the cross section itself has the form
\be
\sigma(E) = {S(E)\over E} \exp[-(E_G/E)^{1/2}] \,,
\ee
where the Gamow energy $E_G=2(\pi\alpha Z_1Z_2)^2 m_R$. The cross
section thus includes the effects of Coulomb repulsion and the net 
cross section $\langle\sigma{v}\rangle$ results from averaging over a
thermal distribuiton of particles at temperature $T$. The other part
of the cross section is written as $\sigma\sim S/E$ to separate out
the primary energy dependence, so that the factor $S$ is a slowly varying
function of $E$. In equation (\ref{avesigma}), ${\cal S}_{\rm eff}$ is
the energy-averaged version of $S(E)$ and the exponential suppression
due to tunneling is encapsulated in the parameter 
\be
\Phi = 3 \left({E_G \over 3T}\right)^{1/3} \approx 20
\left({T\over10^7{\rm K}}\right)^{-1/3}
(Z_1Z_2)^{2/3} \left({A_1A_2\over A_1+A_2}\right)^{1/3} \,. 
\ee
For hydrogen burning in the core of the Sun, the parameter
$\Phi\approx14$. For the reactions that fuse protons into diprotons,
the nuclear burning temperature is about $T_c\sim10^6$, similar to that
of deuterium burning, so that $\Phi\approx40$. This additional
suppression due to tunneling ($\exp[-40]$ compared to $\exp[-14]$)
largely compensates for the increase in cross section (the increase
in the coefficient ${\cal S}_{\rm eff}$ by factors
of $10^{15}-10^{18}$).

\begin{figure}[tbp]
\centering 
\includegraphics[width=1.0\textwidth,trim=0 150 0 150]{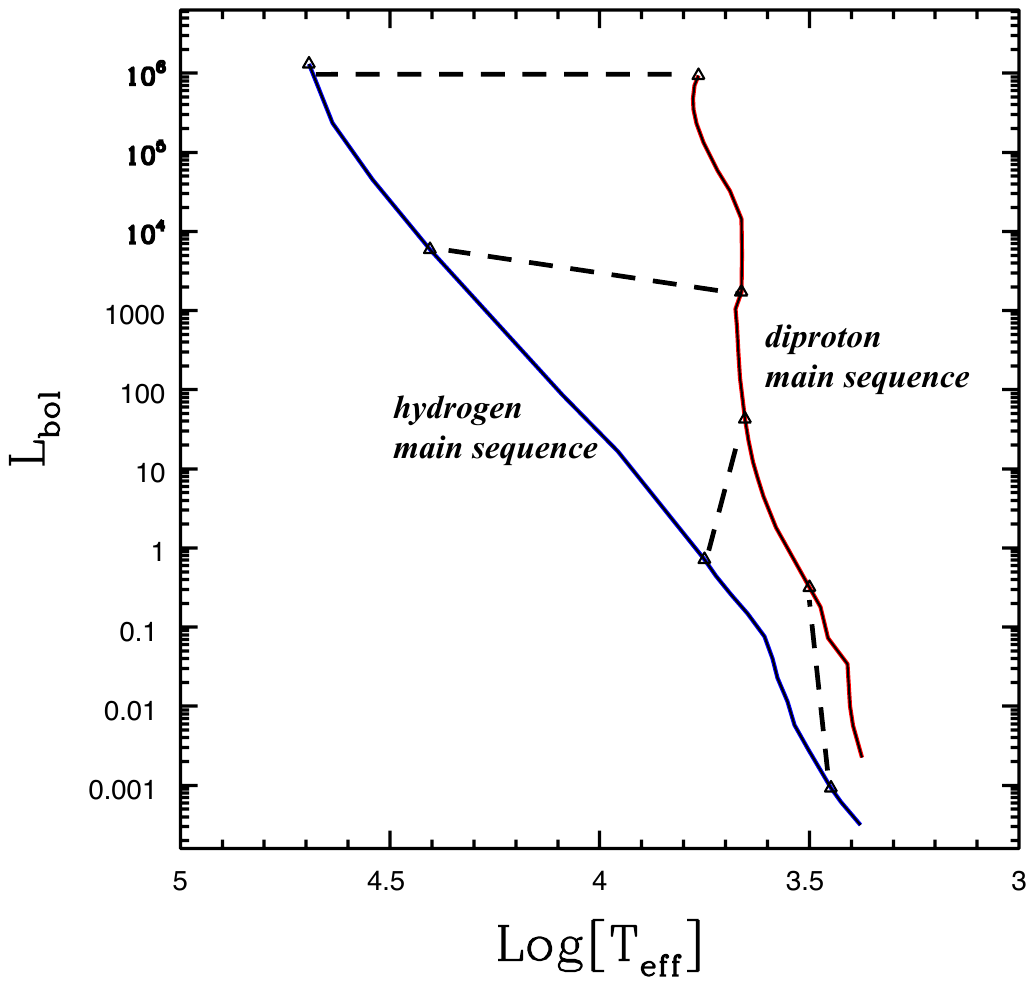} 
\caption{H-R Diagram comparing the main sequence for standard
hydrogen burning (blue curve on the left) and that for universes with
stable diprotons (red curve on the right). The luminosities
$L_{\rm bol}$ are given in solar units and the temperatures
$T_{\rm eff}$ are given in kelvin. (Note that temperature is plotted
backwards in accordance with astronomical tradition.)  The cross
section for producing diprotons is assumed here to be larger than that
of the standard $p$-$p$ chain by a factor of $10^{15}$. The triangular
symbols denote stellar masses of 100, 10, 1, and 0.1 $M_\odot$ from
top to bottom. }
\label{fig:diproton} 
\end{figure} 

Using the modifications for nuclear reactions outlined above,
one can solve the equations of stellar structure
\cite{chandra,kippenhahn,phillips}, using either a semi-analytic
approach \cite{adams,adamsnew,barnes2015} or state-of-the-art
computational modules \cite{mesaone,mesatwo}. One example is shown by 
the Hertzsprung-Russell (H-R) diagram of Figure \ref{fig:diproton}
(from \cite{adamsreview}, see also \cite{adamsdiproton}), which plots
the stellar luminosity $L_{\rm bol}$ as a function of photospheric
temperature $T_{\rm eff}$. The diagram shows the main sequence for
hydrogen burning in our universe (blue curve) and the corresponding
main sequence for a universe with stable diprotons (red curve). In the
latter case, the nuclear reaction cross section is enhanced by a
factor of $X=10^{15}$. The triangles show points on the main sequence
for particular masses, 0.1, 1, 10, and 100 $M_\odot$ from bottom to
top. Note that for high masses, the luminosities are similar. Since
the amount of nuclear fuel is the same, the stellar lifetimes are
comparable. At mass $M_\ast=1 M_\odot$, the luminosity is about 10
times larger for a universe with diprotons, so that such stars only
live about 1 billion years. For slightly lower masses, however, stars
live longer, so that stars with luminosity and lifetimes comparble to
the Sun have masses $M_\ast\sim0.5M_\odot$. In addition, for the
greatly enhanced cross section (with diprotons) the main sequence
extends down to lower masses $M_{\ast min}\sim0.01M_\odot$, and the
longest lived stars shine for trillions of years \cite{adamsdiproton}.
The surface temperatures for diproton stars span a somewhat narrower
range than in our universe, where $T_\ast\approx3000-4000$ K. For a
given mass, the stars are thus cooler, redder, and larger, but only by
modest factors.

Another interesting change arises: In our universe, the normal
operating temperature for a star burning hydrogen (at
$T_c\sim1.5\times10^7$ K) is higher than the temperature for the
reaction $^3$He + $^3$He $\to$ $^4$He + 2p. The chain of nuclear
reactions thus promptly processes protons into helium. For stars with
greatly enhanced cross sections, however, the central temperature (at
$T_C\sim10^6$ K) is too cool to burn the $^3$He. As a result, nuclear
reactions burn protons into helium in two phases. First, the stellar
core produces $^3$He until all of the hydrogen fuel is exhausted. The
star then readjusts its internal structure and contracts, so that the
core heats up and the $^3$He is then burned into $^4$He.

The discussion thus far has considered variations in the nuclear
reaction cross sections, but the values of the gravitational constant
$G$ and the fine-structure constant $\alpha$ have been held fixed. As
considered in the previous section, however, larger cross sections
allow for more of the $\alpha$-$G$ plane to support solutions to the
stellar structure equations and hence working stars
\cite{adams,adamsnew,adamsreview,barnes2015}. 

\medskip 
{\bf Bottom line:} {\sl In universes with stable diprotons, stars can
live for trillions of years. They can form and function over a wider
range of parameter space (including the values of $\alpha$ and $G$, as
well as the stellar mass $M_\ast$) than in universes without stable
diprotons. }

\bigskip 
\section{Unstable Deuterium} 
\label{sec:deuterium} 

In our universe, the deuterium nucleus has a small binding energy of
only $\sim2.2$ MeV. A relatively small decrease in the strength of the
strong force ($[\Delta\alpha_s]/\alpha_s\sim0.1$ \cite{davies1972,
reessix}) would result in unstable deuterium. This scenario, in turn,
appears to be problematic, as deuterium is the first step in the chain
of nuclear reactions that builds heavy elements. Recall that the
universe emerges from its early epochs with only protons and neutrons
so that, in the absence of three-body reactions, the universe must
produce $A=2$ nuclei before forging larger ones. Citing this issue, a
number of authors have claimed that universes without stable deuterium
cannot support working stars and that such regions would not be
habitable \cite{bartip,barnes2017,donoghue,hogan,pochet,lewbarn,
reessix,schellekens,tegmark100,tegmark}. Such claims are premature.
In the absence of stable deuterium, the usual network of nuclear
reactions must be modified, but stars have alternate channels to
produce both energy and heavy elements. Gravitational contraction of
stars, expolosive nucleosynthesis, the CNO cycle, and triple-nucleon
reactions provide potential avenues for stars to generate energy and
synthesize larger nuclei.

We first consider gravitational contraction of stars. In our universe,
stars are born with radii that are too large and central temperatures
that are too cool to sustain nuclear fusion \cite{sal1987}. As a
result, low mass stars (i.e., most stars) are born without engaging
their nuclear power source. They spend millions of years deriving
energy from gravitational contraction until their central cores attain
the proper conditions for fusion to occur.  The onset of fusion
provides the star with energy and hence pressure to hold up the star
and arrest further contraction. In the absence of nuclear reactions,
the phase of pre-main-sequence contraction would continue until the
end of the star's life. Through this channel of evolution, stars can
produce power comparable to the solar luminosity for timescales up to
$\sim1$ Gyr \cite{agdeuterium}. This process produces enough energy to
sustain life, but not the necessary heavy elements. 

At the end of the extended phase of gravitational contraction, the
fate of stars is determined by their mass. For stars with masses below
the modified Chandrasekahr mass at $M_\ast\approx5.6M_\odot$
\cite{chandra}, further contraction is halted by quantum mechanical
degeneracy pressure, and the objects become white dwarfs with a pure
hydrogen composition. For larger stars, the mass is sufficient to
overcome degeneracy pressure, and the stellar cores continue to
collapse.

Numerical simulations \cite{agdeuterium} follow the condensing stars
until the temperatures exceed $T\sim10^{10}$ K and the densities
exceed $\rho\sim10^{11}$ g cm$^{-3}$. The stars will contract even
further, as pressure forces are inadequate to support the structure.
Although detailed nucleosynthesis calculations have not been carried
out, the conditions are favorable, given the following considerations:
These temperatures ($T\sim1$ MeV) are high enough to drive nuclear
reactions without the need for tunneling through the Coulomb barrier.
Under these conditions ($T\sim2\times10^{10}$ K and $\rho\sim10^{10}$
g cm$^{-3}$) the weak interaction rate exceeds the gravitational
collapse rate ($\sqrt{G\rho}$) so that nuclear statistical equilibrium
(NSE) can be maintained. At somewhat lower densities ($\rho\sim10^8$ g
cm$^{-3}$), electron capture reactions ($p+e^{-}\to n+\nu_e$) become
operable, and the stellar gas develops a population of neutrons. This
neutron population allows (unstable) deuterium to be produced via the
strong interaction, and the resulting NSE abundance of deuterium
results in Helium production (via $d+p\to$ $^3$He + $\gamma$).
Moreover, the nuclear reaction rates increase with density faster than
the rate of gravitational collapse.\footnote{The reaction rate for
$^3$He production $\Gamma$ = $n_d\langle\sigma{v}\rangle$, where $n_d$
is the deuterium abundance.  Using the Saha equation, one obtains $n_d
= n_p^2 \lambda_T^3$, where $\lambda_T=(2\pi/m_r T)^{1/2}$, and
where deuterium has zero binding energy. The nuclear reaction rate
thus scales as $\Gamma \propto \rho^2 T^{-3/2}$.  For an $n=3$
polytrope, $\Gamma\propto\rho^{3/2}$. For comparison, the collapse
rate scales as $\sim\sqrt{G\rho}$.} As a result, nuclear reactions
have time to operate and a some amount of nuclear processing will take
place as these stars meet their demise. Note that even if NSE is not
fully achieved, out of equilibrium reactions can also take place.

The collapsing star is roughly analogous to a white dwarf in our
universe that accretes mass and exceeds the Chandrasekhar limit,
thereby leading to a supernova. The nuclei that are produced, such as
$^3$He and $^4$He, can be expelled in the explosion, and then be
incorporated into later generations of stars where they provide
nuclear fuel. In this manner, explosive nucleosynthesis can jump the
$A=2$ barrier and thereby endow the universe with heavy elements.
However, additional work is necessary to understand the conditions
required for such explosions to occur.

The triple-nucleon process \cite{agdeuterium} provides another channel
for stellar nucleosynthesis. This reaction network is roughly
analogous to that of the triple-alpha process, which produces carbon
in our universe (see \cite{clayton,kippenhahn,hoyle,salpeter} and
Section \ref{sec:triplealpha}). Consider the case where deuterium is
unstable, and has a short but finite half-life. The production process
$p+p\to d+e^{+}+\nu_e$ will still take place, although the product
deuterium nucleus will decay. As a result, the stellar core will build
up a standing population of deuterium nuclei, where the abundance is
determined by the competition between the production and decay rates.
In this context, the production rate is expected to be small, because
the weak interaction is required to convert one of the reacting
protons into the neutron found in the deuterium nucleus. When the
deuterium nucleus decays, however, the weak interaction would also be
required for the inverse channel $d\to p + p + e^{-} + {\bar\nu}_e$.
The more common decay channel would produce a both proton and a
neutron, $d\to p+n$, and the remaining neutron can interact ($n+p\to
d$) much more rapidly.  Although the neutron can only live about 10
minutes before decaying, this latter process ($n+p$) takes place more
quickly.  Through this network of reactions, the standing population
of deuterium can be large enough that the next reaction in the chain,
$p+d\to$ $^3$He, can occur fast enough to power the star.

\begin{figure}[tbp]
\centering 
\includegraphics[width=1.0\textwidth,trim=0 150 0 150]{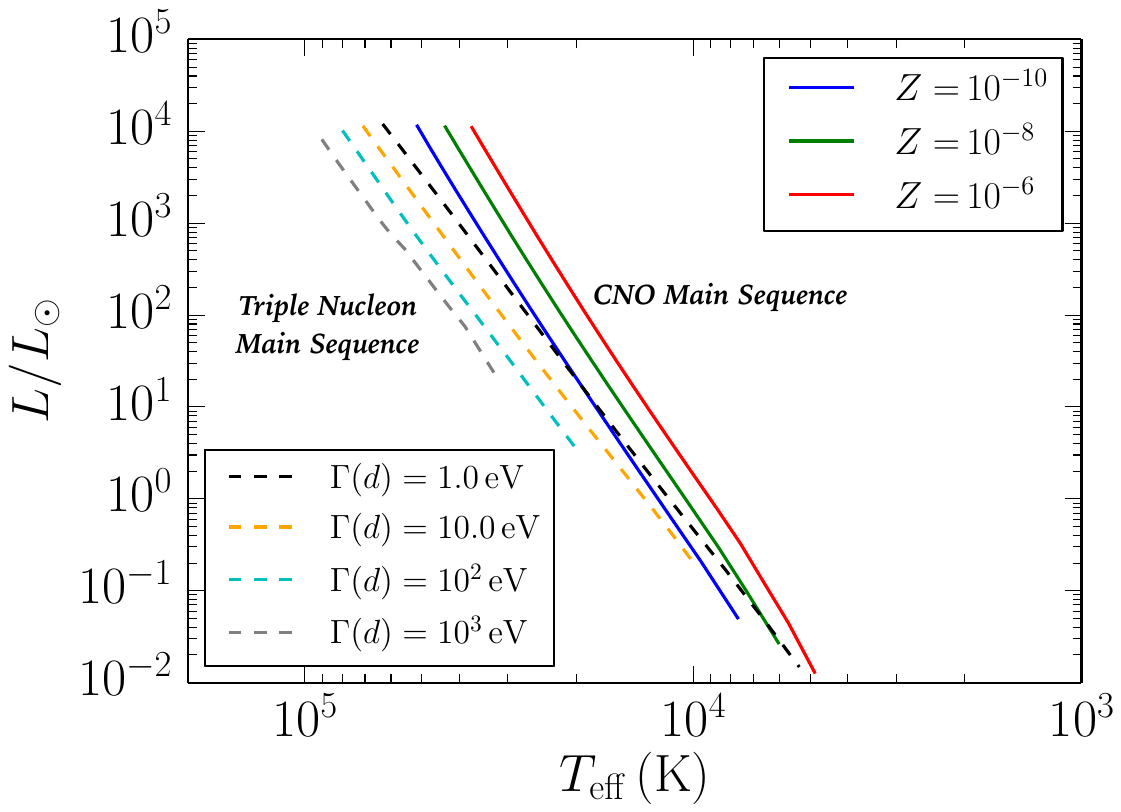} 
\caption{H-R Diagram for stars in universes without stable deuterium
(adapated from \cite{agdeuterium}). Dashed curves (left side of the
diagram) show the main sequences for stars operating through the
triple-nucleon process, where the decay width for unstable deuterium
is given in the inset on the left.  The solid curves show main
sequences for stars operating through the CNO cycle only, where the
metallicity (which determines the carbon abundance) is given in the
inset on the right. }
\label{fig:hrnod} 
\end{figure} 

Numerical simulations of stars with the triple-nucleon process show
that the resulting main sequences are relatively normal, provided that
the lifetime of the (unstable) deuterium nucleus is long enough.
Figure \ref{fig:hrnod} shows the H-R diagram for a collection of stars
operating through the triple-nucleon process \cite{agdeuterium}, where
the relevant reactions were programmed into existing stellar evolution
modules \cite{mesaone}. Results for shown for a range of values for
the decay width for the deuterium nucleus $\Gamma(d)=1-1000$ eV, which
corresponds to lifetimes of order $10^{-16}-10^{-19}$ sec. The longest
lifetime is comparable to that of unstable $^8$Be, which determines
the properties of the triple-alpha process. As the decay width grows,
so that the lifetime decreases, the mass range for stars that can
operate through the triple-nucleon process shrinks. Although not shown
in the figure, the highest mass stars can operate with decay widths up
to $\Gamma(d)\sim10^5$ eV, corresponding to lifetimes of order
$10^{-21}$ sec.\footnote{Note that this lifetime corresponds to an
energy $E\approx$ 0.66 MeV, an energy scale expected for nuclear
reactions.} The stellar luminosities for these objects span a wide
range, from fainter than the Sun up to $L_\ast=10^4L_\odot$, with the
range decreasing with the decay width $\Gamma(d)$. The surface
temperatures are mostly above $T\sim10^4$ K, so that the stars are
hotter than in our universe. The coresponding lifetimes are also
somewhat shorter, with a maximum of about 1 Gyr.

With these stellar properties, the following scenario unfolds: In a
universe without stable deuterium, the first generation of stars can
operate through the triple-nucleon process, with the stellar
properties depicted by the dashed curves in Figure \ref{fig:hrnod}.
These stars will be hotter and have shorter lifetimes than those in
our universe, so they will not be ideal hosts for potentially
habitable planets. Through their nuclear processing, however, these
stars will produce helium, carbon, and other heavier elements. If the
first stars only produce helium, then the next stellar generation can
burn the helium into carbon.  In the limit of extremely short
lifetimes for deuterium, $\sim10^{-21}$ sec, only the most massive
stars will operate. With even shorter deuterium lifetimes, even
massive stars cannot be supported, but nuclear reactions will occur
while such objects evolve, implode, and then explode. As long as trace
amounts of carbon can be produced, subsequent stellar generations can
operate through the CNO cycle and have ordinary stellar properties.
The main sequences for these CNO stars are shown as the solid curves
in the figure, and this process is considered next.

In the CNO process, carbon acts as a catalyst to faciliatate the
fusion of protons into helium. The chain of nuclear reactions can
be written
\be
^{12}{\rm C} + p \to {^{13}{\rm N}} + \gamma \qquad 
^{13}{\rm N} \to {^{13}C} + e^{+} + \nu_e \qquad 
^{13}{\rm C} + p \to {^{14}{\rm N}} + \gamma 
\ee
$$
^{14}{\rm N} + p \to {^{15}{\rm O}} + \gamma \qquad 
^{15}{\rm O} \to {^{15}{\rm N}} + e^{+} + \nu_e \qquad
^{15}{\rm N} + p \to {^{12}{\rm C}} + {^4{\rm He}} 
$$
Note that the carbon nucleus is returned to the stellar core, so that
the net reaction is $4p\to$ $^{4}$He + leptons.  Our Sun generates
only a fraction of its power through this process
\cite{clayton,kippenhahn,phillips}, but this reaction network
dominates energy generation in larger stars. In the absence of stable
deuterium, the usual $p$-$p$ chain does not operate, so stars must
produce energy through either the triple-nucleon reaction (see above)
or through the CNO cycle. As long as the carbon abundance is nonzero,
the CNO cycle tends to dominate over the triple-nucleon process. 

Figure \ref{fig:hrnod} shows the main sequences for stars operating
through the CNO cycle only (solid curves on the right).  These stars
have properties much like those in our universe, with luminosities
ranging from $10^{-2}L_\odot$ to $10^4L_\odot$ and surface
temperatures in the range $T$ = 5000 K to 40,000 K. The corresponding
lifetimes are similar to those in our universe, with the largest stars
burning hydrogen for millions of years and the smaller stars lasting
for trillions of years \cite{agdeuterium}. The finding that the CNO
cycle can operate and allow stars to function is not surprising. What
could be considered counterintutive is that the carbon adundance
required for stars to have normal properties is quite low. With a
metallicity of only $Z=10^{-10}$, stars operate over most of the usual
main sequence. The highest mass stars can sustain nuclear fusion via
the CNO cycle with metallicity as low as $Z\sim10^{-14}$. Although
such high mass stars have short lifetimes, they can produce their
own carbon, and then explode as supernovae, thereby seeding their
galaxy with carbon for future stellar generations.

\medskip 
{\bf Bottom line:} {\sl Universes with unstable deuterium have a
number of pathways to remain viable. Stars can function through the
CNO cycle, and the requisite carbon can be synthesized through the
triple-nucleon process and/or explosive nucleosynthesis. Additional
energy generation can occur through gravitational contraction in
non-nuclear stellar objects. }

\bigskip 
\section{Weak Force} 
\label{sec:weak}

Stellar operations place interesting limits on the strength of the
weak interaction. Nuclear reactions in stars produce copious
quantities of neutrinos, which have a small interaction cross
section. As a result, they stream out of the stellar interiors with
essentially no interaction, thereby providing energy loss with no
pressure support. If the weak interaction were significantly stronger,
however, neutrino propagation would occur under optically thick
conditions and the structure of stars would change accordingly
\cite{howeweakful}. The optical depth for neutrino interactions is
given by the integral
\be
\tau_\nu = \int_0^{R_\ast} n \sigma_\nu dr \approx
\sigma_\nu \langle{n}\rangle R_\ast \,, 
\ee
where the interaction cross section has the form $\sigma_\nu \approx
G_F^2 E_\nu^2$. Using the requirement that neutrinos remain optically
thin, so that $\tau_\nu\simless1$, one can derive an upper bound on
the strength of the weak force. Using standard results from stellar
structure theory \cite{chandra,hansen,phillips}, the optical depth 
expression takes the form \cite{adamsreview}
\be
\tau_\nu \sim G_F^2 E_\nu^2 T_c^2 \alpha_G^{-1/2}\,, 
\ee
where $T_c$ is the temperature of the stellar core (typically about 15
million K), and where we neglect dimensionless constants of order
unity.  Enforcing the constraint $\tau_\nu<1$ implies the corresponding
limit $G_F \simless 24,000 (G_F)_0$, where the latter value is that in
our universe $(G_F)_0\approx$ (293 GeV)$^{-2}$.

An analogous constraint arises in the context of supernova
explosions \cite{carr}.  Although the detonation mechanism remains
under study \cite{janka,jankarev}, neutrinos are thought to play an
important role. In order for neutrino interactions to affect the
explosion, their interaction rate must be roughly comparable to the
local collapse time of the star, so that $n \sigma_\nu v \sim
(G\rho)^{1/2}$. When stars explode, they first collapse to attain
nearly nuclear energy density so that $\rho \sim \mpro m_\pi^3$, where
$m_\pi$ is the pion mass. The neutrino energy $E_\nu$ is given by
nuclear energy scales, so that $E_\nu\sim m_e$ in order of magnitude.
Using these expressions, the requirement that neutrino interactions
are significant has the approximate form
\be
G_F^4 \sim {\mpro \over \mplanck^2 m_e^4 m_\pi^3} \,. 
\ee
This constraint can be written in terms of the dimensionless
coupling constant using $\alpha_w = G_F \mpro^2$, so that we find
\be
\alpha_w^2 \sim \left({\mpro\over\mplanck}\right)^2
\left({\mpro\over m_\pi}\right)^3
\left({\mpro\over m_e}\right)^4\,.
\label{alphaweak} 
\ee
If the weak interaction is too weak, then all of the neutrinos freely
stream out of the collapsing star, so that they cannot contribute to
driving the explosion. If the weak interaction is too strong, then the
neutrinos become optically thick and provide additional pressure
forces which suppress the collapse and could suppress the explosion.
Although supernovae require two sides of equation (\ref{alphaweak})
to be comparable, we do not know how close the condition must be to
equality. These issues are complicated by the two- and three-
dimensional nature of fluid motions during supernova detonation.

In the worst case scenario, core collapse supernovae could fail,
which would shut down one channel for heavy element production for the
universe in question. But explosions of massive stars are not the only
source of heavy elements \cite{frebelbeers}. Supernovae can also be
instigated by mass acccretion onto white dwarfs, when the final mass
exceeds the Chandrasekhar limit, and the resulting explosion produces
large quantities of heavy nuclei \cite{iwamoto,nomoto1997}. For nuclei
with intermediate atomic numbers, such a carbon, winds from asymptotic
giant branch stars \cite{agbstars} represent an important supply chain
(see also \cite{gustafsson,henry}). For the heaviest elements that
must be synthesized by the rapid neutron capture process, neutron star
mergers represent the primary source \cite{freiburghaus}. Finally, we
note that spallation processes also contribute to nucleosynthesis in a
variety of astrophysical environments \cite{adamsalpha,spallradiation}.
One additional complication is that the core collapse supernovae can
produce neutron stars, the bodies that collide during mergers,
although neutron stars can also be produced by white dwarf collisions
\cite{wdmerger}.  The flow chart for producing heavy nuclei is thus
complicated, with multiple branches, so that alternate universes have
many alternate paths to produce the nuclei necessary for life.

Changes in the weak force also affect Big Bang Nucleosynthesis
(BBN). The outcome of BBN determines the inventory of nuclei for
subsequent stellar operations and depends sensitively on the neutron
lifetime (Section \ref{sec:bbn}), where $\tau_n^{-1}\sim G_F^2m_e^5$
(equation [\ref{neutronlife}]). If the neutron lifetime is too short,
roughly $\tau_n\simless10$ sec, most of the neutrons decay before
interacting and BBN yields become vanishingly small (see Figure 15
of \cite{adamsreview}). In this scenario, the universe emerges from
the BBN epoch with all of its baryons in the form of protons.  This
composition is not problematic, as stars would simply burn protons as
usual and eventually produce heavier elements. In the opposite limit
where $\tau_n$ becomes long, the neutron to proton ratio approaches
unity during BBN and most of the baryons are processed into $^4$He.
The requirement that some protons remain thus sets an upper limit on
the neutron lifetime and a lower limit on $G_F$ (equivalently
$\alpha_w$). The bound on $G_F$ depends on the baryon-to-photon ratio
$\eta$ (see Section \ref{sec:bbn}). For the standard value of $\eta$,
the neurton lifetime $\tau_n$ must be increased by a factor of 1000 to
decrease the proton abundance to below 10 percent. Much larger changes
are needed to compromise the universe.

Bounds on the neutron lifetime beome weaker as $\eta$ decreases.
BBN in our universe operates in a regime where almost every neutron
interacts, first by producing deuterium ($p+n\to d$), which is then
processed into helium. In contrast, essentially no protons interact
with each other, as the reaction $p+p\to d + e^{+}+\nu_e$ is highly
suppressed by the small (weak interaction) cross section and by
Coulomb repulsion. If the baryon abundance decreases, not all of
the neutrons can interact during the BBN epoch and not all of the
baryons can be processed into helium. One can consider the extreme
weakless limit where the coupling strength $\alpha_w\to0$ and the
neutron lifetime $\tau_n\to\infty$. In this regime, BBN produces the
same helium abundance as in our universe for baryon-to-photon ratio
$\eta\sim4\times10^{-12}$ (see \cite{grohsweakless,weakless} and
Section \ref{sec:bbn}; see also \cite{gedalia}). 

For completeness, we note that stars operate more or less normally
when the weak interaction is much weaker than its standard value, or
even absent altogether \cite{grohsweakless}. In the extreme weakless
limit, the universe processes a substantial fraction of its baryons
into helium during BBN, with the remaining inventory consisting of
free neutrons, protons, and some deuterium. Both the relative
abundance of these constituents and the astrophysical setting
determine the subsequent evolution. If stars are formed with ample
supplies of neutrons and deuterium (in addition to protons), then the
neutrons will readily form more deuterium, which is the first step in
the standard $p$-$p$ nuclear reaction chain in stars.  Nuclear
processing will then proceed to the next steps in the sequence by
burning the deuterium through the reactions $p+d\to$ $^3$He $+\gamma$, 
$d+d\to$ $^3$He $+n$, $d+d\to$ $^4$He $+\gamma$, and others. The
end result of the nuclear reaction chain is still the production of
$^4$He, as usual. Since the binding energy of deuterium is only
2.2 MeV, whereas the binding energy of helium is $\sim28$ MeV, the
energy supplied by the nuclear reactions is smaller by only $\sim10$
percent, so that stars have access to comparable stores of energy.
Detailed stellar evolution calculations \cite{grohsweakless} show that
the resulting stars in a weakless universe have surface temperatures,
core temperatures, total luminosities, and mass-radius relations that
are comparable to those of stars in our universe. As a result, stars
in a weakless universe are sufficient for habitability. 

\medskip 
{\bf Bottom line:} {\sl Stars can operate in universes with no weak
force, with properties similar to those in our universe. Stellar
evolution changes when the weak interaction increases in strength so
that neutrinos are optically thick, which occurs for values of
$\alpha_{\rm w}\propto G_F$ that are larger by factors of $\sim10^4$.}

\bigskip 
\section{Planets} 
\label{sec:planets} 

Planets represent the smallest astrophysical objects that are
necessary for the development of life \cite{lunine,scharf}, at least
for life in familiar forms. They provide the immediate environments
where life can originate and thrive. Astronomers have detected more
than 6000 planets to date and estimate that the galaxy contains more
planets than stars. If one counts smaller objects, like dwarf planets
or asteroids, the number of minor bodies far exceeds the number of
stars. At first glance, it seems that a universe with an ample supply
of stars can also produce a corresponding inventory of planets. This
section outlines the conditions required for a universe to contain
planets with the properties needed to support habitability.

The properties required for a planet to be habitable are currently
under debate \cite{ikoma,kasting2003,lammer,lunine99,lunine,scharf}.
The general concensus is that habitable planets must have rocky
surfaces, like Earth, instead of extended atmospheres, like
Jupiter. The preferred planetary mass is within an order of magnitude
of Earth, perhaps within a factor of two.  Much smaller planets cannot
hold onto an atmosphere, whereas much larger planets become gaseous
and have overly strong surface gravity.

Planets have three mass scales of interest. Most considerations of
habitability \cite{lunine,scharf}, suggest that viable planets should
have rocky surfaces, which requires them to be non-degenerate.
The maximum mass scale for planets to remain non-degenerate
\cite{bartip,sandora2016,weisskopf} can be writtten in the form 
\be
M_{\rm nd} \approx A^{-3}
\left( {\alpha \over \alpha_G} \right)^{3/2} \mpro \,,
\ee
where $A$ is the atomic weight of the planetary material. This mass
scale is closely related to the Chandrasekhar mass and the typical
stellar mass scale, where the latter scales do not have the factor of
$\alpha^{3/2}$. This factor arises because electromagnetic forces are
required to keep the planet in ``rocky'' form. Since $\alpha\ll1$,
the typical planetary mass scale is much smaller than that of a star. 

Another mass scale of interest is that required for the planet to
retain an atmosphere. This derivation (e.g., \cite{adamsreview})
assumes that the planet resides in the habitable zone, so that the
temperature of the atmosphere is known. Specifically, one assumes that
the temperature is high enough to drive chemical reactions so that 
\be
T > E_{\rm chem} = \epsilon \alpha^2 m_e \,,
\ee
where $\epsilon\approx0.001$ converts the atomic energy scale
($\sim\alpha^2 m_e\sim27$ eV) to the minimum energy required for
chemistry, which takes place at room temperature where $T\sim0.026$
eV. Note that both the fine-structure constant, which determines atomic
energy levels, and the fraction $\epsilon$ required for chemistry can
vary from universe to universe. The resulting minimum mass scale for
atmospheric retention thus has the form 
\be
M_{\rm atm} = \left({\epsilon\over A A_{\rm atm}}\right)^{3/2} 
\left( {\alpha \over \alpha_G} \right)^{3/2} \mpro \,,
\ee
where $A_{\rm atm}$ is the atomic weight of the atmosphere, which is
in general different from the atomic weight of the bulk material that
makes up the planet. Note that $M_{\rm atm}\ll M_{\rm nd}$ due to the
factor of $\epsilon^{3/2}$.

Another mass scale of interest is that of galaxies. The requirement
that planets have masses smaller than their galactic hosts places
a weak limit on the structure constants \cite{adamsreview} of the form
\be
\alpha_G < \alpha^7 \beta^{-1} \,,
\ee
where $\beta$ is the electron-to-proton mass ratio.  Since gravity is
much weaker than the electromagnetic force, this constraint only comes
into play in the limit where $\alpha$ becomes extremely small.

\begin{figure}[tbp]
\centering 
\includegraphics[width=0.80\textwidth,trim=0 150 0 150]{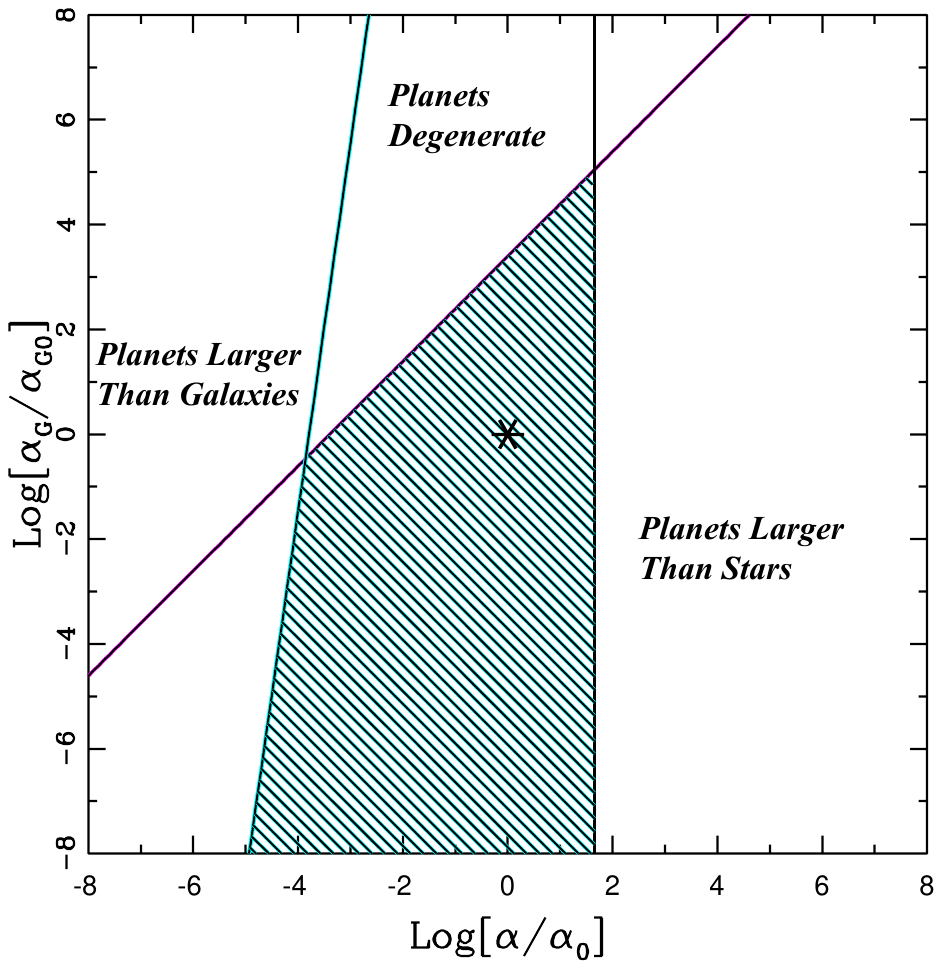} 
\caption{Allowed region for Habitable Planets in the plane of $\alpha$
and $\alpha_G$ structure constants. The region to the right of the
green curve corresponds to large values of the fine-structure
constant, where planets are larger than their host stars.  In the
region to the left of the cyan curve, galaxies have difficulty cooling
rapidly and have smaller masses than the planets.  In the upper region
of the plane, planets cannot be massive enough to support atmospheres
and still remain non-degenerate. }
\label{fig:planetplane} 
\end{figure}

Consistent with the above discussion, Figure \ref{fig:planetplane}
shows the allowed region of the $\alpha$-$\alpha_G$ plane that allows
for habitable planets. On the right side of the diagram, planets
become larger in mass than their host stars. Although this condition
does not necessarily preclude life, it occurs for $\alpha\simgreat1$
and we expect smaller values of the fine-structure constant for
additional reasons.  On the left side of the diagram, for extremely
small $\alpha$, the galactic masses become so small that they contain
few planets. The upper portion of the diagram is disallowed because
planets would be degenerate, and gaseous, and hence not have rocky
surfaces. 

The constraints invoked here are not as definitive as those found
earlier for stars. Stars clearly stop working when no solutions to the
stellar structure equations exist, and these conditions determine 
well-defined boundaries in parameter space. Although the constraints
on planets are not as sharp, the resulting parameter space is
comparable to, but somewhat larger than, that for stars. 

\medskip 
{\bf Bottom line:} {\sl Planets are generally less sensitive to
changes in the fundamental constants than stars, so that universes
with working stars should also have viable planets. However, the
constraints arising from planetary considerations are more uncertain. } 

\begin{figure}[tbp]
\centering 
\includegraphics[width=0.50\textwidth,trim=0 150 0 150]{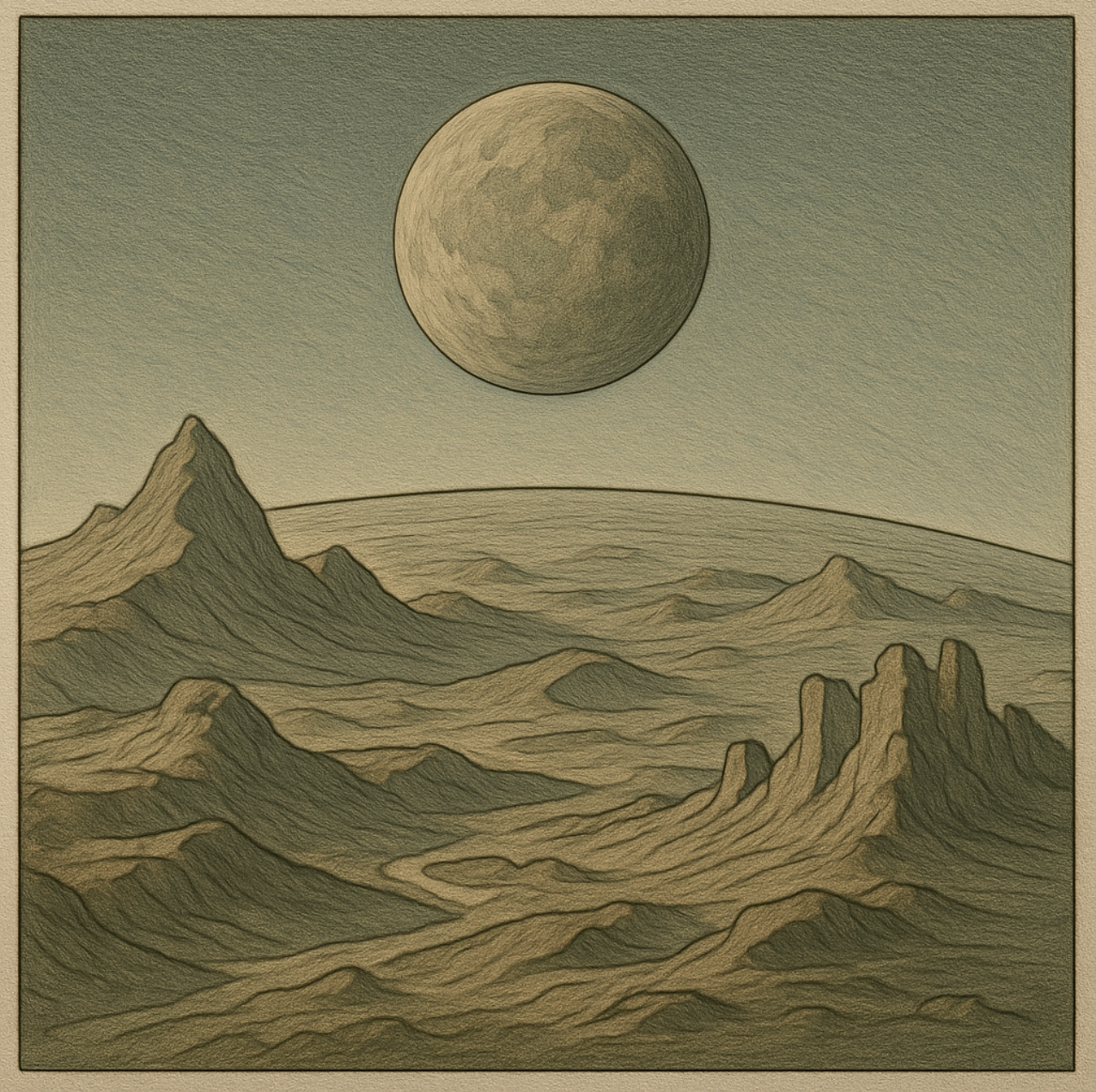} 
\vskip80pt
\caption{Planets provide the most likely environments for life to develop.
The constraints on the fundamental constants required for planets to fill
this role are comparable to those required for successful stars. As a
result, universes with operational stars are likely to have suitable
planets for habitability. [This AI-generated figure will be replaced by
an artist-generated picture in the published version of the article.] } 
\label{fig:planetscape} 
\end{figure}

\newpage 
\bigskip 
\section{Stable Atoms and Nuclei}  
\label{sec:nuclei}

Working our way to smaller scales, we require that atoms and their
nuclei must be stable in order for a universe to be habitable. Atomic
structure depends on the mass of the electron and the fine-structure
contant $\alpha$, which sets the charge on the nuclei.  Here we work
in terms of the ratio $\beta$ of the electron mass to the proton mass.
To start we consider that long-lived stable nuclei exist, and then put
constraints on the values of $(\alpha,\beta)$ that allow for atoms
and hence a working universe. Stability of the nuclei themselves is
considered subsequently. 

The first constraint is that both $\alpha\ll1$ and $\beta\ll1$ in
order for atoms to have structures roughly analogous to those in our
universe \cite{bartip}.  The idea is that this constraint allows for
chemistry to take place in an ordinary manner. If $\alpha$ approaches
unity, the electrons in atoms become relativistic and chemistry is
altered \cite{kingchemistry}. The requirement of a small fine-structure
constant $\alpha\ll1$ also arises from the stability of nuclei (see
below) and the stability of bulk matter
\citep{lieb1975,lieb1990,liebyau,liebyaualt}. Similarly, if $\beta$
approaches unity, then electron orbitals enter a new regime, as the
nucleus would move as much as the electrons. Specifically, as $\beta$
increases, the fluctuation amplitude of nuclei in solid material
increases. If the fluctuations are too large, then every nucleus
shifts out of its place is a solid, thereby destroying the lattice and
melting the solid. By limiting this fluctuation amplitude to be
smaller than the typical distance between atoms, one finds the
approximate constraint $\beta\simless1/81$ \cite{tegmark100}.

\begin{figure}[tbp]
\centering 
\includegraphics[width=1.0\textwidth,trim=0 200 0 200]{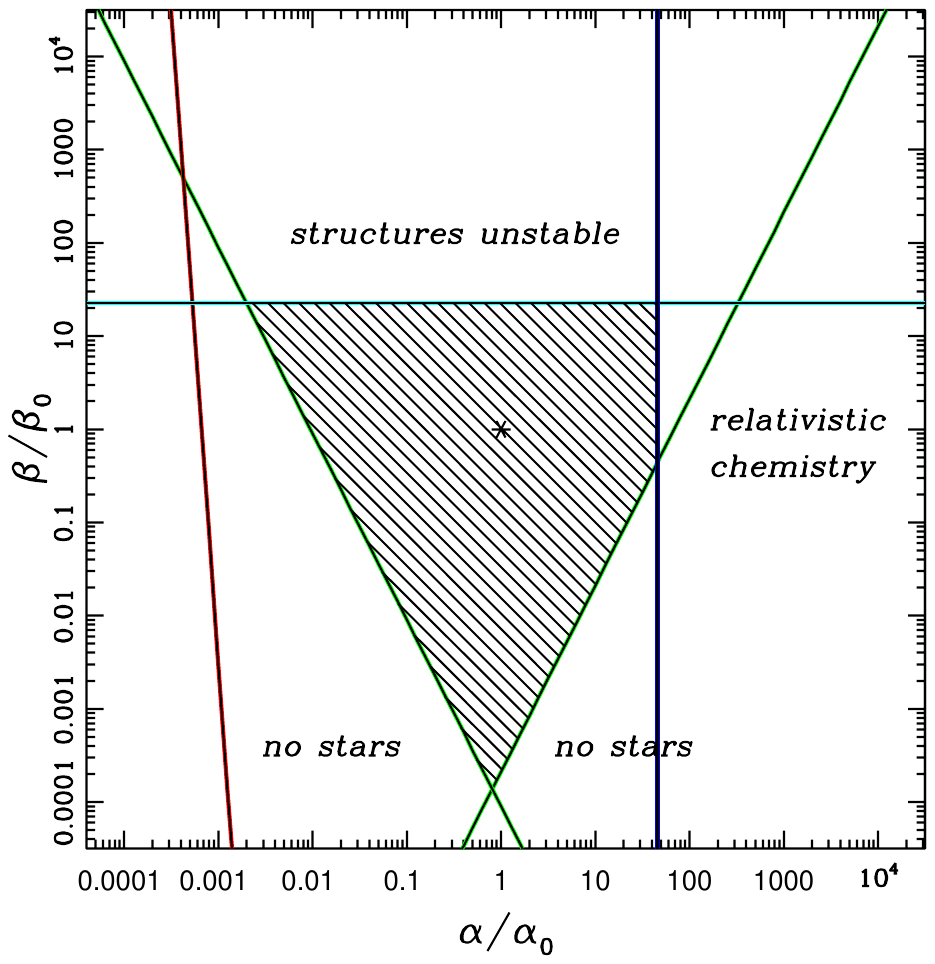}
\vskip12pt 
\caption{Allowed region of parameter space in the $\alpha$-$\beta$ plane.
Both the fine-structure constant $\alpha$ and the electron-to-proton
mass ratio $\beta$ must be smaller than unity to keep chemistry
similar to that of our universe (blue and cyan boundaries).
Additional constraints arise from the necessity of operational stars
(green curves). }
\label{fig:abplane} 
\end{figure}

In addition, a working universe requires working stars. The
constraints on the fundamental constants that allow for solutions to
the stellar structure equations were considered in the previous
sections (see also \cite{adams,adamsnew}).  Here we fix all of the
other quantities (for example the gravitational constant, the nuclear
reaction rates, and nuclear energy yields) and only allow
($\alpha,\beta$) to vary.  If the fine-structure constant $\alpha$
becomes too large, then nuclear reactions are suppressed, and the
minimum stellar mass becomes larger than the maximum. This constraint
sets an upper limit of the form 
\be
\alpha^2 \simless 10^{8/3} \beta \,. 
\ee
On the other hand, if $\alpha$ is too small, then quantum mechanical
tunneling is no longer required for protons to interact, and stable
nuclear fusion is compromised. This constraint sets a lower limit of
the form
\be
\alpha^2 \simgreat 2.6 \times 10^{-12} \beta^{-1} \,. 
\ee
For the electron-to-proton mass ratio $\beta$ is our universe, these
constraints require $\alpha$ to lie in the approximate range
$7\times10^{-5}\simless\alpha\simless0.5$. Equivalently, the fine
structure constant can be larger or smaller by a factor of $\sim100$,
so that the range of possible $\alpha$ values spans four orders of
magnitude. 

The resulting parameter space in the $\alpha$-$\beta$ plane that
allows for stable atoms and working stars is shown in Figure
\ref{fig:abplane}. The green curves show the constraints required
for stars to experience long-lived stable nuclear fusion.  In
addition, both the fine-structure constant $\alpha$ and the electron
to proton mass ratio $\beta$ must be small compared to unity (shown by
the blue and cyan curves, respectively). For completeness, the red
curve shows the limit where galactic masses become smaller than stars,
so that the region to the left of the curve is disallowed. Although
this constraint is weaker than that required for working stars
(green), it plays a role if one considers the case where electrons are
much more massive than protons, so that the protons (nuclei) orbit the
electrons. Although this part of the diagram is not shown, it is
possible in principle for chemistry to operate in this regime (e.g.,
see \cite{tegmark100} and references therein).  Overall, the allowed
range of parameter space spans several orders of magnitude in both
parameters $\alpha$ and $\beta$.

For completeness we note that tighter constraints can be derived based
on theoretical considerations. The three gauge coupling constants for
the strong, weak, and electromagnetic force are energy dependent, and
can attain the same value at a high energy scale in Grand Unified
Theories (GUT). In order for this convergence to occur, so that the
GUT scale exists, the value of $\alpha$ at low energies is constrained
to vary by less than a factor of $\sim2$ (for further discussion, see
\cite{bartip,donoghuethree,ellisnano}). However, these considerations
are based on physics beyond the Standard Model, and their status
remains experimentally undetermined.  

\begin{figure}[tbp]
\centering 
\includegraphics[width=1.0\textwidth,trim=0 200 0 200]{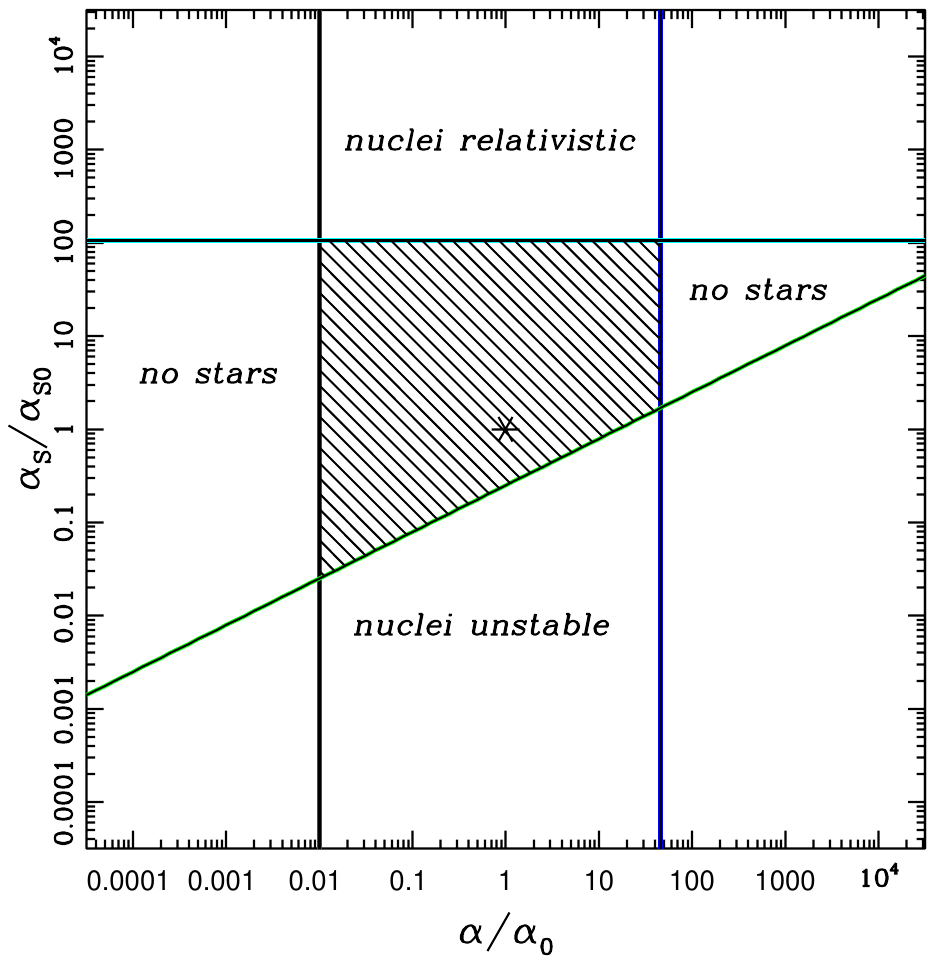}
\vskip12pt
\caption{Allowed region of parameter space in the $\alpha$-$\alpha_{\rm s}$
plane. The strong force must hold nuclei together, so that $\alpha_{\rm s}$
lines above the green curve. If the strong coupling is too large (above
the cyan curve), then energy levels in nuclei become relativistic and
nuclear reactions in stars are altered. The limits on $\alpha$ (vertical
lines) allow for working stars. } 
\vskip12pt 
\label{fig:asplane} 
\end{figure}

The requirement that stable nuclei exist places bounds on the value
of the strong coupling constant $\alpha_{\rm s}$. Using the Semi-Empirical
Mass Function as a starting model \cite{semf}, along with the current
best estimates for the fitting parameters \cite{basdevant}, one can
derive a constraint on the relative size of the strong and
electromagnetic coupling constants,
\be
{\alpha_{\rm s} \over \alpha_{\rm s0}} \simgreat {1 \over 4}
\left( {\alpha \over \alpha_{0}} \right)^{1/2} \,,
\ee
where the subscripts denote the values in our universe, and where the
numerical coefficient depends on the largest atomic nucleus that is
required to be stable \cite{adamsreview,bartip,tegmark100}.  In our
universe, all of the nuclei required for life are safely stable, but 
electromagnetic interactions limit the size of the largest nuclei.
This finding shows the numerical coefficient must be less than but
comparable to unity. The different powers for the coupling constants
arise because the electromagnetic force is long range, whereas the
strong force has a limited range.

In order for chemistry to take place without driving nuclear reactions
and thereby changing the constituent elements, the energy scales for
electrons in atoms must be well separated from nuclear binding
energies. This requirement places a constraint on the coupling
constants of the form 
\be
\alpha^2 m_e \ll \alpha_{\rm s}^2 \mpro \,. 
\ee
A related constraint is that chemistry requires the electrons
to orbit outside the nuclei, which in turn requires atomic sizes
to be much larger than nuclear sizes. This constraint takes
the form
\be
\alpha m_e \ll \alpha_{\rm s} \mpro \,. 
\ee
This result follows from estimating atomic size scales according to
$\ell_e \sim 1/(\alpha m_e)$ and estimating nuclear size scales
according to $\ell_n \sim 1/(\alpha_{\rm s}\mpro)$.

The constraints discussed so far provide lower limits to the strength
of the strong coupling constant. However, the value of $\alpha_{\rm s}$
cannot become arbitrarily large. In our universe, the most tightly
bound nucleus ($^{56}$Fe) has a binding energy per particle of
$E_{56}\sim8.8$ MeV, which is far below the mass energy of the
nucleons ($\mpro\sim m_n\sim1$ GeV). If the strong coupling constant
increases by a factor of $\sim100$, then the energy levels in nuclei
become relativistic. Although nuclear reactions could still take place
in this regime, the manner in which BBN and stellar nucleosynthesis
plays out would be significantly altered. As a result, a conservative
upper bound on the strong coupling constant takes the form 
$\alpha_{\rm s} \simless$ 100 $\alpha_{\rm s0}$. 

With the above considerations, the constraints on the coupling
constants in the $\alpha$-$\alpha_{\rm s}$ plane are shown in Figure
\ref{fig:asplane}. In order for nuclei to remain stable, the strong
coupling constant must lie above the green curve. In order for the
nuclei to remain non-relativistic, $\alpha_{\rm s}$ must lie below the
cyan line. Working universes also require that stars exist, which
limits the value of the fine-structure constant to lie between the red
and blue lines. The resulting area of allowed parameter space thus
spans three to four orders of magnitude.

For completeness, note that some considerations of habitability
require radioactive heating in planets to provide molten cores
and hence magnetic fields, as well as plate tectonics (see
\cite{lunine99,lunine,scharf} for further discussion). Such heating,
in turn, would require long-lived metastable nuclei, and would place
additional constraints on the fundamental parameters. 

\medskip 
{\bf Bottom line:} {\sl In order for stable atoms and stable nuclei to
exist, with sufficiently large atomic numbers, the fine-structure
constant $\alpha$, the strong force analog $\alpha_{\rm s}$, and the
electron-to-proton mass ratio $\beta$ can vary by a few orders of
magnitude. The range of the strong force must also be large enough.}

\bigskip 
\section{Quarks} 
\label{sec:quarks} 

Quarks (and leptons) represent the smallest scales for which we have
physical experiments. The masses of the light quarks are constrained
by the requirements that atoms, nuclei, and protons remain stable
\cite{barrkhan,hogan,jaffe,donoghue,donoghuemass,donoghuethree,hallnomura}.
Complementary constraints have also been placed on the Higgs mass
\cite{hallnomura2010}, Higgs Vacuum Expectation Value \cite{donoghuetwo},
the proton mass \cite{page,pagenew}, and additional aspects of the
Standard Model of Particle Physics.  Related work has constrained the
time variation of quark masses in our universe, primarily through
their effects on Big Bang Nucleosynthesis \cite{bedaque,berengut,damour}. 

The first, and weakest, constraint on the quark masses is that protons
and neutrons must be stable. If the mass difference between the up
quark and the down quark is too large, then it would become
energetically favorable for the heavier quark to decay inside a hadron
and render it unstable. For example, down quark could decay into an up
quark inside a proton unless the light quark masses obey the constraint
\be
m_d < m_u + m_e + {\cal E} \,,
\ee
where the decay requires the production of an anti-symmetric
state of three quarks, which in turn requires energy ${\cal E}$.
Similarly, for the case where the up quark has too much mass,
the opposite type of decay could occur unless the masses obey
the complementary constraint
\be
m_u < m_d + m_e + {\cal E} \,. 
\ee
Since ${\cal E}\approx300$ MeV in our universe, so that ${\cal E}$ is
much larger than both quark masses $m_u$ and $m_d$, this constraint is
relatively weak.

A stronger set of constraints arises from the requirement that protons
and neutrons cannot decay within bound states of nuclei. For example,
a neutron could undergo beta decay within a nucleus unless the
following constraint is met 
\be
m_d < m_u + m_e + B + \Delta_{em} \,,
\ee
where $B$ is the binding energy of the nucleon within the nucleus
and $\Delta_{em}$ is the contribution of the electomagnetic
interaction to the mass difference between the proton and neutron.
Similarly, going the other way, the stability of protons within
nuclei requires the corresponding constraint 
\be
m_u < m_d + m_e + B - \Delta_{em} \,. 
\ee
In our universe, the electromagnetic contribution $\Delta_{em}$
$\sim1.7$ MeV and the binding energy per nucleon $B\sim10$ MeV.

A related type of constraint arises from the requirement that protons
cannot decay into neutrons, as such a decay would not allow for the
existence of hydrogen atoms (or water molecules). The constraints for
free protons and hydrogen atoms take the forms 
\be
m_d > m_u - m_e + \Delta_{em} \qquad {\rm and} \qquad
m_d > m_u + m_e + \Delta_{em} \,. 
\ee
The second constraint, where hydrogen atoms decay when the central
proton absorbs the orbiting electron, is stronger because the
electron mass is already provided. In the first constraint, the mass
of the positron must be provided by the decaying down quark.

\begin{figure}[tbp]
\centering 
\includegraphics[width=0.90\textwidth,trim=0 200 0 200]{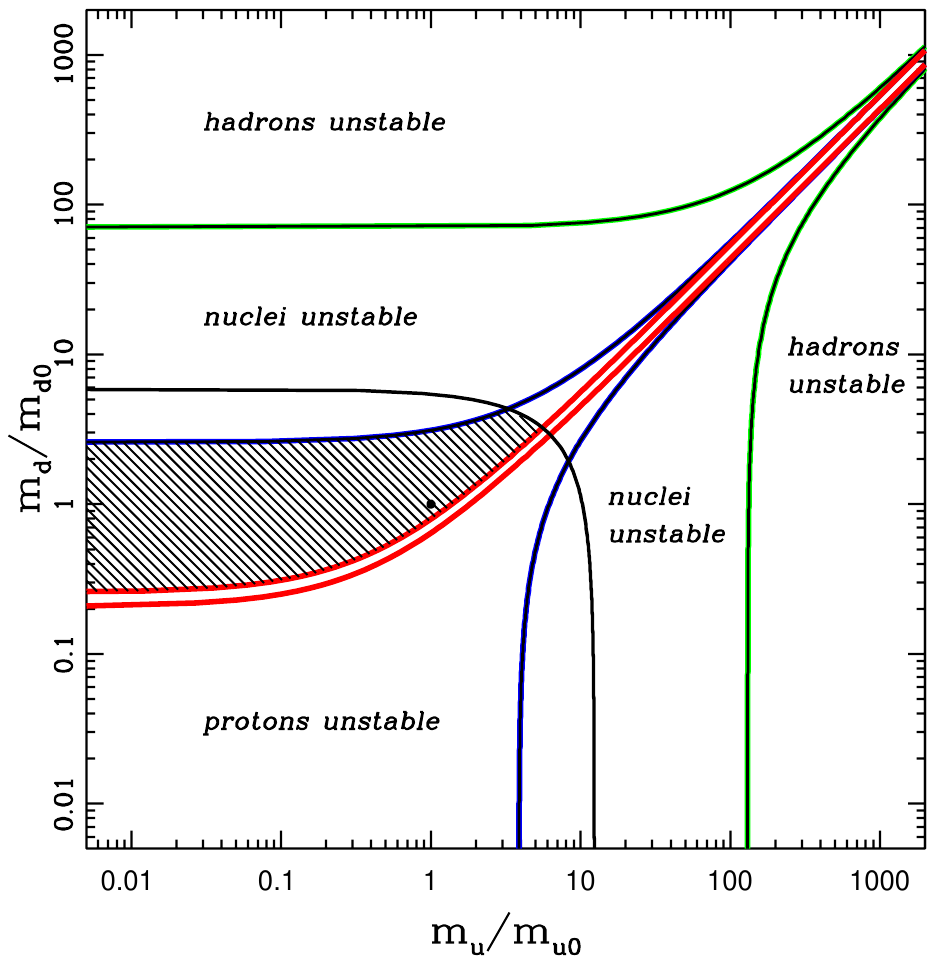} 
\vskip12pt
\caption{Allowed masses for the light quarks $(m_u,m_d)$. For large
departures of the quark masses from their observed values, constituent
quarks decay inside hadrons (beyond the green curves). For more modest
departures (beyond the blue curves), protons and neutrons could decay
within bound nuclei. In the lower left part of the diagram, protons
could decay into neutrons, either as free particles (lower red curve),
or within hydrogen atoms (upper red curve). These constraints are
derived under the assumption that the electron mass scales with the
mass of the down quark (see text, and
\cite{adamsreview,barrkhan,hogan}). }
\label{fig:quarks} 
\end{figure}

The above constraints involve the electron mass in addition to the
masses of the lightest quarks. In order to evaluate the constraints,
and thereby gain some understanding of how precisely quark masses must
be specified to obtain a working universe, the electron mass must be
specified. Due to symmetry considerations, a common assumption is that
the ratio of the electron mass to the mass of the down quark is held
constant \cite{barrkhan} so that $f=m_e/m_d\approx0.107$ =
{\sl constant}. With this specification, the resulting bounds on
the quark masses $(m_u,m_d)$ are shown in Figure \ref{fig:quarks}. 

For completeness, we note that many previous treatments of this topic
invoke the additional constraints that the diproton cannot be bound
and that the deuterium nucleus must be bound \citep{hogan,reessix}.
As outlined in previous sections, however, viable stars exist in the
presence of stable diprotons (see Figure \ref{fig:diproton} and
\cite{adamsreview,adamsdiproton,barnes2015}) and for universes with
unstable deuterium (see Figure \ref{fig:hrnod} and \cite{agdeuterium}).
On a related note, however, the light quark masses are also constrained
by the requirement that the range of the strong force does not become
too small. The quark masses determine the pion mass, which in turn
determines the range. The black curve in Figure \ref{fig:quarks} marks
the contour where the sum of the quark masses is four times that in 
our universe. Although the required range, and hence the sum of the
masses, is not precisely known, this additional range requirement
decreases the allowed parameter space by disallowing the high-mass
tail.

\medskip 
{\bf Bottom line:} {\sl The masses of the light quarks can vary by an
order of magnitude, but the masses have the smallest allowed ranges
among the fundamental parameters considered herein.}

\bigskip
\section{Number of Space-Time Dimensions} 
\label{sec:dimensions}

Another property of the universe that could vary from region to region
is the number of space-time dimensions. Our universe has the familiar
${\cal D}=3$ spatial dimensions, with one time dimension, resulting in
the standard $3+1$ structure of our space-time. Since one can readily
write down a space-time metric with different numbers of dimensions,
it is reasonable to ask whether other possible cases could support
complex structures and life \cite{ehrenfest1,ehrenfest2}.  The
development of string theory, M theory, and their generalizations
suggest that the fundamental space-time manifold could have at least
ten dimensions \cite{greenwitten}. Although only three of the spatial
dimensions have macroscopic size in our universe, other regions of
space-time could have greater numbers of large dimensions. In this
context, the scale that divides `small' and `large' spatial dimensions
is the size of the structure(s) of interest. As a result, macroscopic
dimensions correspond to scales that are relevant for life to arise,
roughly over the range from the size of hadrons to the size of
galaxies.

Although fundamental physical theories have ${\cal D}\ge3$, they all
invoke exactly one dimension of time. Multiple time dimensions would
lead to severe problems with causality, stability, and predictability
\cite{barrow1983,borstnik,tegmarkdimension}. With just one time axis,
events follow a well-defined sequence, from past to present to future.
This feature allows for familiar cause-and-effect relationships to be
consistent. For a space-time with two or more temporal dimensions,
particles could move along the different directions in time and
execute closed time-like loops.  Such loops would then allow for
violations of causality and would give rise to paradoxes --- like
effects preceding their causes. As a result, any space-time with
multiple time dimensions would not only be significantly different
from our own, but would most likely not be habitable.

The argument for why three large spatial dimensions are preferred
arises from the necessity of both stable structures and sufficient
complexity. If the universe had less than three spatial dimensions, it
would be far too simple \cite{abbott,fostermuller}: For ${\cal D}=2$,
the space would not support knots, stable 3D molecules, or the complex
geometries needed for life. For ${\cal D}=1$, only points would exist
and they would be confined to a line, with no possibility for
branching, extended networks, or organisms.

If the universe had more than three large dimensions, it would be too
unstable \cite{barrow1983,borstnik,tangherlini,tegmarkdimension,whitrow}.
For three spatial dimensions, forces due to gravity and
electromagnetism follow an inverse square law, which makes stable
orbits possible. In higher dimensions, the forces would decrease with
distance more quickly and orbits would be unstable for both
astronomical bodies and for electrons in atoms. This claim can be
illustrated by considering the effective potential that governs
orbital motion. For a given specific angular momentum $L$, the
effective potential in ${\cal D}$ spatial dimensions becomes 
\be
V_{\rm eff} = -\frac{K_{\rm c}}{r^{{\cal D}-2}}
+ \frac{L^2}{2 r^2}\,,
\ee
where the first term represents the attractive central force
(gravitational or electromagnetic), and the second term is the
centrifugal barrier that resists collapse (and where we assume
spherical symmetry). For ${\cal D}<4$, the competing terms can balance
so that a minimum in the potential exists. In turn, stable orbits can
persist, leading to ellipses in the case of a point potential for
gravity and quantized shells in the case of atoms.

For larger numbers of dimensions, ${\cal D}\ge4$, the potential no
longer has a minimum and no stable orbits exist.  Without stable
atoms, chemistry becomes impossible.  Without stable planetary orbits,
solar systems become impossible.  As a result, in regions of space
with higher dimensions, matter would not be able to reside in the type
of long-lived, structured environments that life requires.  For
completeness, note that if the laws of physics change significantly,
for example maintaining an $r^{-2}$ force law in higher dimensions
(instead of the expected form $r^{-({\cal D}-1)}$), then stable atoms
can be constructed (see \cite{andrew1990,caruso,chavanis,bures,
burgbacher,gurevich} for further discussion).

The stability argument given above applies for the potential of point
masses or point charges in higher dimensions. It remains possible for
stable orbits to exist within extended mass/charge distributions in
regions with ${\cal D}>3$. Such spaces could conceivably give rise to
structures of some kind, but they would be quite different (and most
likely more transient) than the atoms and solar systems of our own
universe.

While the choice ${\cal D}=3$ is preferred, as outlined above (see
also \cite{reessix}), the number of large dimensions that are expected
remains unknown (however, see \cite{karchrandall}). Theories of
quantum gravity suggest that the total number of dimensions should be
10 or 11, but the number of large spatial dimensions is not yet
calculable from first principles
\cite{greenwitten,schellekens,susskind,susskind2005}. For these
scenarios, the odds of obtaining a working universe are not
necessarily long. For example, if the number of total dimensions is of
order 10 and if the selection of the number of large spatial
dimensions is random and uniformly distributed, then a given universe
would have a $\sim10\%$ chance of obtaining the desired outcome
${\cal D}=3$.

\medskip 
{\bf Bottom line:} {\sl The conventional wisdom that the number of
spatial dimensions must be ${\cal D}=3$ is robust.}

\bigskip
\section{Anthropic Arguments} 
\label{sec:anthropic}

The disucssion so far has considered the ranges of parameter values
that are required for a universe to be viable, and thus quantifies the
required degree of fine-tuning. In order for a universe to develop
interesting structures such as galaxies, stars, planets, and complex
nuclei, the fundamental constants and cosmological parameters must lie
within specified ranges, as delineated in this review. These
structures are necessary for the universe in question to develop and
support observers. As a result, the existence of observers places
constraints on the allowed values of the fundamental constants.
Anthropic arguments take this idea one step further: Since the
universe must have the proper version of physical law for observers to
arise, one can argue that the presence of observers provides at least
a partial explanation for why the laws of physics have their
particular form and/or why the constants of nature have their
particular values.  Anthropic arguments have been developed using a
wide range of definitions, and make these claims with varying degrees
of strength (classic references include
\cite{bartip,carr,carter1974,carter1983}), but the basic point is
that the presence of observers explains, in part, the form taken by
the laws of physics (see also \cite{balashov,barrow1981,weinstein}).

In order for anthropic arguments to carry weight, the parameters they
seek to explain must be fine-tuned to some degree.  For the sake of
definiteness, consider the following example: Let $\chi$ be a
fundamental parameter and suppose that the value of $\chi$ must fall
within the range $\chi_1\le\chi\le\chi_2$ in order for the universe to
be viable (e.g., suppose that values of $\chi$ outside this range
would render all atomic nuclei unstable). If the allowed range is
extremely small, then one can argue that since nuclei are stable, and
stable nuclei are necessary for observers to exist, then our existence
as observers `explains', in part, the value of $\chi$ that we observe.
If the range of allowed values is sufficiently large, however, then
the finding that the value of $\chi$ in our universe falls within that
range becomes less profound, and the corresponding anthropic argument
is weaker. As a result, a narrow allowed range for the fundamental
parameters is a prerequisite for anthropic arguments to be useful. Of
course, one is still left with the issue of how narrow the range of
parameters must be in order for anthropic arguments to be effective.

Anthropic arguments face a number of difficulties: One issue is that
there is no concensus regarding how fine-tuned a parameter must be for
the argument to be meaningful \cite{friederich}. In order to tune a
radio to receive a particular station, the frequency must be tuned so
that $(\Delta{f})/f\simless0.01$. As shown in this review, however,
the allowed range for most of the fundamental parameters spans several
orders of magnitude. If any value of the parameter in question works,
then the presence of observers has no predictive value, of course, so
the question becomes: how narrow must the range be in order to be
useful?  A related issue is that more than one parameter can vary, and
coupled variations can provide more opportunities for structure
formation.  For example, the allowed range of a parameter could be
small if all other constants are fixed, but could be much larger for
other (allowed) values of those parameters. Another point of
contention is the degree to which anthropic arguments actually predict
the values of the fundamental parameters.  If a more fundamental
derivation of the parameter values were available, most researchers
would find it preferable to an anthropic explanation.  Yet another
criticism is that the manner in which such arguments can be falsified
(if they can be falsified) does not follow the usual implementation of
the scientific method \cite{ellissilk,freivogel,kaneperry}.

In spite of the complications outlined above --- and others --- a wide
range of anthropic arguments have been used to explain the values of
both the cosmological parameters and fundamental constants observed in
our universe. One important study was the original constraint on the
cosmological constant \cite{weinberg87}, which has been subsequently
generalized to include variations of additional parameters and other
complications \cite{adamsrhovac,efstathiou,garriga2006,graesser,
hartle,kallosh,liviorees,mersini,peacock}. Additional work has
constructed anthropic constraints on the energy scale of electroweak
symmetry breaking \cite{agrawalprl,agrawal,barrkhan,jeltema}, big bang
nucleosynthesis \cite{coc2007,coczero,cocone,macdonaldmullan}, the
triple-alpha resonance for stellar nucleosynthesis
\cite{jeltema,livio,meissner}, the parameters of aluminum-26 decay
\cite{sandora2017}, the proton mass \cite{page}, the neutrino masses
\cite{pogosian,tegmarkneutrino}, the age of the universe
\cite{cirkovic2000,garriga1999}, the dark matter abundance
\cite{boussohall,wilczek} and for the existence of three generations in
the Standard Model of particle physics \cite{gould,ibe,schellekens}. 

Anthropic arguments have amassed a vast literature, and at least 30
definitions of the anthropic cosmological principle have been put
forth \cite{bostrom}. As a result, a proper review of this topic is
beyond the scope of this paper (the reader is referred to many
existing treatments, including \cite{balashov,barrow1981,bartip,
bostrom,boussoetal,carr,davies1983,liviorees,meissner,weinstein} and
the references therein).  Nonetheless, anthropic arguments place
constraints on the universe: Our location in the universe and its
intrinsic properties are necessarily privileged in that they must be
consistent with our existence as observers \cite{carter1974}.

\bigskip 
\section{Summary} 
\label{sec:summary}

This review has presented a broad overview of the degree of
fine-tuning that is required for our universe -- and others -- to be
habitable. Approximately one dozen fundamental and cosmological
parameters (Table \ref{table:parameters}) affect the ability of any
given universe to develop the complex structures necesary for life to
develop, including galaxies, stars, planets, and heavy nuclei.  Under
the assumption that they can vary from universe to universe (Figure
\ref{fig:machine}), these parameters are constrained by a number of
physical and astrophysical considerations as considered herein. The
main results can be summarized as follows:

The coupling strengths of the four fundamental forces provide one
class of parameters that could vary from universe to universe. In
order for stars to function, the coupling constants that determine the
strength of gravity and electromagnetism can vary by several orders of
magnitude (Figure \ref{fig:starplane}).  In spite of its large allowed
range of variation, the gravitational force must always be much weaker
than the electromagnetic force $(\alpha_G\ll\alpha)$.  In order for
atomic nuclei to stay bound, the strong and electromagnetic forces can
vary by only a few orders of magnitude (Figure \ref{fig:asplane}).
Big Bang Nucleosynthesis places only modest constraints on the
strength of gravity, as well as the baryon-to-photon ratio $\eta$
(Figure \ref{fig:bbnplane}). In general, it's difficult for BBN to
render the universe inhospitable.  Constraints on the weak force are
generally weaker (Section \ref{sec:weak}). Scenarios exist where a
universe can function without any weak interactions, although low
values of $\eta$ are required \cite{grohsweakless,weakless}.

The particle masses are also constrained.  The ratio $\beta$ of the
electron mass to the proton mass can vary by a few orders of magnitude
(Figure \ref{fig:abplane}), limited by the requirement that atoms
remain bound and non-relativistic. The quark masses are among the most
highly contrained quantities of the parameters considered here
(Figure \ref{fig:quarks}). Although the allowed mass of the up quark
spans several orders of magnitude, the mass of the down quark can only
vary by a factor of $\sim7$. Moreover, the closest failure point of
our universe involves the mass of the down quark. If the mass $m_d$
were smaller by an increment of $\Delta{m}\sim0.3$ MeV, then hydrogen
atoms could collapse into neutrons through the reaction
$p+e\to{n}+\nu_e$ (again see Figure \ref{fig:quarks}), and chemistry
would be significantly different.

Stars are remarkably robust, in that the gravitational constant, the
fine-structure constant, and the nuclear reaction cross sections can
vary by many orders of magnitude and still allow for solutions to the
stellar structure equations (Section \ref{sec:stars}). A sizeable
fraction of this possible parameter space also allows for stars to be
sufficiently long-lived and support surface temperatures hot enough to
potentially host habitable planets (Figure \ref{fig:starplane}).
Constraints on planet properties are comparable to, but somewhat less
restrictive than, the constraints required for working stars (Figure
\ref{fig:planetplane}). Universes with working stars are thus likely
to also have viable planets (although more work should be carried out
on this latter topic).

Stars are mostly impervious to the challenges provided by several
classic issues that have been claimed as instances of fine-tuning. If
the strong force is only about 10 percent stronger, then diprotons are
stable and nuclear reaction cross sections can increase by factors of
$\sim10^{18}$. In the face of such increases, however, stars are
relatively unchanged (Figure \ref{fig:diproton}), and burn their
nuclear fuel at the lower core temperature of about one million kelvin
(instead of 15 million kelvin). Contrary to many previous claims,
stars in universes with stable diprotons are not short-lived, but
rather can burn for trillions of years. The related issue of unstable
deuterium is more complicated. If the strong force is about 10 percent
weaker, so that the deuterium nucleus is unbound, then the first
stellar generations can operate through triple-nucleon reactions if
the half-life for (unstable) deuterium is long enough (Figure
\ref{fig:hrnod}).  After trace amounts of carbon are produced,
through either triple-nucleon burning or explosive nucleosynthesis,
subsequent stellar generations can operate through the CNO cycle and
have their usual properties.  Finally, stars are less sensitive to the
triple-alpha resonance than previous claims. The Hoyle resonance can
vary in energy over a span of $\sim800$ keV and still allow for carbon
production (Figure \ref{fig:metals}).  In contrast, an energy
increment of only 92 keV would lead to bound $^8$Be nuclei, which
would obviate the need for the triple-alpha process.

On larger scales, galaxies provide important constraints on
cosmological parameters. If the energy density of the vacuum is too
large, then the formation of galaxies and large scale structure is
suppressed. The energy density, essentially the cosmological constant
$\Lambda$, can be larger by many orders of magnitude. More
specifically, the value of $\Lambda$ can vary upward by factor of
$\sim10^{10}$ if the fluctuation amplitude $Q$ can also vary and much
more if $\eta$ can be larger (Figure \ref{fig:qvlambda}). In all
cases, however, the energy scale of $\Lambda$ is always constrained to
be much smaller than the Planck scale. The observed value of the
cosmological constant $\Lambda$ thus represents a significant
hierarchical fine-tuning problem.  The fluctuation amplitude $Q$
affects galaxy properties, especially their mean density, and is
confined to the approximate range $10^{-6}\simless{Q}\simless10^{-2}$
(Figure \ref{fig:galsurvive}). Larger values of $Q$ lead to dense
galaxies that provide highly destructive environments, whereas
galaxies with lower $Q$ have difficulty cooling and making stars.

As summarized above, most of the fundamental and cosmological
parameters can vary by at least a few orders of magnitude and still
accommodate a working universe. Compared to tuning a radio, where
frequency variations must be less than one percent, the allowed
parameter variations are large and suggest minimal tuning. However, a
full assessment of fine-tuning requires that we specify the underlying
probability distributions from which the parameters are selected.
These distributions are currently unknown and their consideration is
complicated (see the Appendix for further discussion). Finally, we
note that some quantities (e.g., the cosmological constant) are
subject to hierarchy problems: They require values that are far from
those expected, in spite of having wide ranges of possible values.

Viable universes must have additional properties.  In order to be
habitable, a universe must live for a long time, which in turn
requires spatial flatness with density parameter $\Omega=1$ (or
perhaps some other fortuitous expansion history). On one hand,
inflation provides an attractive mechanism to drive universes to
$\Omega\rightsquigarrow1$. On other hand, inflation could pose its own
tuning issues (see Section \ref{sec:omega} and references therein).
The universe must also have three and only three macroscopic spatial
dimensions ${\cal D}=3$, with larger ${\cal D}$ leading to stability
problems and smaller values leading to simplicity issues (Section
\ref{sec:dimensions}). Another important property of the universe is
its zero (or extremely small) net charge $\quni$.  Because of the
relative strength of electromagnetic forces and gravity, a viable
universe must be extremely close to charge neutrality
\cite{caprini,foot,gamow,lyttleton,oritoyoshi,rees1972}. The laws of
physics in our universe are built with charge quantization, charge
conservation, and charge neutrality as defining features, but we do
not know if these properties are required in all cases.  Although
habitable universe must have $\Omega=1$, ${\cal D}=3$, and $\quni=0$,
it remains difficult to assess how much tuning is required for these
properties to be realized. The first requirement ($\Omega=1$) could be 
satisfied through a dynamical mechnaism. The other two requirements
are `built in' to the laws of physics, at least in our universe. As a
result, although these constraints are important, they stand on a
different footing the other issues outlined above -- they are not
described by a range of parameter values that are needed for a working
universe.

In spite of many complications, considerations of fine-tuning and the
range of allowed parameters provide us with important insight into how
our own universe operates. Theoretical physics has a long history of
using thought experiments as a means to further our understanding,
even when the `experiment' cannot be physically realized. For example,
Einstein is said to have considered what happens when an observer
travels alongside a beam of light \cite{einstein}, even though his own
work shows the impossibility of such an observation. In the present
context, the collection of alternate universes that make up the
multiverse cannot be experimentally probed. By definition of what it
means to be in another universe, those space-times are causally
disconnected from our own. Nonetheless, by considering how our
universe changes in response to changes in its fundamental constants,
we obtain a better understanding of how astrophysical objects
operate as they trace through their life cycles.\footnote{We can 
also gain insight into the related problem of how the fundamental
constants might change with cosmic time within our own universe
\cite{uzan,uzantwo}, and the consequences of those variations.}
Consider the evolution of stars: One can increase the nuclear reaction
cross section by a factor of $10^{15}$ or more
(Figure \ref{fig:diproton}) and yet stellar properties are only
modestly changed. If the carbon resonance level is increased so that
no net carbon is produced in massive stars, then other alpha elements
are forged instead (Figure \ref{fig:metals}). If the standard $p$-$p$
chain for nuclear reactions is unavailable because deuterium is
unstable, then the CNO cycle takes over and powers the stars
(Figure \ref{fig:hrnod}).  The universe is thus remarkably robust to
some changes and rather sensitive to others (like the mass of the down
quark in Figure \ref{fig:quarks}). These examples, along with many
others, illustrate how considerations of fine-tuning provide insight
into how our universe works.

\bigskip
\noindent
{\bf Acknowledgments}
\medskip

This manuccript was requested by the Foundational Questions Institute,
as part of a collection of reviews about open problems in foundational
physics and related fields. The author (FCA) is supported in part by
the Leinweber Institute for Theoretical Physics at the University of
Michigan. The author is grateful to the three referees (including
R. Alves Batista) who provided a host of useful suggestions, as well
as the editor (N. Rutherford), who provided a fourth review.

\bigskip
\appendix

\bigskip
\appendix 
\section{Probability Distributions}
\label{sec:prob}

Most of this review has focused on determining how much the
fundamental parameters of physics and cosmology can change and still
allow for a universe to be viable. In addition to specifying the range
of parameter values that allow for working universes, and/or the
required hierarchies, we would also like to know both the range of
parameters values that could be realized and the probability with which
they arise.  While important, the relevant probability distributions
are not known at the present time (e.g., see
\cite{donoghuethree,hartlesrednicki,hossenfelder}). These
distributions are not measured, but they are also not measurable, in
that other realizations of the parameters arise (by definition) only
in other universes that are not experimentally accessible. As a
result, considerations of the probability distributions are
necessarily on a different footing than the rest of this review and
are thus discussed in this Appendix.

The goal of this section is to illustrate some of the complications
that arise in assigning probability distributions to the fundamental
parameters. Toward that end, we first consider the selection of quark
masses. The shaded region of Figure \ref{fig:quarks} shows the masses
for the up and down quarks needed for a viable universe.  The mass of
the up quark can vary by at least three orders of magnitude, whereas
the mass of the down quark can vary less than one order of magnitude.
Suppose that the light quark masses vary from universe to universe and
that their values are drawn from a probability distribution. For the
sake of definiteness, suppose further that we limit our discussion to
the probability that the mass of the down quark falls in the range
$m_1 \le \dmass \le m_2$ (where the limits $m_k$ are calculated as
shown in Figure \ref{fig:quarks}).

To determine the probability, we need to know the range of parameters
that can occur, i.e., the minimum $m_{min}$ and the maximum $m_{max}$.
We also need the probability distribution $p(m)$ defined on the interval 
$[m_{min},m_{max}]$ where $\int p(m)dm=1$. Here we consider two common
choices \cite{aguirrecarr}, a uniform (flat) probability distribution 
\be
p_{uni}(m) = {1 \over \Delta} \qquad {\rm where} \qquad
\Delta \equiv m_{max} - m_{min} \,,
\ee
and a log-uniform (power-law) probability distribution
\be
p_{log}(m) = {1 \over m \ln {\cal R}} \qquad {\rm where} \qquad
{\cal R} \equiv {m_{max}\over m_{min}} \,.
\ee
For these two example cases, the probability of the down quark mass 
falling in the target range $[m_1,m_2]$ is given by 
\be
P_{uni} = {m_2 - m_1 \over \Delta} \qquad {\rm and} \qquad
P_{log} = {\ln(m_2/m_1) \over \ln {\cal R}} \,.
\ee
Note that the masses $(m_1,m_2)$ that specify the target range are
determined by known physics (Figure \ref{fig:quarks}).  In this
example, we can take $m_2\approx12$ MeV and $m_1\approx2$ MeV as a
starting estimate. In contrast, we have no means of calculating the
masses $(m_{min},m_{max})$ that specify the range of the probability
distribution. A plausible choice for the upper bound is the Planck
mass (e.g., see Section \ref{sec:lambda}), so here we take $m_{max}$
= $10^{19}$ GeV.  The lower bound must be finite. Otherwise the quarks 
could be massless, so they would move at the speed of light, which
would make it difficult for them to be confined into hadrons. For
purposes of illustration, we take the lower limit $m_1$ = 1 eV,
but note that the results can be readily computed for other choices. 
The resulting probabilities then have approximate numerical values  
\be
P_{uni} \approx 10^{-21} \qquad {\rm and} \qquad P_{log} =
{\ln 6 \over \ln 10^{28}} \approx 0.03 \,. 
\ee
This exercise thus shows the following: The target range $[m_1,m_2]$
is determined by known physics, whereas the allowed range
$[m_{min},m_{max}]$ is not. The estimate for $P_{uni}$ is nearly
independent of $m_{min}$, whereas $P_{log}$ is only logarithmically
sensitive. In addition --- and significantly --- different choices for
the probability distributions lead to wildly different estimates for
the probability of a successful outcome.

Notice also that the probabilities depend on what question is being
asked. We could, for example, consider the probablity that both the up
quark and the down quark fall within specified mass ranges, or the
probability that the masses of the six quarks span a range as large as
that observed, and so on. 

Next we can consider what a Bayesian approach tells us. Here we
consider the $data$ to be the fact that we live in a working universe,
along with the (calculable) results that a working universe requires
$\dmass \in [m_1,m_2]$. Let the $model$ be our working assumption for
the probability distribution from which the quark masses are
drawn. Then Bayes' theorem says that the probabilities obey the
relation 
\be
P(model\,|\,data) = { P(data\,|\,model) P(model) \over P(data)} \,.
\ee
The quantity $P(data\,|\,model)$ is the likelihood function, the
probability of obtaining the $data$ (a quark mass in the right range)
given the $model$ (the choice of prior probability distribution, here
uniform [denoted $uni$] or log-uniform [denoted $log$]). In our
example, the quantities $P_{uni}$ and $P_{log}$ are the likelihood
functions evaluated for the two choices of prior, and $P(model)$ is
the prior probability assigned to the model itself. In the limiting
case where our characterization of the universe only involves the mass
of the down quark, and where we know that the prior is uniform, then
the probability $P(data\,|\,model)=P\sim10^{-21}$, which most people
would consider as fine-tuned. For the same problem, except where we
use a log-uniform prior, the probability $P\sim0.03$, which is
significantly less fine-tuned. Bayesian reasoning makes explicit that
the two probabilities that we calculated before are conditional, i.e.,
based on the specific assumption of how the masses are distributed
(uniform or log-uniform in this example). In other words, the
probabilities $P_{uni}\sim10^{-21}$ and $P_{log}\sim0.03$ are not
really statements about the universe, but rather statements about how
the assumed priors (and hence our two models) distribute the quark
masses.

One can compare the probabilities of the two models as follows: 
\be
{P(uni\,|\,data) \over P(log\,|\,data)} =
{ P(data\,|\,uni) \over P(data\,|\,log)} \cdot 
{P(uni) \over P(log)} \approx  \left(3\times10^{-19}\right) 
{P(uni) \over P(log)} \,. 
\ee
If the two priors for the models are equally likely, $P(uni)=P(log)$,
then the evidence overwhelmingly favors the log-uniform model. If, on
the other hand, the log-uniform model is disallowed (or disfavored by
a factor greater than $\sim10^{20}$), then the uniformly distributed
model could still be favored.  On a related note, previous work has
shown that the six known quarks have a mass distribution that is close
to log-uniform \cite{adamsreview,donoghuemass,jaffe}, which also
argues in favor of the log-uniform model. In any case, these
considerations pose the question: What is the correct prior from which
to choose the prior?  One key point here is that the Bayesian
framework does not obviate the need for knowing the prior, i.e., the
underlying probability distribution (in this case telling us how
possible quark masses are distributed). The framework does provide a
means to evaluate different possible models, but the results depend on
knowing the probabilities $P(uni)$ and $P(log)$ that the models are
valid. Keep in mind that the uniform and log-uniform priors used here
are two common examples, but many other distributions are possible.

Another complication arises from the fact that our observations are
conditioned on the existence of observers. In principle, one should
distinguish between
\be
P(\dmass \in [m_1,m_2] \,|\, model) \quad {\rm and} \quad
P(\dmass \in [m_1,m_2] \,|\, model + {\rm observers}).
\ee 
Since observers can only exist for $\dmass$ in the viable range, the
latter probability is unity by construction. As a result, if we use
the existence of observers as part of the $model$, then the
requirement of the quark mass in the proper range provides no
discrimination between models. Instead, the question shifts to become:
how is the parameter $\dmass$ distributed within the viable range
$[m_1,m_2]$ under the different models?
 
Next we consider the cosmological constant $\Lambda$. This issue more
complicated for a number of reasons. First, one can work in terms of
the value of $\Lambda$ itself, which is an energy density, or in terms
of the energy scale given by $\lambda^4 = \Lambda$. Next we note that
the cosmological constant could be either positive or negative.  One
can deal with this issue by considering the sign and the magnitude of
the constant separately. A related issue is that the case $\Lambda=0$
should be included, whereas the log-uniform distribution requires a
lower (non-zero) cutoff. 

One way forward is to note that $\Lambda$ must either be zero,
negative, or positive, so that these options must arise with
probabilities $\{p_{-},p_0,p_{+}\}$, where $p_{-}+p_0+p_{+}=1$. The
case $\Lambda=0$ can arise if some symmetry leads to a perfect
cancellation of the zero-point energy fluctuations that give rise to
the cosmological constant problem. Given that the cosmological
constant is a problem, however, the probability $p_0$ is not expected
to be extremely close to unity (although it could be small and, at
present, remains unknown). With the sign (including zero) chosen, the
magnitude of the cosmological constant is chosen next. For the case of
the log-uniform distribution, one can then introduce a lower cutoff.
The Planck mass provides a benchmark upper limit (Section
\ref{sec:lambda}), so we use $\lambda_{max}=\mplanck$. 
For purposes of illustration, the current horizon size suggests a low
energy cutoff, so we use the corresponding minimum energy scale
$\lambda_{min}\sim10^{-34}$ eV and note that the results are at most
logarithmically sensitive to this choice.\footnote{Notice also that
any smaller value of $\lambda$ would be indistinguishable from zero,
in terms of its impact on the universe to date, and that even smaller
values are not problematic.}  The main text shows that universes
remain viable for energy scales up to $\lambda_2\approx1$ eV
(specifically when $Q$ is allowed to vary, but $\eta$ is fixed),
whereas no lower bound arises (so that $\lambda_1=\lambda_{min}$).
With these specifications, and for the case where $\Lambda>0$, the
probability for the energy scale of the cosmological constant to fall
in the range required for a working universe is given by
\be
P_{uni} = p_{+}{\lambda_2-\lambda_{min} \over \mplanck-\lambda_{min}}
\approx 10^{-28} p_{+}
\qquad {\rm and} \qquad
P_{log} = p_{+} {\ln(\lambda_2/\lambda_{min}) \over
\ln(\mplanck/\lambda_{min})} \approx {1\over2} p_{+}\,.
\label{lambprob} 
\ee
We again find that a model with a uniform probability distribution is
apparently highly fine-tuned, whereas a model with a log-uniform prior
is less so.  If we work in terms of energy density $\Lambda$ instead
of energy scale $\lambda$, then the first estimate decreases to
$P_{uni}\sim p_{+}10^{-112}$, but the second estimate for $P_{log}$
remains the same.  However, even if we somehow `know' that the
distribution is log-uniform, so that a working value of $\lambda$
arises with reasonable odds, we still need a solution to the
cosmological constant problem. In other words, we would like to have
a first-principles calculation that explains our observed value of
$\Lambda=\lambda^4$ with $\lambda\approx0.003$ eV (see Section
\ref{sec:lambda} and references therein). For completeness, note
that a working universe (but not our own) could have $\Lambda=0$ or
a (smaller) range of values with $\Lambda<0$.  As a result, the total
probability of obtaining a working universe is greater than the
expressions given above, e.g., $P_{log}>p_0+p_{+}/2$.

The above discussion considers only the quark masses and the
cosmological constant, whereas a full assessment requires the
probability distributions for all of the parameters listed in Table
\ref{table:parameters}. For each variable $\chi$, we need to know
the minimum and maximum values, as well as the distribution $p(\chi)$
for $\chi\in[\chi_{min},\chi_{max}]$. Additional complications arise:
A full treatment of the problem requires not only the consideration of
all of the probability distributions simultaneously, but also must
include all possible correlations between/among the variables. This
present discussion highlights some of the difficulties associated with
considering the probability distributions from which the fundamental
parameters might be selected, but a full treatment is beyond the scope
of this present work.

\medskip
{\bf Bottom Line:} {\sl In order to assess the odds of obtaining a
successful universe, one needs to consider the underlying probability
distributions, which remain unknown. Different choices for these
distributions lead to extremely different estimates for the
probabilities of obtaining a working unvierse and hence for the degree
of fine-tuning.}

\vskip1.0truecm 


\bigskip 
\noindent

\end{document}